\documentclass[fleqn,usenatbib]{mnras}

\usepackage[T1]{fontenc}
\usepackage{ae,aecompl}

\usepackage{color}
\usepackage{xcolor}
\usepackage{ulem}

\usepackage{graphicx}	
\usepackage{amsmath}	
\usepackage{amssymb}	
\usepackage{longtable}
\usepackage{lscape}

\title[Binarity and the abundance discrepancy]{Confirmation of the link between central star binarity and extreme abundance discrepancy factors in planetary nebulae}

\author[Wesson et al.]{
R. Wesson,$^{1,2}$\thanks{E-mail: rw@nebulousresearch.org}
D. Jones,$^{3,4}$
J. Garc\'ia-Rojas$^{3,4}$
H.M.J. Boffin$^{5}$
\newauthor and R.L.M. Corradi$^{6,3}$
\\
$^{1}$Department of Physics and Astronomy, University College London, Gower St, London WC1E 6BT, UK\\
$^{2}$European Southern Observatory, Alonso de C\'ordova 3107, Casilla 19001, Santiago, Chile\\
$^{3}$Instituto de Astrof\'isica de Canarias, E-38205 La Laguna, Tenerife, Spain\\
$^{4}$Departamento de Astrof\'isica, Universidad de La Laguna, E-38206 La Laguna, Tenerife, Spain\\
$^{5}$European Southern Observatory, Karl-Schwarzschild-Str. 2, 85738 Garching bei M\"unchen, Germany\\
$^{6}$Gran Telescopio Canarias S.A., c/ Cuesta de San Jos\'e s/n, Bre\~na Baja, E-38712 Santa Cruz de Tenerife, Spain
}

\date{Accepted XXX. Received YYY; in original form ZZZ}

\pubyear{2018}

\begin{document}
\label{firstpage}
\pagerange{\pageref{firstpage}--\pageref{lastpage}}
\maketitle

\begin{abstract}

It has recently been noted that there seems to be a strong correlation between planetary nebulae with close binary central stars, and highly enhanced recombination line abundances. We present new deep spectra of seven objects known to have close binary central stars, and find that the heavy element abundances derived from recombination lines exceed those from collisionally excited lines by factors of 5--95, placing several of these nebulae among the most extreme known abundance discrepancies. This study nearly doubles the number of nebulae known to have a binary central star and an extreme abundance discrepancy.

A statistical analysis of all nebulae with measured recombination line abundances reveals no link between central star surface chemistry and nebular abundance discrepancy, but a clear link between binarity and the abundance discrepancy, as well as an anticorrelation between abundance discrepancies and nebular electron densities: all nebulae with a binary central star with a period of less than 1.15~d have an abundance discrepancy factor exceeding 10, and an electron density less than $\sim$1000\,cm$^{-3}$; those with longer period binaries have abundance discrepancy factors less than 10 and much higher electron densities. We find that [O~{\sc ii}] density diagnostic lines can be strongly enhanced by recombination excitation, while [S~{\sc ii}] lines are not.

These findings give weight to the idea that extreme abundance discrepancies are caused by a nova-like eruption from the central star system, occuring soon after the common-envelope phase, which ejects material depleted in hydrogen, and enhanced in CNONe but not in third-row elements.

\end{abstract}




\section{Introduction}
\label{sec:intro}

The abundance discrepancy problem in ionized nebulae was first observed over 70 years ago, when \citet{wyse1942} found, with primitive atomic data and measurements from photographic plates, that the abundance of oxygen calculated from emission lines formed by recombination exceeded that calculated from lines formed by collisional excitation by large factors in the planetary nebulae (PNe) NGC~7009 and NGC~7027. Subsequent improvements in observational techniques and atomic data calculations confirmed that this discrepancy was real. Its magnitude is now known to be typically of the order of 2-3 in planetary nebulae (\citealt{tsamis2004}, \citealt{wesson2005}, \citealt{garcia-rojas2013}), but reaching 2--3 orders of magnitude in the most extreme cases -- e.g. 70--80 in Hf\,2--2 (\citealt{liu2006}), 150--300 in Abell~46 (\citealt{corradi2015}, and $\sim$700 in Abell~30 (\citealt{wesson2003}).

Mechanisms proposed for the discrepancy include temperature fluctuations in a chemically homogeneous gas (\citealt{peimbert1967}), abundance gradients (\citealt{torres-peimbert1990}), density fluctuations (\citealt{viegas1994}), X-ray illumination of quasi-neutral material (\citealt{ercolano2009}), hydrogen-deficient clumps (\citealt{liu2000}), non-thermal electron energy distributions (\citealt{nicholls2012}) and most recently resonant temperature fluctuations caused by varying ionization fields in nebulae with binary central stars (\citealt{bautista2017}). Strong indirect evidence for hydrogen-deficient material has been found in a number of objects, and such material is directly observed in a small number of unusual planetary nebulae such as Abell 30 (\citealt{wesson2003}) and Abell 58 (\citealt{wesson2008a}).

Surveys aimed at determining the binary fraction of PN central stars have in recent years revealed binaries inside planetary nebulae with periods of hours or days, which would have placed the companion inside the Asymptotic Giant Branch (AGB) progenitor (e.g. \citealt{miszalski2009b}, \citealt{demarco2013}, \citealt{jones2017}). These indicate that the nebula was ejected in a common envelope phase. \citealt{corradi2015} pointed out that several of the nebulae with the most extreme abundance discrepancies were post-common envelope nebulae, suggesting that the two phenomena were linked. Further support for this scenario was obtained by \citet{jones2016}, who studied the planetary nebula NGC~6778, and confirmed the prediction of a high abundance discrepancy based on its central star being a binary system with a period of 3.7~h; the measured abundance discrepancy factor ({\textit{adf}) was 18 in an integrated spectrum, and a spatial analysis showed that it peaked at over 40 close to the central star. With a clear association between binarity and extreme recombination line abundances, one can ask whether binarity is a necessary, sufficient and unique condition for an extreme abundance discrepancy.

Here we present the results of a study in which we investigate this question, by measuring new chemical abundances in a sample of nebulae known to host close binary central stars. Before the current work was carried out, nine nebulae with a confirmed short period ($<$10~d) binary central star had had their abundances determined from recombination lines, out of a total of nearly 150 nebulae for which the abundance discrepancy has been measured. 2 had a ``normal" \textit{adf}, 2 had an ``elevated" \textit{adf}, and 5 had an ``extreme" \textit{adf}\footnote{Throughout this paper, we consider an \textit{adf}$<$5 to be ``normal", a value between 5 and 10 to be ``elevated", and values greater than 10 to be ``extreme"} With studies in recent years having revealed close binaries in a number of planetary nebulae, we targeted objects known to have binary central stars, with the aim of further investigating the link between the two phenomena and building a statistically useful sample of binary objects with well-determined chemistries. We have identified six new objects that have both an elevated or extreme abundance discrepancy and a close binary central star. We describe our observations in Section~\ref{observations}, our line measurements in Section~\ref{reduction}, and abundance analysis in Section~\ref{results}.

\section{Sample selection and observations}
\label{observations}

\begin{figure*}
	\includegraphics[width=\textwidth]{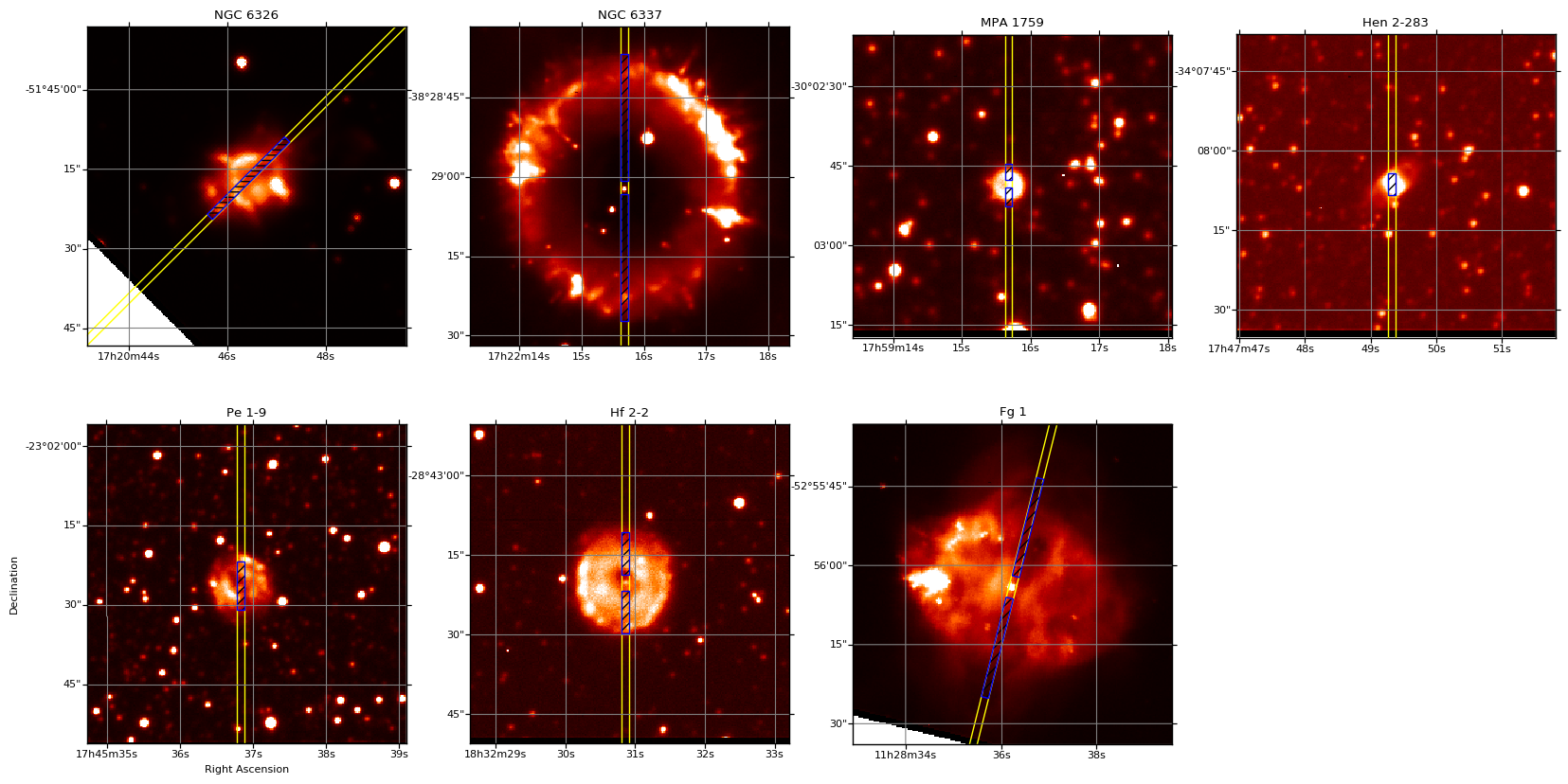}
	\caption{FORS2 H$\alpha$ images of the objects with new \textit{adf} measurements with the position of the 0.7 arcsec slit superimposed. Blue hatchings indicate the regions from which the integrated spectra were extracted. North is up and east left in all. Our slits were aligned north-south for all except for Fg~1 (-14$^{\circ}$) and NGC~6326 (-45$^{\circ}$).  The nebulae are all shown to the same spatial scale, with each image being 1 arcmin $\times$ 1 arcmin. The brightness scaling for each object was chosen to highlight the observed morphology. The white area visible in two images is the FORS2 chip gap. Top row (l to r): NGC~6326, NGC~6337, MPA~1759, Hen~2-283. Bottom row (l to r): Pe~1-9, Hf~2-2, Fg~1.}
	\label{forsimages}
\end{figure*}

We obtained observations of 42 planetary nebulae in ESO programmes 093.D-0038(A) and 096.D-0080(A). The nebulae in our sample were southern hemisphere objects selected from a list of PNe with close binary central stars\footnote{\url{http://www.drdjones.net/bCSPN}}. The sample included Hf~2-2, already known to exhibit an extreme abundance discrepancy (Liu et al. 2006), both as an opportunity to check our methodology and the quality of our observations, and to permit a spatially resolved study of the object. For all the other objects in the sample, there were no previously published recombination line abundances.

The observations were carried out during service time with the FORS2 instrument mounted on the ESO VLT's UT1 Antu telescope \citep{appenzeller1998}. The programmes were designated as filler programmes, designed to be performed under any conditions, including periods of bad weather (seeing greater than 2" and/or clouds). Because of this, a number of our targets were observed during times of poor seeing and transparency, and nebular emission was sometimes only very weakly detected. We detected strong recombination line emission in 8 objects, while in 20 objects, the spectra were deep enough to obtain collisionally excited line abundances but not recombination line abundances. In 14 objects, no abundances could be measured. The effect of sub-optimal conditions on our observations is further discussed in Section~\ref{results}.

The instrument set-up consisted of a long slit measuring 0.7\arcsec $\times$6.8\arcmin{}, and a mosaic of two 4k$\times$2k MIT/LL CCDs binned 2$\times$2 ($\equiv$0.25~arcsec per pixel). The slit was orientated along the major axis of the nebula, where images were available to determine this, and north-south otherwise. For all objects, a single exposure was taken using the GRIS\_1200B grism, followed by a further exposure with the GRIS\_1200R grism and GG435 filter (in the case of those objects observed during period P93 as part of programme 093.D-0038(A)) or the GRIS\_600RI and GG435 filter (in the case of objects observed during period P96 as part of programme 096.D-0080(A)), resulting in a resolution of 1.5--3\AA{} across the observed wavelength range of $\sim$3600--5000\AA{} and $\sim$5800--7200\AA{}/5010--8300\AA{} (GRIS\_1200R/GRIS\_600RI), for the blue and red grisms respectively. A log of the observations is shown in Table \ref{tab:obs}.

\section{Data reduction and analysis}
\label{reduction}
The two-dimensional spectra were bias subtracted, wavelength calibrated and flux calibrated against exposures of standard stars taken using the same set-up (as part of the standard ESO calibration plan), all using standard \textsc{starlink} routines \citep{shortridge2014,warren-smith2014}. Cosmic ray removal was achieved using a combination of \textsc{starlink}'s figaro routines and a python implementation of the LAcosmic algorithm \citep{vandokkum2001}. The spectra were then sky subtracted and extracted to one-dimensional spectra. The nebular windows for the extraction are indicated in Figure~\ref{forsimages}; we excluded a region around the central star when the star was bright enough to potentially contaminate the nebular spectrum.

Emission line fluxes were measured using the automated line-fitting algorithm, {\sc alfa} \citep{wesson2016a}, which has been used by \citet{jones2016} and \citet{sowicka2017} to perform similar analyses for the high \textit{adf} PN NGC~6778 and low \textit{adf} PN IC~4776 respectively. {\sc alfa} derives fluxes by optimizing the parameters of Gaussian fits to line profiles using a genetic algorithm, after subtracting a globally fitted continuum. The observed and dereddened fluxes for each object, along with their 1$\sigma$ uncertainties, are given in the Appendix in section \ref{sec:tables}. {\sc alfa} calculates uncertainties from the RMS of the residuals after subtracting the fitted lines and continuum from the observations, and does not include any contribution from systematic uncertainties, such as those originating from the flux calibration. As such, the quoted uncertainties are lower limits. {\sc alfa} calculates the spectral resolution, and reports lines as blended if they are separated by less than the half width at half-maximum at the calculated resolution. Figure~\ref{orlspectra} shows the fits for the objects with prominent recombination line spectra, in the region around 4650{\AA} where there are numerous recombination lines. The complete list of measured emission lines for each object is given in Appendix A.

\begin{figure*}
\includegraphics[width=0.85\textwidth]{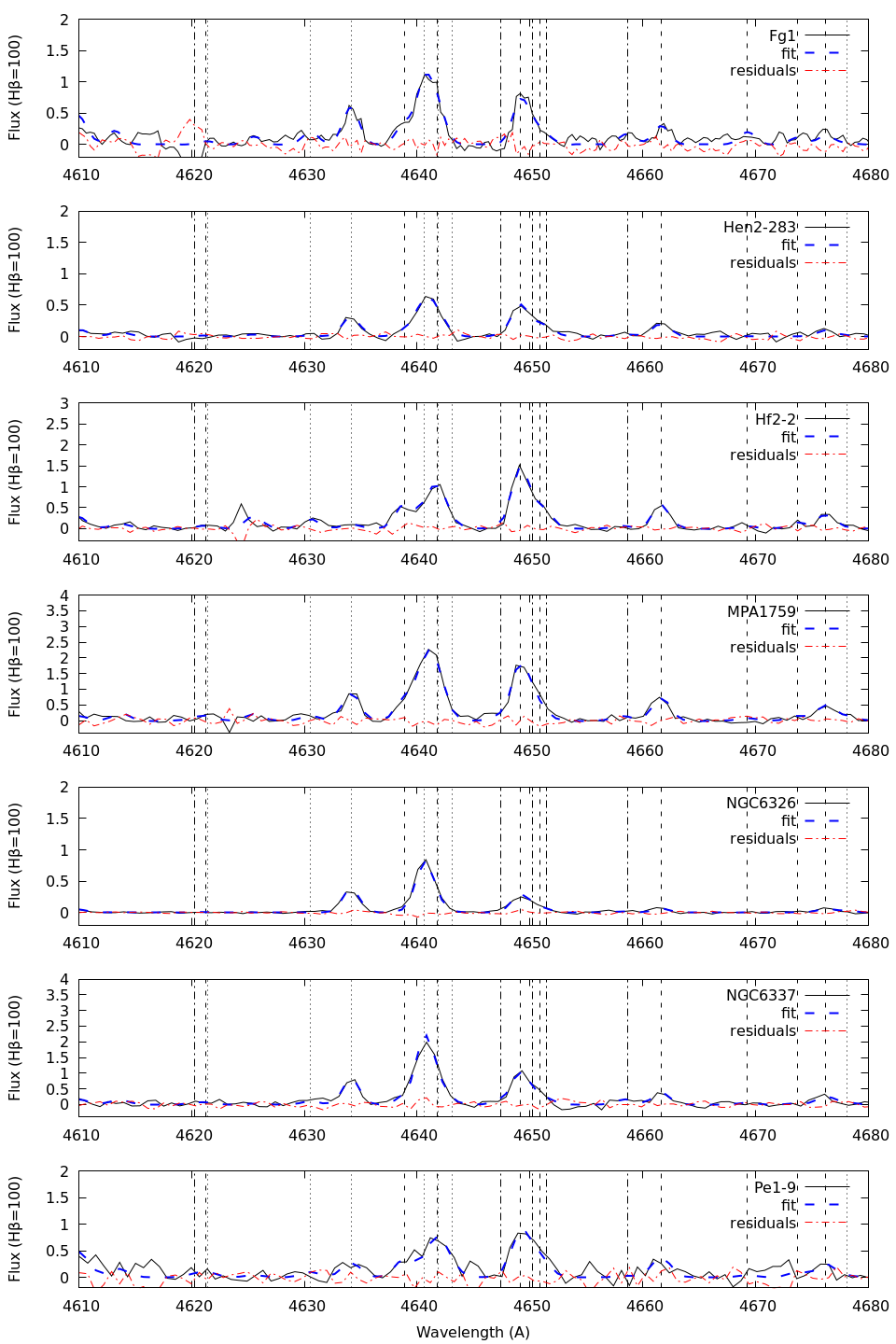}
\caption{Spectra of the seven extreme objects in our sample, covering the region around 4650{\AA} which contains recombination lines due to O~{\sc ii}, N~{\sc ii} and C~{\sc iii}. The scale is such that the integrated flux of H$\beta$=100. The observed spectra, the {\sc alfa} fits and the residuals are plotted. Vertical lines indicate atomic species: dotted for O~{\sc ii}, dashed for N~{\sc ii} and dot-dashed for C~{\sc ii}.}
\label{orlspectra}
\end{figure*}

\section{Physical conditions and abundances}
\label{results}

The code {\sc neat} \citep[Nebular Empirical Analysis Tool;][]{wesson2012} was then used to derive final ionic and elemental abundances from these emission line fluxes. {\sc neat} corrects for interstellar extinction using the flux-weighted ratios of H$\gamma$ and H$\delta$ to H$\beta$ and the Galactic extinction law of \citet{howarth1983}. The H$\alpha$/H$\beta$ ratio was not used to calculate the extinction but rather as a check to ensure that line fluxes measured from the non-overlapping red and blue spectra were consistent. In a few objects, we scaled the red spectrum such that the extinction derived from H$\alpha$ was consistent with the values from H$\gamma$ and H$\delta$. The corrections applied were up to 22\%. The necessity for such corrections may arise from several factors. First, for many of our observations, the object was reacquired between red and blue exposures, resulting in a small shift in the position of the slit (see \citealt{jones2016} for details of this in relation to NGC\,6778). Second, due to the filler nature of our programme, many targets were observed in poor conditions. This may have given rise to variation in the sky transparency between red and blue spectra, while changing seeing conditions could also result in red and blue spectra experiencing differing degrees of slit losses. For example, Hf\,2-2 was observed on the night of 2014-06-19, with the blue spectrum starting at 07:08:22UT, and the red spectrum starting at 07:29:54UT. Archived data for Paranal from ESO's Astronomical Site Monitor\footnote{\url{http://www.eso.org/asm}} show that the seeing during the blue exposure varied between 1.1 and 1.8 arcsec, while during the red exposure it exceeded 2 arcsec. Although these measurements overestimate the actual image quality\footnote{Seeing measurements at Paranal are recorded by a Differential Image Motion Monitor (DIMM), which measures a larger seeing than measured by the VLT Unit Telescopes, due to a turbulent surface layer at the observatory site that the DIMM is exposed to but the UTs are not (\citealt{sarazin2008})}, they indicate that variations in slit losses between red and blue exposures are very likely to have occurred.

This rescaling has only a small effect on our results and does not affect any of our conclusions. Temperatures derived from He~{\sc i} lines depend on the ratio of lines in the red spectra to lines in the blue, while abundances of N$^+$, S$^{+}$ and Ar$^{2+}$ are derived from lines in the red spectra only. These measurements are thus affected by the scaling, which may introduce an additional uncertainty into their values.

We determined the physical conditions in our sample nebulae from traditional collisionally excited line diagnostics, from He~{\sc i} line ratios, and from O~{\sc ii} recombination line ratios. For most of our nebulae we were not able to estimate a temperature from the Balmer jump, as it lies very close to the blue end of our wavelength coverage and is subject to large flux calibration uncertainties. However, it was well enough detected in both Hf~2-2 and NGC~6326 for a temperature to be estimated. In both objects, we find very low temperatures. For Hf~2-2, we find T$_{BJ}$=840$^{+330}_{-250}$\,K, in excellent agreement with the value of 890\,K reported by \citet{liu2006}.

For the O~{\sc ii} recombination line temperatures, we also show their values in Figure~\ref{figure:oii_rl_temperature_density}. Temperatures and densities were derived using the data of \citet{storey2017}, and the figure shows the locuses of possible values of the ratios of line pairs $\lambda$4649/$\lambda$4089 and $\lambda$4649/$\lambda$4662, for given combinations of temperatures and densities. Similarly to \citet{storey2017}, we find that some of our objects have line ratios implying temperatures below the lowest value for which the atomic data are available. Hen~2-283 in particular has ratios that appear spuriously low, perhaps indicating that the $\lambda$4089 line is overestimated. This may simply be due to noise: this object has the lowest \textit{adf} and thus the weakest recombination lines. It has also been suggested that a [Si~{\sc iv}] line coincident with the O~{\sc ii} line could result in an overestimate of the O~{\sc ii} flux (\citealt{peimbert2013}), although in that case, [Si~{\sc iv}] emission should also be present at 4116 {\AA} with twice the flux of the $\lambda$4089 emission. We see no feature at this wavelength and so we attribute the extremely low ratio in Hen 2-283 to noise alone.

\begin{figure}
\includegraphics[width=0.47\textwidth]{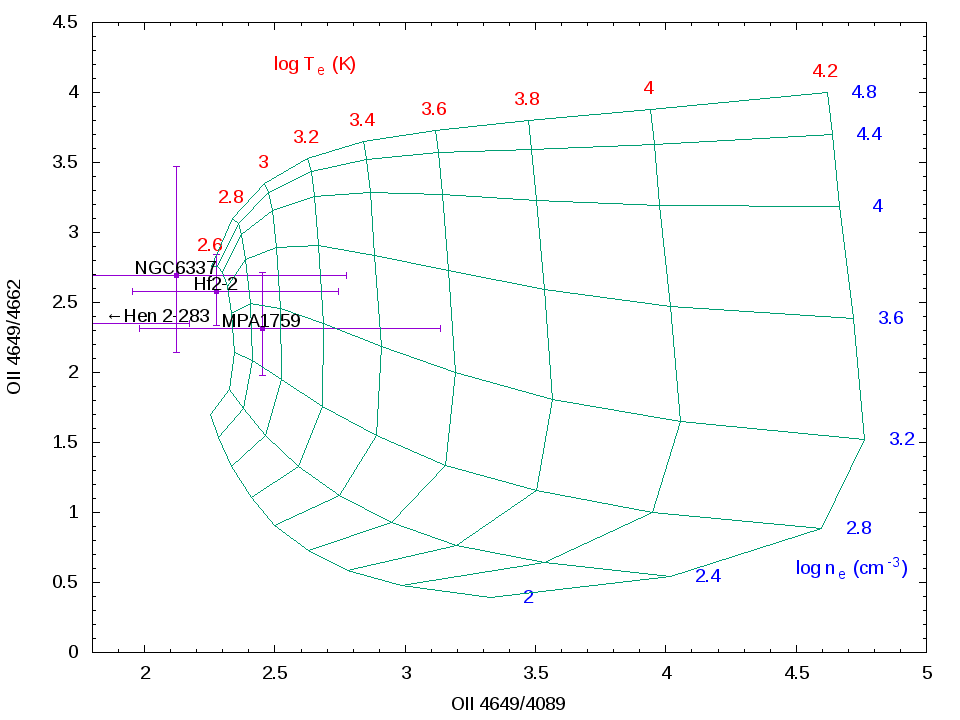}
\caption{Temperatures and densities derived from O~{\sc ii} recombination line ratio pairs, using atomic data from \citet{storey2017}.}
\label{figure:oii_rl_temperature_density}
\end{figure}

Abundances were derived using {\sc neat} version 2.1. {\sc neat} version 1.0's abundance analysis scheme was described in \citet{wesson2012}; updates since then have extended the analysis to report many new diagnostics and abundances where the relevant lines are available, including the O~{\sc ii} temperature and density diagnostics shown in Figure~\ref{figure:oii_rl_temperature_density}. Changes to the original analysis scheme other than extensions are: the incorporation of the ionization correction scheme of \cite{delgado-inglada2014} as the default scheme, replacing \citet{kingsburgh1994}; the incorporation of He~{\sc i} recombination coefficients from \cite{porter2012} and \cite{porter2013} as default; and a change to the default abundance weighting scheme. Originally, recombination line abundances for O$^{2+}$ and N$^{2+}$ were determined using the flux-weighted average of lines within a multiplet to determine the abundance for each multiplet, with the final ionic abundance being the equally weighted average of all multiplet abundances. In the latest version of {\sc neat}, the multiplet abundances are instead weighted by the co-added intensity of the multiplet. {\sc neat} allows the user to set any weight where the default is not suitable; we set the weight of the V5 multiplet of O$^{2+}$ to zero as it frequently gives higher abundances than other well detected multiplets. Line flux uncertainties are propagated through the analysis into the derived quantities.

{\sc neat} calculates abundances using a three-zone model; lines from ions with ionization potential (IP) $<$20\,eV are assigned to the low ionization zone, those with 20\,eV$<$IP$<$45\,eV to the medium ionization zone, and those with IP$>$45\,eV to the high ionization zone. As our spectral coverage does not include the [N~{\sc ii}] $\lambda$5754 line, and no high ionization zone temperature diagnostics are observed, the medium ionization zone temperature derived from [O~{\sc iii}] (4959+5007)/4363 is also used for the low and high  ionization zones. The atomic data used in {\sc neat} v2.1 is listed in Table~\ref{atomicdatatable}, while Table~\ref{neatzones} shows the diagnostics used in each zone, and the ions for which they are used. The electron temperatures and densities derived for the sample objects are given in Table~\ref{diagnosticstable}, while the abundances are given in Table~\ref{abundancestable}.

\begin{table}
\begin{tabular}{lll}
\hline
Ion & Transition probabilities & Collision strengths \\
\hline
O$^{+}$ & \citet{zeippen1982} & \citet{pradhan1976} \\
S$^{2+}$ & \citet{mendoza1982a} & \citet{mendoza1983} \\
\end{tabular}
\begin{tabular}{ll}
\hline
Ion & Recombination coefficients \\
\hline
H$^+$ & \citet{storey1995} \\
He$^{+}$ & \citet{porter2012,porter2013} \\
He$^{2+}$ & \citet{storey1995} \\
C$^{2+}$ & \citet{davey2000} \\
C$^{3+}$ & \citet{pequignot1991} \\
N$^{2+}$ & \citet{escalante1990} \\
N$^{3+}$ & \citet{pequignot1991} \\
O$^{2+}$ (3s--3p) & \citet{storey1994} \\
O$^{2+}$ (3p--3d, 3d--4f) & \citet{liu1995} \\
Ne$^{2+}$ (3s--3p)& \citet{kisielius1998} \\
Ne$^{2+}$ (3d-4f) & Storey (unpublished) \\
\hline
\end{tabular}
\caption{Atomic data used in {\sc neat}. For collisionally excited lines, collision strengths and transition probabilities for ions not listed in this table were taken from version 7.1 of the {\sc chianti} database (\citealt{dere1997}, \citealt{landi2012})}
\label{atomicdatatable}
\end{table}

\begin{table*}
	\begin{tabular}{llll}
		\hline
		Zone & Temperature diagnostics & Density diagnostics & Ions \\
		\hline
		Low    & [O~{\sc iii}] (4959+5007)/4363 & [O~{\sc ii}] 3727/3729 & N$^+$, O$^+$, S$^+$ \\
		       &                                & [S~{\sc ii}] 6717/6731 & \\
		Medium & [O~{\sc iii}] (4959+5007)/4363 & [Ar~{\sc iv}] 4711/4740 & C$^{2+}$, O$^{2+}$, Ne$^{2+}$, Ar$^{2+}$, Ar$^{3+}$, S$^{2+}$ \\
		High   & \multicolumn{3}{c}{(none observed)} \\
		\hline
	\end{tabular}
	\caption{The three-zone model applied by {\sc neat}; no high-ionization zone diagnostics or ions are observed in our sample.}
	\label{neatzones}
\end{table*}

\subsection{Recombination line abundances and abundance discrepancies}

We have detected strong heavy element recombination line emission in Fg~1, Hen~2-283, Hf~2--2, MPA~J1759-3007 (abbreviated hereafter as MPA~1759), NGC~6326, NGC~6337, and Pe~1-9. In Figure~\ref{orlspectra}, we show the 4650 {\AA} region of the spectra of each of these objects, on a scale where F(H$\beta$)=100, with the {\sc alfa} fit and residuals also shown.

{\sc alfa} identified between 11 and 33 lines of O~{\sc ii} in these seven objects. Lines of multiplets V1, V2, V10, V20 and 3d--4f transitions generally give abundances in good agreement with each other in good-quality spectra; see for example \citet{garcia-rojas2013}. The V1 and V10 multiplets are the best detected and thus received the most weight in our abundance determinations. Other 3-3 transitions and a few 3d-4f lines are detected in the deeper spectra, and are also used for the abundance calculation but with lower weight according to their observed flux.

Where RLs are detected, we obtain the \textit{adf} for O$^{2+}$/$H^+$ and O/H. In three objects, we can also estimate the \textit{adf} for N/H, although the uncertainty is always large. N$^+$/H$^+$ can only be derived from CELs, while N$^{2+}$ can only be derived from RLs. RLs of N$^{3+}$ are also detected in some cases but cannot be used for abundance measurements as they are excited by continuum fluorescence as well as recombination (\citealt{ferland1992}). Thus, no \textit{adf} can be derived for any single ion of nitrogen, and the total abundance comparison relies on the ionization correction, which is very uncertain for CELs as only a small fraction of N is in the form of N$^+$. Fluorescence excitation can also be significant for N$^{+}$ and N$^{2+}$ lines in very low excitation nebulae, such as IC\,418 (\citealt{escalante2012}). However, this is not likely to be the case for the nebulae observed here; He$^{2+}$ emission is not detected in very deep spectra of IC\,418 where lines with fluxes of 10$^{-5}\times$F(H$\beta$) are observed (\citealt{sharpee2003}), whereas He$^{2+}$ $\lambda$4686 has a flux of at least 0.03$\times$F(H$\beta$) in all our nebulae.

The \textit{adfs} we derive from the seven objects are all high. The lowest is that of Hen\,2-283, which at a factor of $\sim$6 is higher than at least 80\% of PNe but far lower than the most extreme objects. For the other six objects, the \textit{adfs} are between 18 and 95, all counting as among the highest known.

\begin{table*}
\begin{tabular}{llllllll}
\hline
Diagnostic                       & Fg~1                   & Hen~2-283                 & Hf~2-2                    & MPA~1759                  & NGC~6326                  & NGC~6337                  & Pe~1-9 \\
\hline
\multicolumn{8}{l}{\textbf{Density diagnostics}}\\
{}[O~\sc ii] 3727/3729           & ${780}^{+700}_{-420}$  &  ${3660}^{+2600}_{-1220}$ &  ${1010}^{+480}_{-340}$   &  ${750}^{+1120}_{-530}$   &  ${880}^{+200}_{-170}$    &  ${520}^{+360}_{-250}$    &  ${1060}^{+1210}_{-600}$  \\
{}[Ar~\sc iv] 4711/4740          & ${600}^{+670}_{-590}$  &  ${3150}^{+1300}_{-1070}$ &  --                       &  ${800}^{+1220}_{-800}$   &  ${550}\pm{220}$          &  ${450}\pm{230}$          &  --                       \\
{}[S~\sc ii] 6717/6731           & ${290}^{+150}_{-120}$  &  ${3550}^{+340}_{-310}$   &  ${290}^{+150}_{-130}$    &  ${710}^{+620}_{-360}$    &  ${750}\pm{30}$           &  ${330}^{+80}_{-70}$      &  ${410}^{+110}_{-80}$     \\
O~{\sc ii} 4649/4089,4649/4662        & --                     &  --                       &  ${1250}^{+1500}_{-1250}$ &  ${1200}^{+2000}_{-500}$   &  --                       &  --                       &  --                       \\
\multicolumn{8}{l}{\textbf{Temperature diagnostics}}\\
{}[O~{\sc iii}] (4959+5007)/4363 &  ${11000}\pm{300}$     &  ${8540}\pm{100}$         &  ${9870}\pm{490}$         &  ${11400}\pm{200}$        &  ${14600}\pm{100}$        &  ${12000}\pm{200}$        &  $<$8700                  \\
Balmer Jump                      & $<$6000                &  $<$810                   &  ${840}^{+330}_{-250}$    &  --                       &  ${40}^{+220}_{-40}$      &  $<$950                   &  --                       \\
He~{\sc i} 5876/4471             & ${220}^{+890}_{-220}$  &  ${2760}^{+1060}_{-790}$  &  ${3050}^{+1480}_{-1060}$ &  ${3050}^{+2050}_{-3050}$ &  $<$6450                  &  ${2670}^{+1780}_{-1300}$ &  ${1420}^{+1460}_{-1000}$ \\
He~{\sc i} 6678/4471             & ${360}^{+2030}_{-360}$ &  ${2940}^{+1500}_{-1000}$ &  ${2650}^{+1890}_{-1030}$ &  ${7450}^{+4830}_{-3780}$ &  ${7370}^{+1700}_{-1380}$ &  ${2670}^{+3070}_{-1320}$ &  ${1640}^{+2020}_{-1180}$ \\
O~{\sc ii} 4649/4089,4649/4662        & --                     &  $<$400                   &  $<$2000                  &  $<$3500                  &  --                       &  $<$2000                  &  --                       \\
\hline
\end{tabular}
\caption{Estimates of the electron density and temperature from all available diagnostics in the sample objects}
\label{diagnosticstable}
\end{table*}

\begin{table*}
\begin{tabular}{llllllll}
\hline
Abundance & Fg~1 & Hen~2-283 & Hf~2-2 & MPA~1759 & NGC~6326 & NGC~6337 & Pe~1-9 \\
\hline
\multicolumn{8}{l}{\textbf{CEL abundances}}\\
N$^{+}$/H&6.26$\pm$0.05&7.41$\pm$0.02&6.72$^{+0.07}_{-0.06}$&6.16$\pm$0.04&6.61$\pm$0.02&7.03$^{+0.06}_{-0.04}$&7.10$\pm$0.01\\
icf(N) &15.00$^{+2.00}_{-1.80}$&14.60$^{+1.10}_{-1.40}$&7.30$^{+0.64}_{-0.58}$&23.00$^{+4.40}_{-3.40}$&8.52$^{+0.32}_{-0.31}$&23.50$^{+2.20}_{-2.00}$&5.87$^{+0.84}_{-0.67}$\\
N/H    &7.44$\pm$0.05&8.57$^{+0.03}_{-0.04}$&7.59$\pm$0.07&7.52$\pm$0.09&7.54$\pm$0.03&8.40$\pm$0.05&7.87$^{+0.06}_{-0.05}$\\
O$^{+}$/H&7.33$\pm$0.11&7.70$^{+0.05}_{-0.04}$&7.20$^{+0.11}_{-0.09}$&6.83$\pm$0.08&7.08$\pm$0.02&7.31$^{+0.10}_{-0.08}$&7.53$^{+0.06}_{-0.07}$\\
O$^{2+}$/H&8.46$\pm$0.06&8.71$\pm$0.02&7.90$^{+0.09}_{-0.07}$&8.16$\pm$0.03&7.92$\pm$0.02&8.57$^{+0.07}_{-0.06}$&8.11$\pm$0.00\\
icf(O) &1.13$\pm$0.01&1.01$\pm$0.01&1.01$\pm$0.01&1.17$\pm$0.01&1.28$\pm$0.01&1.54$\pm$0.02&1.01$\pm$0.01\\
O/H    &8.54$^{+0.07}_{-0.06}$&8.76$\pm$0.02&7.98$^{+0.09}_{-0.08}$&8.25$\pm$0.03&8.08$\pm$0.02&8.78$^{+0.07}_{-0.06}$&8.22$\pm$0.01\\
Ne$^{2+}$/H&8.08$^{+0.08}_{-0.07}$&8.22$\pm$0.02&7.55$^{+0.10}_{-0.09}$&7.69$\pm$0.03&7.59$\pm$0.02&8.03$^{+0.09}_{-0.07}$&7.79$\pm$0.03\\
icf(Ne)&1.13$\pm$0.01&1.13$^{+0.03}_{-0.02}$&1.08$\pm$0.01&1.17$\pm$0.01&1.29$\pm$0.01&1.54$\pm$0.02&1.12$^{+0.03}_{-0.02}$\\
Ne/H   &8.13$^{+0.08}_{-0.07}$&8.27$\pm$0.03&7.58$^{+0.11}_{-0.09}$&7.76$\pm$0.03&7.70$\pm$0.02&8.22$^{+0.09}_{-0.07}$&7.84$\pm$0.03\\
Ar$^{2+}$/H&6.04$\pm$0.05&6.48$\pm$0.01& -- &5.89$\pm$0.04&5.68$\pm$0.02&6.42$^{+0.05}_{-0.04}$&6.00$^{+0.02}_{-0.03}$\\
Ar$^{3+}$/H&6.14$\pm$0.06&5.97$\pm$0.03&5.34$\pm$0.13&5.71$\pm$0.04&5.60$\pm$0.01&6.48$^{+0.07}_{-0.06}$&5.53$^{+0.10}_{-0.12}$\\
icf(Ar)&1.00$\pm$0.00&1.13$\pm$0.02&1.00$\pm$0.00&1.12$\pm$0.05&1.00$\pm$0.00&1.01$^{+0.04}_{-0.01}$&1.00$^{+0.01}_{-0.00}$\\
Ar/H   &6.39$\pm$0.05&6.64$^{+0.01}_{-0.02}$&5.34$\pm$0.13&6.16$\pm$0.04&5.97$\pm$0.02&6.81$^{+0.06}_{-0.04}$&6.13$\pm$0.03\\
S$^{+}$/H&4.99$\pm$0.06&6.06$^{+0.06}_{-0.04}$&5.35$^{+0.08}_{-0.07}$&4.94$^{+0.07}_{-0.06}$&5.41$\pm$0.02&5.81$^{+0.06}_{-0.05}$&5.78$^{+0.06}_{-0.03}$\\
S$^{2+}$/H& -- & -- & -- & -- & -- & -- & -- \\
icf(S) &27.10$^{+3.70}_{-3.20}$&22.70$^{+1.70}_{-2.10}$&12.04$^{+0.99}_{-0.92}$&40.70$^{+7.80}_{-6.10}$&13.93$^{+0.56}_{-0.54}$&34.80$^{+3.40}_{-3.10}$&9.65$^{+1.31}_{-1.05}$\\
S/H    &6.42$^{+0.07}_{-0.06}$&7.42$\pm$0.03&6.43$\pm$0.07&6.56$^{+0.10}_{-0.09}$&6.55$\pm$0.03&7.35$\pm$0.05&6.78$^{+0.06}_{-0.05}$\\
\multicolumn{8}{l}{\textbf{RL abundances}}\\
He$^{+}$/H&11.07$\pm$0.01&11.14$\pm$0.01&11.22$\pm$0.01&11.12$^{+0.03}_{-0.02}$&10.85$^{+0.02}_{-0.01}$&10.94$^{+0.01}_{-0.02}$&11.17$\pm$0.01\\
He$^{2+}$/H&10.43$\pm$0.02&9.40$^{+0.02}_{-0.03}$&9.62$\pm$0.01&10.61$\pm$0.01&10.55$\pm$0.00&10.90$\pm$0.00&9.54$\pm$0.03\\
He/H   &11.16$^{+0.01}_{-0.02}$&11.15$\pm$0.01&11.23$\pm$0.01&11.24$\pm$0.02&11.03$\pm$0.01&11.22$\pm$0.01&11.18$\pm$0.01\\
C$^{2+}$/H&9.23$\pm$0.05&9.27$\pm$0.04&9.89$\pm$0.02&9.85$\pm$0.03&8.71$\pm$0.04&9.35$\pm$0.04&9.75$\pm$0.04\\
C$^{3+}$/H&8.61$^{+0.08}_{-0.09}$&8.21$^{+0.12}_{-0.17}$&8.43$^{+0.11}_{-0.15}$& -- & -- & -- & -- \\
icf(C) &1.10$\pm$0.05&1.10$\pm$0.03&1.15$\pm$0.01&1.44$\pm$0.01&1.54$\pm$0.01&1.85$\pm$0.02&1.19$\pm$0.00\\
C/H    &9.36$\pm$0.05&9.35$\pm$0.04&9.96$\pm$0.02&10.01$\pm$0.03&8.89$\pm$0.04&9.62$\pm$0.04&9.83$^{+0.03}_{-0.04}$\\
N$^{2+}$/H&9.90$^{+0.10}_{-0.12}$& -- &10.53$^{+0.08}_{-0.10}$& -- & -- & -- & -- \\
N$^{3+}$/H&8.47$\pm$0.02&8.31$\pm$0.03&8.03$^{+0.08}_{-0.10}$&8.79$^{+0.03}_{-0.04}$&8.43$\pm$0.01&8.81$\pm$0.02& -- \\
icf(N) &1.00$\pm$0.00&1.00$\pm$0.00&1.00$\pm$0.00&1.00$\pm$0.00&1.00$\pm$0.00&1.00$\pm$0.00&1.00$\pm$0.00\\
N/H    &9.95$^{+0.09}_{-0.11}$&8.31$\pm$0.03&10.53$^{+0.08}_{-0.10}$&8.79$^{+0.03}_{-0.04}$&8.43$\pm$0.01&8.81$\pm$0.02& -- \\
O$^{2+}$/H&10.11$\pm$0.03&9.42$\pm$0.04&9.85$\pm$0.03&9.96$\pm$0.03&9.28$\pm$0.06&9.83$\pm$0.05&9.90$^{+0.10}_{-0.06}$\\
icf(O) &1.23$^{+0.02}_{-0.01}$&1.11$\pm$0.01&1.22$\pm$0.02&1.25$\pm$0.02&1.50$\pm$0.02&1.62$\pm$0.02&1.28$\pm$0.04\\
O/H    &10.20$^{+0.05}_{-0.06}$&9.47$^{+0.03}_{-0.04}$&9.94$\pm$0.03&10.06$\pm$0.03&9.46$\pm$0.06&10.04$\pm$0.04&10.00$^{+0.08}_{-0.10}$\\
Ne$^{2+}$/H& -- & -- & -- & -- &9.28$^{+0.12}_{-0.17}$& -- & -- \\
icf(Ne)&1.23$^{+0.02}_{-0.01}$&1.11$\pm$0.01&1.22$\pm$0.02&1.25$\pm$0.02&1.50$\pm$0.02&1.62$\pm$0.02&1.28$\pm$0.04\\
Ne/H   & -- & -- & -- & -- &9.45$^{+0.13}_{-0.17}$& -- & -- \\
\hline
\end{tabular}
\caption{Ionic and total abundances for the sample objects. Abundances are given on a scale where log(H)=12.}
\label{abundancestable}
\end{table*}

\subsection{Nebulae with extreme abundance discrepancies}
\label{objectssection}

FORS2 H$\alpha$ images of our seven newly observed nebulae are shown in Figure~\ref{forsimages}. We discuss the analysis of the individual objects below.

\subsubsection{Fg 1}

Fg~1 (PN~G290.5+07.9) was shown by \citet{boffin2012a} to host a double-degenerate central binary system with an orbital period of 1.20~d, while the nebula is held as an archetype of the influence of a central binary star with an expanding equatorial ring and bipolar, rotating episodic jet ejections \citep{lopez1993a,palmer1996}.

The value of c(H$\beta$) calculated from H$\alpha$/H$\beta$ is higher than the values obtained from H$\gamma$/H$\beta$ and H$\delta$/H$\beta$; the latter two ratios are in excellent agreement with each other. As H$\alpha$ was observed using a different grating to the other three lines, we adopt c(H$\beta$)=0.52$\pm$0.18 from the three blue lines, and scale the red spectrum by a factor of 0.85 such that the H$\alpha$ line flux gives the same extinction.

All density diagnostics are in good agreement, indicating an electron density of around 500\,cm$^{-3}$. The temperature from the [O~{\sc iii}] ($\lambda\lambda$4959+5007)/$\lambda$4363 line ratio is 11\,000$\pm$300\,K. The Balmer jump is detected but with a large uncertainty due to being close to the edge of our spectral coverage; it yields an upper limit to the temperature of about 6\,000\,K. He~{\sc i} line ratios are consistent with much lower temperatures of $<$1kK.
~\\~\\
The extremely strong recombination line emission in this object and its high excitation -- I(He$^{2+}$ $\lambda$4686)$\sim$ 0.35$\times$I(H$\beta$) -- means that recombination contributes significantly to the flux of the temperature-sensitive [O~{\sc iii}] $\lambda$4363 line. Neglecting this effect, the abundance discrepancy is calculated to be a factor of 80. Using Equation~3 from \citet{liu2006b}, and taking the value of O$^2+$/H$^+$H derived from recombination lines, we find that the recombination contribution is 40\% of the total flux of the line. Subtracting this, the electron temperature is revised to 9430$\pm$290\,K, yielding an abundance discrepancy factor for O$^{2+}$ of 46$^{+10}_{-8}$. 

The abundance discrepancy for N/H may also be extremely high; the abundance measured for N/H from recombination lines exceeds that for collisionally excited lines by a factor of nearly 150. However, the value of N/H for recombination lines is determined from a single line of N~{\sc ii} and thus could be subject to systematic effects such as line blending, imperfect cosmic ray removal and hot pixels.

\subsubsection{Hen~2-283}
\label{subsec:Hen2-283}
Hen~2-283 (PN~G355.7-03.0) is a canonical bipolar nebula shown to host a 1.13~d period central binary star showing photometric variability due to the irradiation of a main-sequence companion \citep{miszalski2009b,miszalski2009a}.

No scaling is applied to the red spectra in this object as c(H$\beta$) values from all Balmer lines are consistent, giving a mean c(H$\beta$) of 1.48$\pm$0.04. Diagnostics of electron density yield values in excellent agreement with each other: ${  3550}^{+   340}_{  -310}$\,cm$^{-3}$ from [S~{\sc ii}], ${  3660}^{+  2600}_{ -1220}$\,cm$^{-3}$ from [O~{\sc ii}], and ${  3150}^{+  1300}_{ -1070}$\,cm$^{-3}$ from [Ar~{\sc iv}].

We could not measure a Balmer jump temperature in this object. He~{\sc i} lines, though, are well detected and indicate very low temperatures: He~{\sc i} $\lambda$5876/$\lambda$4471 line ratio gives T$_e$=${  2760}^{+  1060}_{  -790}$\,K, while He~{\sc i} $\lambda$6678/$\lambda$4471 gives T$_e$=${  2940}^{+  1500}_{ -1000}$\,K. O~{\sc ii} lines give low upper limits to the temperature and the density, implying T$_e$<700\,K. In contrast to Fg~1, the recombination contribution to the $\lambda$4363 line is negligible.

This object has the lowest measured \textit{adf}(O/H) of our sample at 5.1$\pm$0.5. Recombination lines of N$^{2+}$ are not detected. We note that the orbital period of its binary lies very close to the apparent threshold of $\sim$1.15~d approximately separating extreme from non-extreme \textit{adfs} (Section~\ref{sec:binaryperiod}). It also has a markedly higher electron density than the rest of the sample; the mean and standard deviation of their densities is 650$\pm$120\,cm$^{-3}$, compared to 3430\,cm$^{-3}$ in this object. And it has the highest O/H and N/H abundances of the sample. This is discussed further below.

\subsubsection{Hf~2-2}
\label{subsec:Hf2-2}
Hf~2-2 (PN~G005.1-08.9) has previously been shown to present an extremely high \textit{adf} \citep[$\sim$70-80,][]{liu2006a,mcnabb2016} and enters our sample as a control, in order to verify our methodology and compare our results with the previous analysis. The central star of Hf~2-2 was shown to be a post-CE binary with a period of 0.40~d by \citep{bond2000}, where the form of the light curve indicates variability due to an irradiated main-sequence secondary. Deep narrow-band imagery of Hf~2-2 displays a roughly circular profile consistent with an apparently double-shelled structure \citep{miszalski2009a}. Unpublished morpho-kinematical observations confirm this multiple-shelled structure with both shells likely having a nearly pole-on elliptical morphology (Jones et al. in preparation), similar to that of the post-CE PN Abell~65 \citep{huckvale2013} and those predicted by the hydrodynamic models of \citet{garcia-segura2018}.

The extinction we derive towards Hf~2-2 is higher from H$\alpha$/H$\beta$ than from other Balmer line ratios, which yield more consistent values of c(H$\beta$). We thus rescale line fluxes measured from the red spectrum by $\times$0.95, which results in consistent values from H$\alpha$-$\delta$ line fluxes. The weighted mean extinction is then c(H$\beta$)=0.33$\pm$0.02, slightly higher than the value of 0.20 found by \citet{liu2006}. Their slit was aligned at 45$^{\circ}$ east from north, while for our observations, the slit was aligned N-S.

From the integrated spectrum of Hf~2-2, we derive T$_e$([O~{\sc iii}])=9870$\pm$490\,K, and an \textit{adf} for O$^{2+}$ of 95$\pm$15. This is in reasonably good agreement with the values found by \citet{liu2006a} of 8740\,K and a factor of 70, and by \citet{mcnabb2016} of 9650\,K and a factor of 80. A correction to our measured $\lambda$4363 line may be necessary, as it may be significantly enhanced by recombination; taking O/H from the recombination line value, and using Equation 3 from \citet{liu2000}, we find that 11\% of the line flux is due to recombination. Subtracting this, the measured temperature is 9410\,K, and \textit{adf}(O/H) is reduced somewhat to 83$\pm$15.

Electron temperature estimates from several recombination line diagnostics are available in Hf~2-2, thanks to the strength of its recombination line spectrum. The hydrogen Balmer jump is well detected and gives a temperature of T$_{BJ}$=840$^{+330}_{-250}$\,K, in excellent agreement with the value of 890\,K reported by \citet{liu2006a}. He~{\sc i} line ratios yield temperatures of 3050$^{+1480}_{-1060}$\,K ($\lambda$5876/$\lambda$4471) and ${  2650}^{+  1890}_{-1030}$\,K ($\lambda$6678/$\lambda$4471). O~{\sc ii} recombination lines, meanwhile, give an electron temperature of $<$3100\,K and an electron density of $<$2750\,cm$^{-3}$, as shown in Figure~\ref{figure:oii_rl_temperature_density}.

\subsubsection{MPA~J1759-3007}
\label{subsec:MPA1759}
MPA~J1759-3007 (PN~G000.5-03.1a, hereafter MPA~1759) was classified by \citet{miszalski2009a} as a likely bipolar nebula, while the acquisition images acquired as part of this programme show the nebula to have a roughly circular appearance (see Figure \ref{forsimages}). The central star of  MPA~1759 was shown by \citet{miszalski2009b} to be a 0.50~d period binary where the companion is most likely a degenerate star.

c(H$\beta$) values from Balmer lines $\alpha$-$\delta$ were all consistent, giving c(H$\beta$)=1.46$\pm$0.07, and thus we applied no scaling to our red or blue spectra. Density diagnostics are all in excellent agreement with all three available giving an average of 740\,cm$^{-3}$, with which each individual diagnostic agrees to within its 1$\sigma$ uncertainty.

No Balmer jump temperature could be measured. He line ratios imply much lower temperatures than the [O~{\sc iii}] line ratio, albeit with large uncertainties (in the case of T$_e$(6678/4471), large enough that the 1$\sigma$ uncertainty encompasses the value of T$_e$([O~{\sc iii}]). O~{\sc ii} line ratios imply an extremely low temperature of ${  1430}^{+  3160}_{ -1430}$\,K, and a density of ${   960}^{+  1630}_{  -960}$\,cm$^{-3}$, in excellent agreement with the values from CEL diagnostics.

\textit{adf}(O/H) is estimated at 68$\pm$7. If the [O~{\sc iii}] $\lambda$4363 line is corrected for the recombination contribution, which we estimate could be up to 40\%, the temperature would be revised from ${ 11400}\pm{   200}$\,K to ${  9740}\pm{   220}$\,K, and \textit{adf}(O/H) to 62$\pm$8.

\subsubsection{NGC~6326}
\label{subsec:NGC6326}
NGC~6326 (PN~G338.1-08.3) is a bright, irregular nebula whose central star is a binary system with orbital period 0.37~d \citep[the exact composition of which is uncertain, but the secondary is most likely a main-sequence star,][]{miszalski2011b}.

[O~{\sc iii}] $\lambda$5007 is saturated in the spectrum of this bright nebula, but the $\lambda$4959 line is not. We thus calculate the electron temperature using the $\lambda$4959 line only. We apply no scaling to our line fluxes, although there is some scatter among the values of c(H$\beta$) derived from the various Balmer lines. The ratios H$\alpha$, H$\gamma$ and H$\delta$ to H$\beta$ give c(H$\beta$)=0.29$\pm$0.01, 0.53$\pm$0.05 and 0.38$\pm$0.06 respectively. The mean, weighted by the observed line fluxes, is 0.47$\pm$0.04.

The electron density diagnostic lines of [O~{\sc ii}], [S~{\sc ii}] and [Ar~{\sc iv}] are detected, giving an average density of 750\,cm$^{-3}$. The values of the individual diagnostics are each consistent with this average, to within their estimated 1$\sigma$ uncertainties.

The electron temperature from collisionally excited lines is ${ 14600}\pm{   100}$K. The recombination contribution to $\lambda$4363 could be up to 6\%, which would result in a correction of $-$300\,K to the electron temperature. He~{\sc i} line ratios give temperatures consistent within their uncertainties: an upper limit of 6450\,K from $\lambda$5876/$\lambda$4471, and ${  7370}^{+  1700}_{ -1380}$\,K from $\lambda$6678/$\lambda$4471. The O~{\sc ii} $\lambda$4089 line may be present in our spectrum but its flux falls below the estimated 3$\sigma$ noise level, and we are thus not able to estimate a temperature or density from the O{~\sc ii} line ratios.

We estimate an \textit{adf}(O/H) of ${ 23}\pm$3 for this object. No recombination lines of N$^{2+}$ are detected, but we detect recombination lines of Ne$^{2+}$, and are also able to estimate \textit{adf}(Ne/H), which we find to be even higher than \textit{adf}(O/H) at 56$\pm$20.

\subsubsection{NGC~6337}
\label{subsec:NGC6337}

NGC~6337 (PN~G349.3-01.1) consists of a ringed waist viewed pole-on (with little or no evidence of nebular lobes extending away from this waist) and a pair of precessing jets oriented more or less in the line of sight \citep{garcia-diaz2009}, while the central star has a main sequence companion in a 0.17~d orbit \citep{hillwig2010}.

We scale our red spectra of NGC~6337 by $\times$0.78 to obtain a consistent value of c(H$\beta$) from the Balmer line ratios, obtaining c(H$\beta$)=0.50$\pm$0.02. Density diagnostics indicate an electron density of around 500\,cm$^{-3}$, with [S~{\sc ii}], [O~{\sc ii}] and [Ar~{\sc iv}] diagnostics in excellent agreement with each other.

Before correcting for the recombination contribution to $\lambda$4363, the [O~{\sc iii}] temperature is ${ 12000}\pm{   100}$\,K. The He~{\sc i} line ratios give temperatures consistent with each other, of ${  2670}^{+  1780}_{ -1300}$\,K ($\lambda$5876/$\lambda$4471) and ${  2670}^{+  3070}_{ -1320}$\,K ($\lambda$6678/$\lambda$4471). The O~{\sc ii} line ratios, shown in Figure~\ref{figure:oii_rl_temperature_density}, give upper limits to the temperature of $<$3250\,K and to the density of $<$3700\,cm$^{-3}$. The abundance discrepancy for O/H is 40$\pm$5.

The recombination contribution to $\lambda$4363 could be up to 55\%, taking O/H from the recombination line abundances. Correcting the line flux accordingly, the temperature would be reduced to ${  9370}\pm{   170}$\,K and \textit{adf}(O/H) to 18$\pm$2.

\subsubsection{Pe~1-9}
\label{subsec:Pe1-9}
Pe~1-9 (PN~G005.0+03.0) is a multiple-shelled PNe with a slightly elongated morphology while its central star has an eclipsing main-sequence secondary with an orbital period of 0.14~d \citep{miszalski2009b,miszalski2009a}.

The \textit{adf} for this object is very uncertain but nevertheless almost certainly extreme. The [O~{\sc iii}] $\lambda$4363 line is not detected; it appears to be present when visually inspecting the spectrum but its estimated flux is within 3$\sigma$ of the noise limit. The important recombination lines, though, are statistically significantly present. Using the upper limit to the line flux reported by {\sc alfa}, the electron temperature is 8700\,K and \textit{adf}(O/H) is then 60. To reduce the \textit{adf} to less than 10, the electron temperature would need to be less than 6000\,K, an implausibly low value given the observed strengths of the collisionally excited lines. Furthermore, the density estimated from [O~{\sc ii}] in this object exceeds that from [S~{\sc ii}] by a factor of $\sim$2.5: 1060 and 410\,cm$^{-3}$ respectively. As discussed below, this discrepancy is associated with the most extreme \textit{adfs}.

\subsection{Spatially resolved analyses}

As can be seen in Figure~\ref{forsimages}, Hen~2-283 and MPA~1759 have small angular radii, of approximately 8 and 10 arcsec in diameter respectively, and our observations are not deep enough to permit a spatially resolved analysis. In Pe\,1-9, the [O~{\sc iii}] line is not detected in the spatially integrated spectrum, and so we did not attempt a spatially resolved analysis for this object either. For the other four extreme-\textit{adf} objects, we extracted spectra divided into spatial bins covering the visible extent of the nebula along the slit. In measuring spatially resolved abundances, we used only the V1 multiplet of O$^{2+}$, as it is the brightest and most easily measured. For each of these nebulae, we show in Figures \ref{spatial-fg1}-\ref{spatial-ngc6337} the variation of \textit{adf}(O$^{2+}$), [O~{\sc ii}] and [S~{\sc ii}] electron densities, and the emission line fluxes of the [O~{\sc iii}] $\lambda$4363 and $\lambda$5007 collisionally excited lines, and the C~{\sc ii} $\lambda$4267 and O~{\sc ii} $\lambda$4649 recombination lines. For NGC\,6326, the $\lambda$5007 line is saturated and we plot instead the intensity of the $\lambda$4959 line.

Generally, as has been observed previously for a number of nebulae with both extreme and non-extreme abundance discrepancies, higher values of the \textit{adf} are seen closer to the central star. This is particularly clear for Hf~2-2, where it peaks at 150, and NGC\,6326 with a peak of 35. Slightly different behaviour is seen in NGC\,6337, which is a bipolar nebula viewed pole-on. In this case, an \textit{adf} is not measured in the central gap where recombination lines are not detected, but the highest values are seen at the inner edges of the ring. Different behaviour is also seen in Fg~1, where a central peak is observed, but higher values are also seen at the outer edge of the bright central region.

A spatial comparison of the electron densities measured from [O~{\sc ii}] and [S~{\sc ii}] shows that the two diagnostics behave differently: the density estimated from [O~{\sc ii}] is almost always higher than that from [S~{\sc ii}], with a greater discrepancy seen in regions with higher \textit{adfs}. In Hf~2-2, for example, the density calculated from the [S~{\sc ii}] line ratio is close to constant at 200$\pm$200\,cm$^{-3}$, while from [O~{\sc ii}], a higher density is estimated at all positions, with the greatest discrepancy occurring in the central regions, where [S~{\sc ii}] gives a density of $\sim$150\,cm$^{-3}$ and [O~{\sc ii}] gives $\sim$2500\,cm$^{-3}$. This indicates that the [O~{\sc ii}] line ratio is affected by recombination of O$^+$, while the [S~{\sc ii}] ratio is not affected, or far less affected, by recombination of S$^{+}$. In a chemically homogeneous nebula, there would be no reason for this difference in behaviour, and this thus provides further support for the existence of chemical inhomogeneities. We observe the same discrepancy in the integrated spectra of other high-\textit{adf} nebulae, and discuss this further in Section~\ref{discussion}.

\citet{corradi2015}, \citet{jones2016} and \citet{garcia-rojas2016} have noted that in a number of objects, the spatial profile of the collisionally excited [O~{\sc iii}] $\lambda$4363 line is very similar to that of the O~{\sc ii} 4649+50{\AA} recombination line, rather than that of the other collisionally excited lines. In most cases, the recombination contribution to $\lambda$4363 should be small, and in any case should trace the distribution of O$^{3+}$, so its spatial similarity to the doubly ionized recombination lines is unexpected.

In Hf~2-2, this pattern is again seen, with the profile of the [O~{\sc iii}] $\lambda$4363 line closely resembling the profiles of the C~{\sc ii} $\lambda$4267 and O~{\sc ii} $\lambda$4649 lines, all three dropping more rapidly with distance from the central star than the [O~{\sc iii}] $\lambda$5007 line. However, in the other three objects the spatial profiles are less similar. In Fg~1, the flux of [O~{\sc iii}] $\lambda$5007 is quite constant across the nebula, while the recombination lines are strongest at the centre and edges, with a minimum at $\pm$8 arcsec from the central star. The $\lambda$4363 line shows a minimum at a greater radial distance of $\pm$12 arcsec. In NGC6326, the spatial profiles are quite distinct, with the recombination lines strongly centrally peaked, while [O~{\sc iii}] $\lambda$4959 and $\lambda$4363 vary more gradually. While also centrally peaked, [O~{\sc iii}] $\lambda$4363 is only $\sim$25\% lower at the edges than in the centre, whereas O~{\sc ii} $\lambda$4649 has an intensity at the edges of about 20\% of its central value. In NGC\,6337, too, the spatial profiles differ, with the recombination lines having their highest intensities towards the inner edge of the ring, while the CELs peak at the outer edge of the ring.

Thus, the spatial profile of [O~{\sc iii}] $\lambda$4363 most closely resembles that of the recombination lines in the nebulae with the highest \textit{adfs}. This suggests that recombination excitation of the line must be significant. To understand how this arises, and whether recombination of O$^{3+}$ really can excite the [O~{\sc iii}] $\lambda$4363 line to such a degree that it dominates the spatial profile will require detailed three-dimensional photoionization models to be created.

\begin{figure*}
 \includegraphics[width=0.32\textwidth]{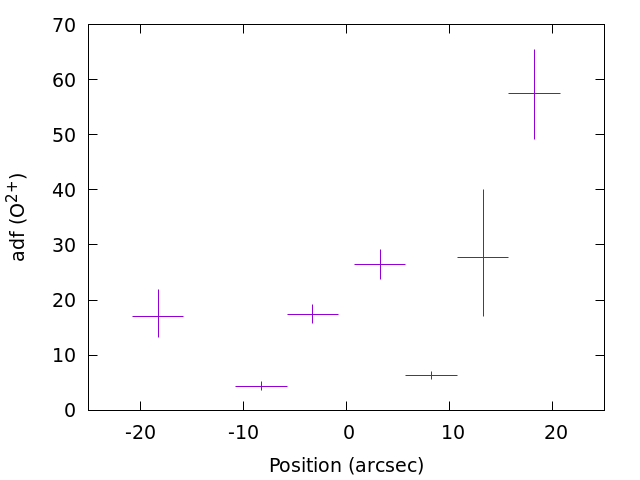}
 \includegraphics[width=0.32\textwidth]{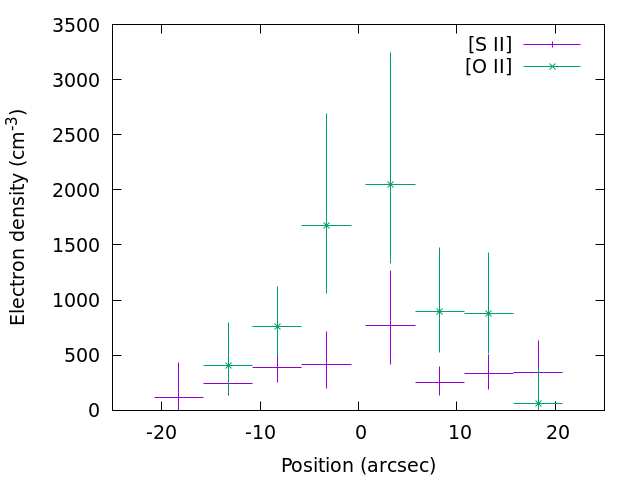}
 \includegraphics[width=0.32\textwidth]{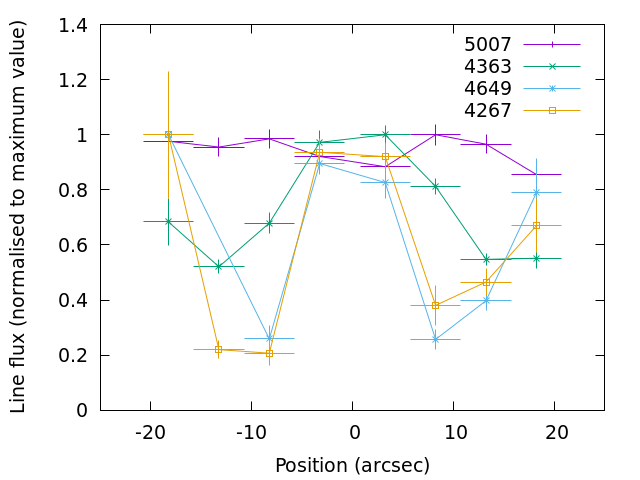}
 \caption{Spatially resolved analysis of Fg~1, showing the variation with slit position of (l) \textit{adf}(O$^{2+}$) (c) electron density from [S~{\sc ii}] and [O~{\sc ii}] (r) emission line intensities}
 \label{spatial-fg1}
\end{figure*}

\begin{figure*}
 \includegraphics[width=0.32\textwidth]{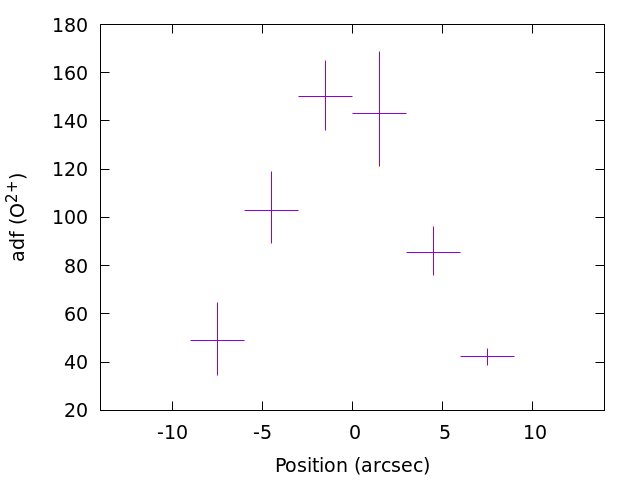}
 \includegraphics[width=0.32\textwidth]{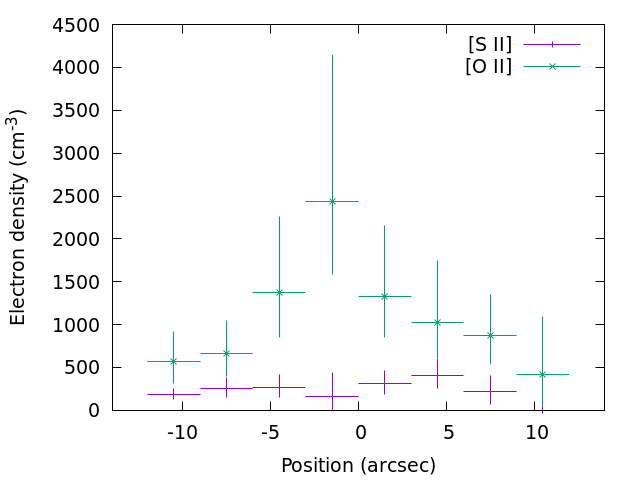}
 \includegraphics[width=0.32\textwidth]{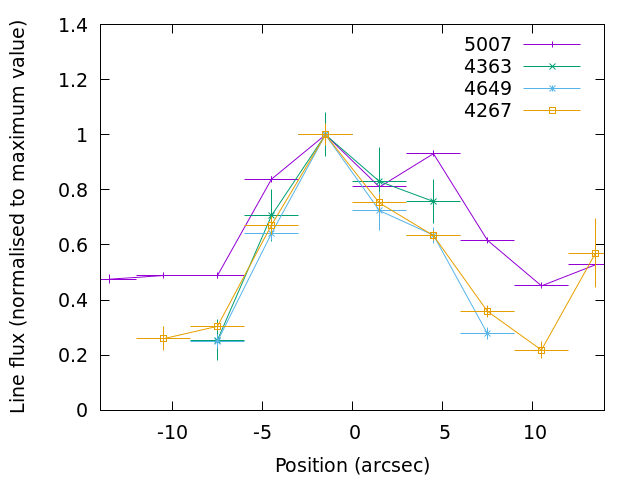}
 \caption{Spatially resolved analysis of Hf~2-2, showing the variation with slit position of (l) \textit{adf}(O$^{2+}$) (c) electron density from [S~{\sc ii}] and [O~{\sc ii}] (r) emission line intensities}
 \label{spatial-hf2-2}
\end{figure*}

\begin{figure*}
 \includegraphics[width=0.32\textwidth]{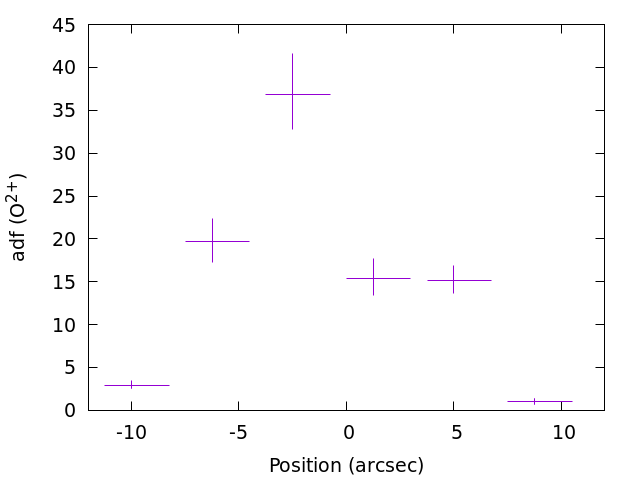}
 \includegraphics[width=0.32\textwidth]{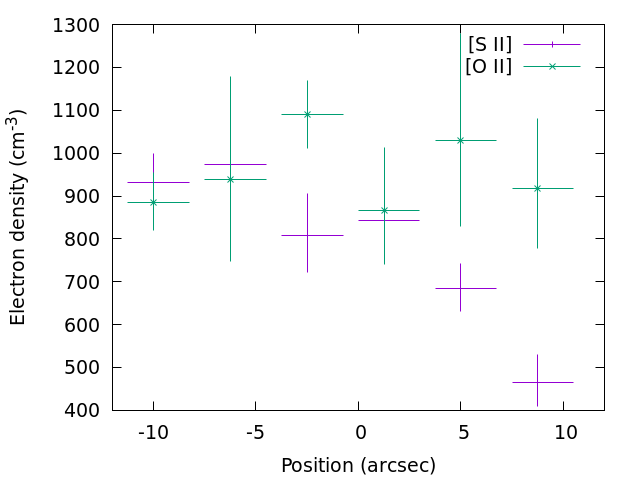}
 \includegraphics[width=0.32\textwidth]{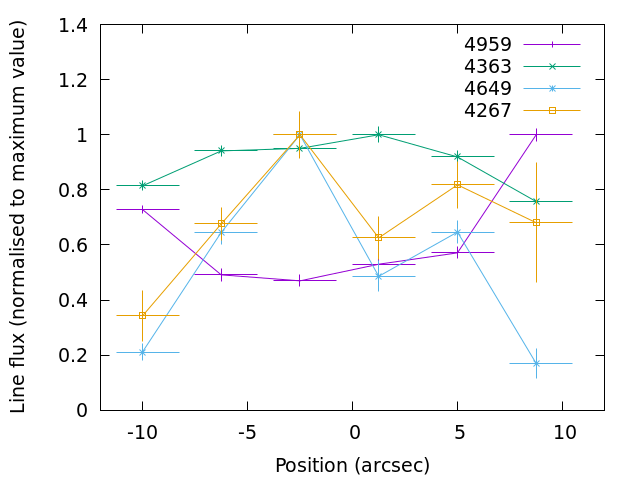}
 \caption{Spatially resolved analysis of NGC~6326, showing the variation with slit position of (l) \textit{adf}(O$^{2+}$) (c) electron density from [S~{\sc ii}] and [O~{\sc ii}] (r) emission line intensities}
 \label{spatial-ngc6326}
\end{figure*}

\begin{figure*}
 \includegraphics[width=0.32\textwidth]{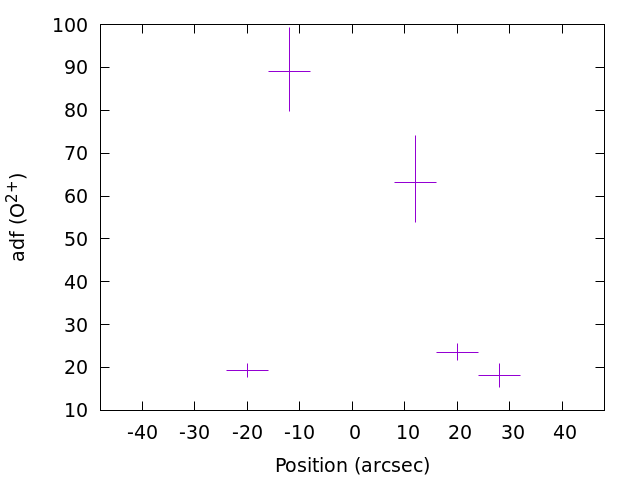}
 \includegraphics[width=0.32\textwidth]{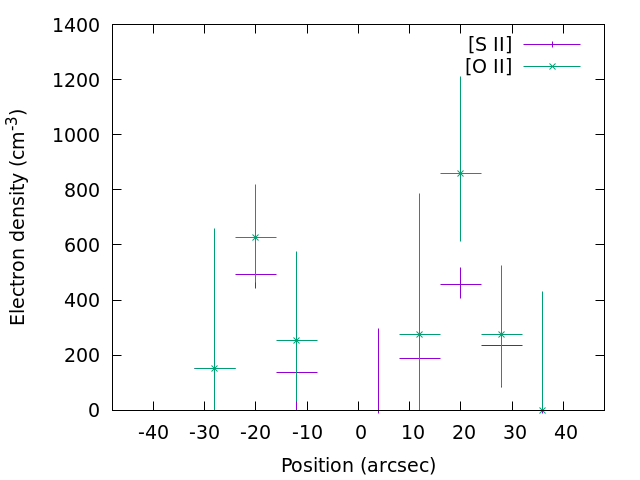}
 \includegraphics[width=0.32\textwidth]{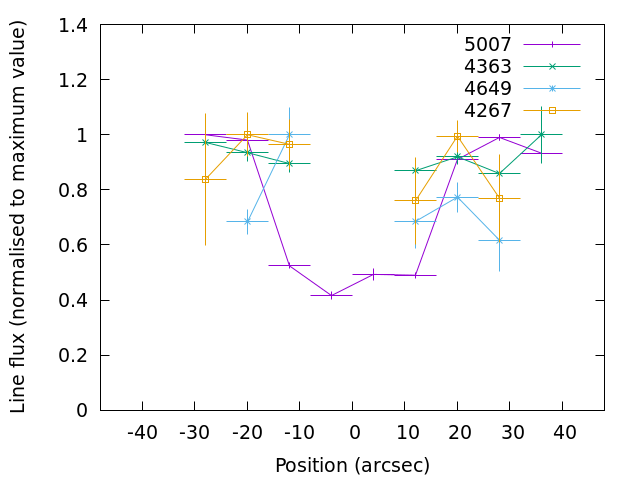}
 \caption{Spatially resolved analysis of NGC~6337, showing the variation with slit position of (l) \textit{adf}(O$^{2+}$) (c) electron density from [S~{\sc ii}] and [O~{\sc ii}] (r) emission line intensities}
 \label{spatial-ngc6337}
\end{figure*}


\subsection{Abundance discrepancy upper limits}
Given that where RLs are detected, we have a 100\% strike rate of elevated to extreme \textit{adfs}, it is of interest to determine upper limits for the remaining objects in the sample, to see if we can rule out an \textit{adf}$>$5 for any objects in the sample.

For the 20 objects where recombination lines are not detected but abundances from collisionally excited lines can be calculated, we establish upper limits for the RL abundances of O$^{2+}$/H${+}$ by calculating synthetic recombination line spectra, matching the observed H$\beta$ line flux and scaling the abundance to the point where the lines would have been detected. We show an example of this technique for one of the nebulae in Figure~\ref{upperlimitfigure}, and list the upper limits we obtained in Table~\ref{upperlimittable}. The limits obtained are generally not very stringent, and we cannot in any case rule out an abundance discrepancy factor greater than 5. We thus do not find in our current sample any examples of nebulae with a close binary central star but a non-extreme abundance discrepancy. Of course, deeper spectra of our recombination line non-detections could yet yield such objects; one or two are known to exist (see discussion).

The most stringent upper limits are found for M~2-19, M~3-16, MPA~1508 and NGC~2346, which are all constrained to have an abundance discrepancy factor of less than about 10. In NGC~2346, we may detect a single O~{\sc ii} line - $\lambda$4649, the strongest line of the V1 multiplet, with a flux that would indicate an \textit{adf} of $\sim$10. However, the line flux only just exceeds the estimated 3$\sigma$ limit, and in the absence of any other O~{\sc ii} line detections, we do not consider that a reliable abundance can be derived. Both M~2-19 and M~3-16 have a high extinction with c(H$\beta$)$\sim$1.6 for each, and the weakest lines detected in either object have fluxes of about 0.5, where H$\beta$=100. The central star of NGC~2346 has an orbital period of 16~d (\citealt{mendez1981}, Brown et al. in preparation), while that of MPA~1508 has a period of 12.5~d (\citealt{miszalski2011a}), according to which we would predict a non-extreme \textit{adf} for both (see discussion below). M~2-19 and M~3-16 have orbital periods of 0.670 and 0.574~d respectively (\citealt{miszalski2009}), but have the highest measured electron densities of the objects in which we do not detect recombination lines, at 1120 and 1250\,cm$^{-3}$ respectively. As discussed in Section~\ref{section:electrondensity}, the most extreme \textit{adfs} are generally seen in the objects with the lowest densities, a finding consistent with the moderate upper limit for these two objects.

Based on our findings from the objects with measured \textit{adfs} (see below), we would predict non-extreme \textit{adfs} for the objects with binary orbital periods greater than $\sim$1.15~d, elevated discrepancies for M~2-19 and M~3-16 given their higher densities, and extreme abundance discrepancies for the rest of the nebula. Deeper spectra of all these objects would thus provide important constraints on the relationships between binary orbital period, electron density and the abundance discrepancy factor.

\begin{figure}
	\includegraphics[width=0.5\textwidth]{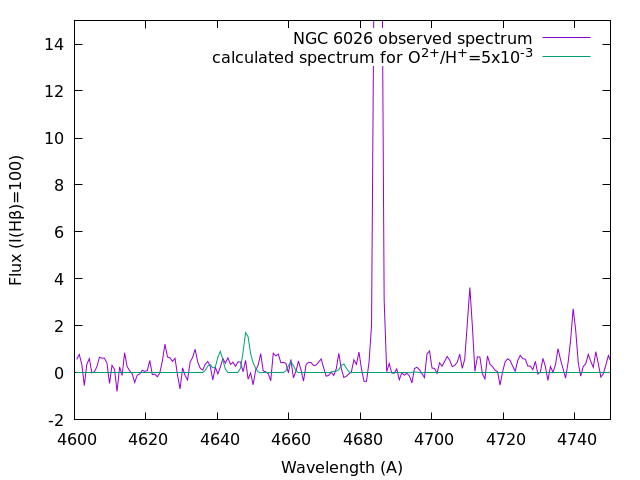}
	\caption{An example of a determination of upper limits to recombination line abundances. The plot shows the observed spectrum of NGC~6026, together with a synthesized spectrum containing O$^{2+}$ and H$^{+}$ only, with O$^{2+}$/H$^{+}$ scaled to the point where it would just have been detected at 3$\sigma$ above the continuum noise level.}
	\label{upperlimitfigure}
\end{figure}
\begin{table*}
	\begin{tabular}{lllllll}
		\hline
		Object     & \multicolumn{2}{c}{O$^{2+}$/H$^+$ (x10$^{-3}$)} & \textit{adf} & Orbital period (days) & \multicolumn{2}{c}{Electron density (cm$^{-3}$)} \\
		           & RLs (upper limit) & CELs (observed) & & & [S~{\sc ii}] & [O~{\sc ii}] \\
		\hline
		
A~41             & 2.0   & 0.07 &  $<$25 & 0.226 & 300$\pm$100 & 140$\pm$60 \\
A~65             & 15.0  & 0.19 &  $<$80 & 1     & --          & --         \\
BMP~1801         & 20.0  & 0.02 & $<$800 & 0.322 & --          & --         \\
H~2-29           & 5.0   & 0.03 & $<$170 & 0.244 & 290$\pm$150 & 1060$^{+1600}_{-650}$\\
HaTr~4           & 12.0  & 0.14 &  $<$90 & 1.74  &  --         & --         \\
Hen~2-11         & 20.0  & 0.35 &  $<$60 & 0.609 & 230$\pm$150 & --         \\
Hen~2-428        & 5.0   & 0.02 & $<$220 & 0.176 & 580$\pm$100 & 460$\pm$300 \\
K~1-2            & 80.0  & 0.38 & $<$210 & 0.676 &  --         & --          \\
K~6-34           & 8.0   & 0.01 & $<$650 & 0.39  & 700$^{+1200}_{-500}$ & -- \\
Lo~16            & 10.0  & 0.20 &  $<$50 & 0.49  & 200$\pm$150 & --          \\
M~2-19           & 1.0   & 0.10 &  $<$10 & 0.670 & 1390$\pm$70 & 1100$\pm$200\\
M~3-16           & 2.0   & 0.20 &  $<$10 & 0.574 & 1120$\pm$80 & 1100$\pm$200\\
MPA~1508         & 2.0   & 0.18 &  $<$10 & 12.50 & 820$^{+2200}_{-690}$ & 600$^{+950}_{-500}$\\
NGC~2346         & 2.0   & 0.23 &   $<$8 & 15.99 & 265$\pm$60 & 260$\pm$50 \\
NGC~6026         & 5.0   & 0.07 &  $<$70 & 0.528 &  --  \\
PHR~1756         & 15.0  & 0.06 & $<$270 & 0.266 &  --  \\
PHR~1804         & 25.0  & 0.22 & $<$110 & 0.626 &  --  \\
PM~1-23          & 40.0  & 0.20 & $<$200 & 1.26 &  --  \\
Sab~41           & 20.0  & 0.20 & $<$100 & 0.297 &  140$\pm$80 & 470$^{+800}_{-400}$ \\
Sp~1             & 6.0   & 0.08 &  $<$70 & 2.91  &  --  \\
		\hline
	\end{tabular}
	\caption{Estimated upper limits to the abundance discrepancy factor in the cases where CEL abundances could be calculated but no recombination lines were detected.}
	\label{upperlimittable}
\end{table*}

\section{Discussion}
\label{discussion}

Figure~\ref{adfcompilation} shows 202 values available in the literature for the abundance discrepancy of O$^{2+}$/H$^{+}$ at the time of submission of this paper, ranked and with close binary PNe highlighted in blue, and H~{\sc ii} regions in purple. For all objects, the median discrepancy is 2.2, and 37 nebulae have a discrepancy larger than a factor of 5.

H~{\sc ii} regions, as found by e.g. \citet{garcia-rojas2007}, show behaviour distinct from PNe, with no evidence of elevated or extreme abundance discrepancies. \citet{tsamis2003} found an elevated abundance discrepancy in the LMC H~{\sc ii} region LMC N11B, but \citet{toribiosancipriano2017} found that this was due to an overestimate of the $\lambda$4089 O~{\sc ii} line, revising its \textit{adf} down to 1.58. Excluding the higher value from the statistics, H~{\sc ii} regions have a mean \textit{adf} of 1.9$\pm$0.5, and a median value also of 1.9. A Shapiro-Wilk test returns a $p$-value of 0.59, well above the value of 0.05 below which the hypothesis of a normal distribution could be questioned.

For PNe, the mean \textit{adf}, strongly skewed by the tail of extreme values, is 12$\pm$59. The median, 16th percentile and 84th percentile values are 2.5, 1.7 and 7.3 respectively. Fully disentangling post-CE and single star PNe from their \textit{adf} alone is not possible, and thus this ``mean" of objects formed by two different evolutionary channels may be unrepresentative of either. Considering only the known close binaries, the median is 18, and the mean is 28$\pm$24. The association between extreme discrepancies and close binarity is clear, and indeed several of the extreme objects not highlighted are nevertheless strongly suspected to have a close binary central star (e.g. Abell 30, Abell 58, NGC~1501).

\begin{figure*}
	\includegraphics[width=\textwidth]{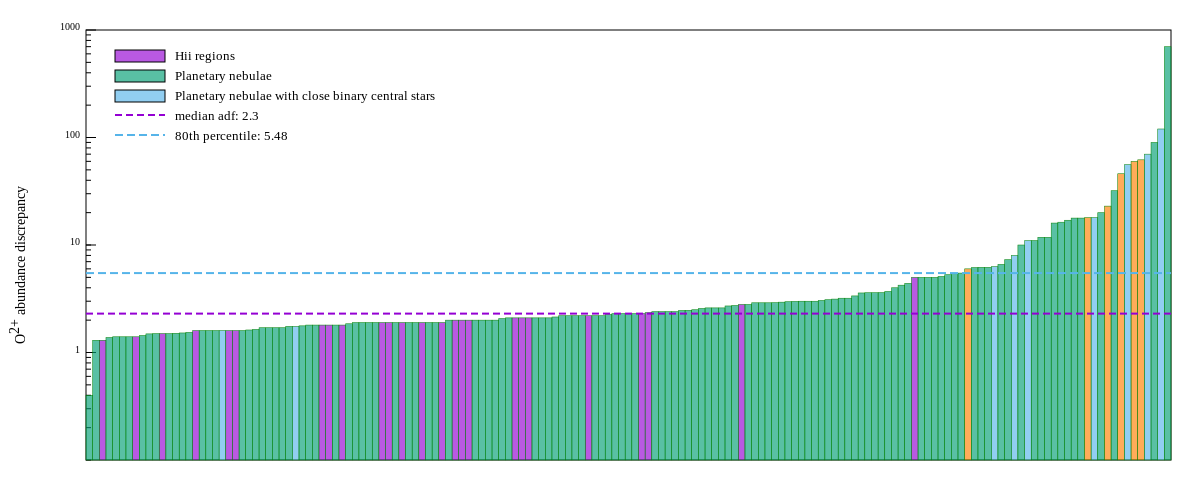}
	\includegraphics[width=\textwidth]{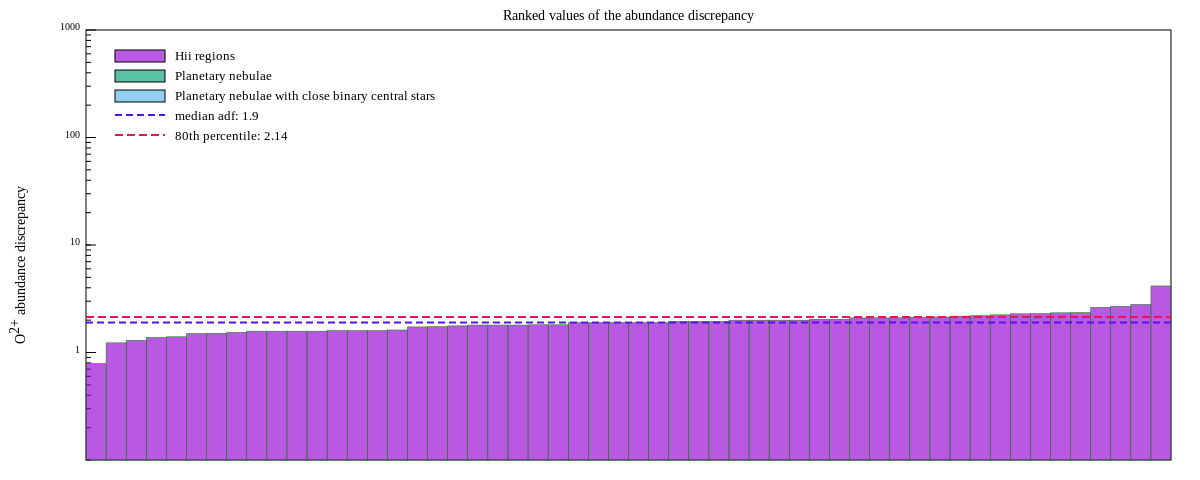}
	\includegraphics[width=\textwidth]{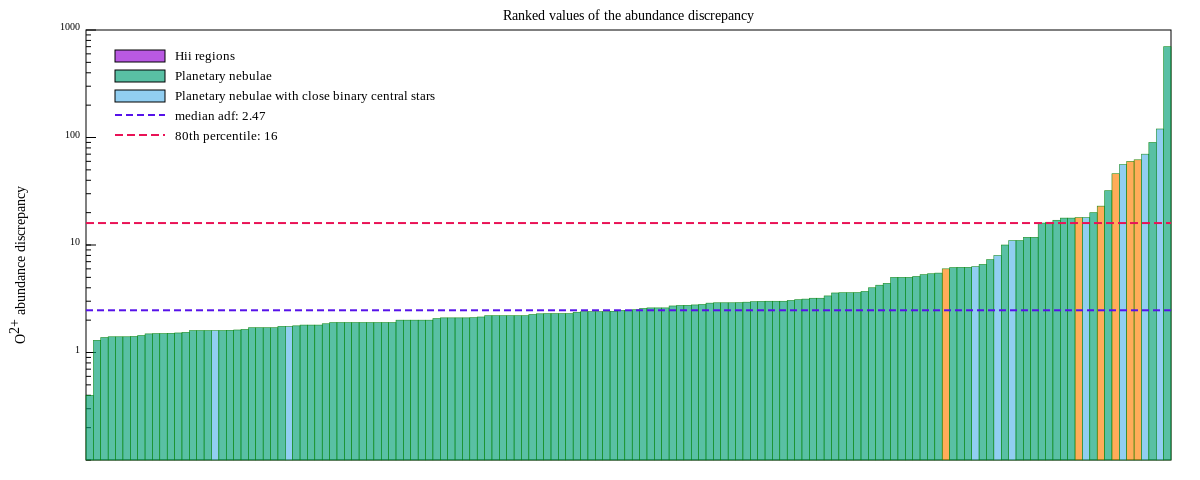}
	\caption{196 measurements of \textit{adf}(O$^{2+}$) available in the literature as of 2018 April, plus the six new measurements presented in this paper. The top panel shows all values, presented in rank order. Objects with close binary central stars are highlighted in blue, while H~{\sc ii} regions are highlighted in purple. Objects studied for the first time in the current work are shown in orange. The middle panel shows the ranked values of H~{\sc ii} region \textit{adf} measurements, excluding that of LMC N11B, which was found to be spuriously large by \citet{toribiosancipriano2017}. The bottom panel shows the ranked values of PN \textit{adf} measurements. A full list of the individual objects and references used to compile this figure is available at \href{https://www.nebulousresearch.org/adfs}{https://www.nebulousresearch.org/adfs}}
	\label{adfcompilation}
\end{figure*}

The number of objects with a measured \textit{adf} and for which binarity and other properties are known allows us to make some preliminary statistical inferences. Physically, there is evidence for (at least) two mechanisms that cause \textit{adfs}: one that can cause moderate \textit{adfs}, as seen ubiquitously in all ionized regions where recombination lines have been detected, including planetary nebulae, H~{\sc ii} regions (\citealt{tsamis2003}, \citealt{garcia-rojas2007}) and Wolf-Rayet ejecta nebulae (\citealt{mesa-delgado2014}); and a second that causes the elevated and extreme \textit{adfs} only seen in planetary nebulae. The two mechanisms show some differences in their effect on heavy element ratios (\citealt{garcia-rojas2007}). To disentangle the two mechanisms, we created synthetic probability distributions, tailored to broadly reproduce the features of the planetary nebula and H~{\sc ii} region \textit{adf} probability distributions. For H~{\sc ii} regions, this is straightforward; the \textit{adfs} closely follow a normal distribution, with the distribution well fitted using $\mu$=1.92 and $\sigma$=0.45. With these parameters, the lowest measured \textit{adf} for an H~{\sc ii} region, although almost unique in being below unity (which must be due to abundance measurement uncertainties, as no known mechanism can account for \textit{adf}$<$1), is in fact only 2.5$\sigma$ from the mean. Meanwhile, the value of 4.16 derived for Mrk 1271 (\citealt{esteban2014}) is a nearly 5$\sigma$ outlier, albeit with a significant uncertainty of $\pm$0.15 dex.

For planetary nebulae, we model the distribution with two components, one normally distributed and the second log-normally distributed, with the second distribution occuring only in a fraction \textit{F} of all nebula. The distribution we settled upon is:

\begin{eqnarray}
\rm{adf}_{\rm{normal}}&=& 2.0 + 0.45*R_1\\
\rm{adf}_{\rm{extreme}}&=&\rm{adf}_{\rm{normal}} + 10^{(0.6+0.9R_2)}
\label{syntheticequation}
\end{eqnarray}

where R$_1$ and R$_2$ are random numbers drawn from a Gaussian distribution with mean zero and variance unity. We initially took \textit{F} as 0.2, to reflect the fraction of PNe known to have close binary central stars, but found that the observed distribution could be better approximated with \textit{F}=0.5. Figure~\ref{syntheticcomparison} shows 154 values generated by this compound probability distribution, ranked and compared with the 154 actual literature values for the \textit{adf} in planetary nebulae.

\begin{figure}
  \includegraphics[width=0.47\textwidth]{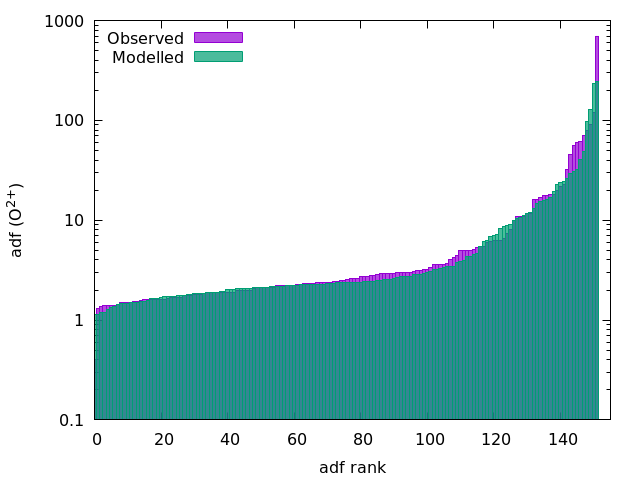}
  \caption{Synthetic distribution of \textit{adfs} generated by equation~\ref{syntheticequation}, compared to the distribution of 154 literature values.}
  \label{syntheticcomparison}
\end{figure}

The ``normal" \textit{adf} may be interpreted as the same mechanism that gives rise to \textit{adfs} in H~{\sc ii} regions, operating in planetary nebulae. A slightly higher mean was required to fit the observed distribution of \textit{adfs} at the low end of the range. A natural interpretation would be to attribute this to temperature fluctuations of the kind envisaged by \citet{peimbert1967}, with their magnitude being greater in planetary nebulae.

The ``extreme" \textit{adf} distribution may then be interpreted as due to a second mechanism, operating only in nebulae with close binary central stars. Though we set \textit{F} to 0.5, this does not imply that the `true' close binary fraction is necessarily significantly higher than the commonly estimated fraction; \textit{adfs} will of course be more easily measured in objects with elevated or extreme abundance discrepancies and thus the overall sample of objects with measured \textit{adfs} will be likely to contain more close binaries than the overall planetary nebula population. As an illustration of the potential biases, this paper has increased the number of PNe with \textit{adf}$>$5 by about 20\%, specifically by studying only PNe known to host close binary central stars.

This synthetic distribution is intended only to be illustrative and is clearly not a unique solution, but the fact that H~{\sc ii} region \textit{adfs} can be well modelled with a Gaussian distribution, while those of PNe clearly cannot, supports the hypothesis that a single mechanism gives rise to H~{\sc ii} region \textit{adfs}, while an additional mechanism needs to be invoked to account for the extremely skewed distribution of planetary nebula \textit{adfs}.

We now consider the relations between nebular \textit{adf} and properties of the central star and nebula.

\subsection{Relationship of \textit{adfs} with central star properties}

\subsubsection{The binary period}
\label{sec:binaryperiod}
In this work and the study by \citet{jones2016}, we have increased the number of close binary CSPNe known to be surrounded by a high-\textit{adf} nebula from 6 to 13. Two further PNe have a close binary but a low \textit{adf}, for a total of 15 objects with a known binary CSPN and a measured \textit{adf}. We can thus start to investigate correlations between the \textit{adf} and the central star properties in these objects.

First, we consider whether there is a relationship between the period of the central binary and the abundance discrepancy. If some kind of continuous relationship existed between the two, this would imply that whatever mechanism gives rise to extreme \textit{adfs} operates regardless of the binary period, but its magnitude is determined by it. Alternatively, if a threshold period dividing elevated from normal \textit{adf} objects exists, it would suggest that the mechanism is triggered only for shorter period binaries.

We list in Table~\ref{sampleproperties} the values of \textit{adf} and binary orbital period for all objects where both are known, and plot \textit{adf} against period in Figure~\ref{adf-period-figure}. The number of points is still too small to draw firm conclusions as the distribution of periods in the sample is strongly concentrated at smaller values, but some hints of a relationship between \textit{adf} and orbital period begin to emerge. All of the objects in the current sample have periods of 1.20~d or less and all have elevated or extreme abundance discrepancies. Of the eight previous objects with known chemistry and orbital period, two have a ``normal" \textit{adf}: NGC~5189 has an \textit{adf} of 1.6 (\citealt{garcia-rojas2013}), and a period of 4.04~d (\citealt{manick2015}), and IC~4776 has \textit{adf}(O$^{2+}$)=1.75, and a period that is poorly constrained but thought to be around 9~d (\citealt{sowicka2017}). Additionally, the Necklace nebula is a definite post-CE object; there are no published recombination line abundances but \citet{corradi2015} report that the non-detection of recombination lines in deep spectra sets a low upper limit to its abundance discrepancy. The orbital period of the central star is 1.16~d (\citealt{corradi2011b}).

\begin{figure}
	\includegraphics[width=0.5\textwidth]{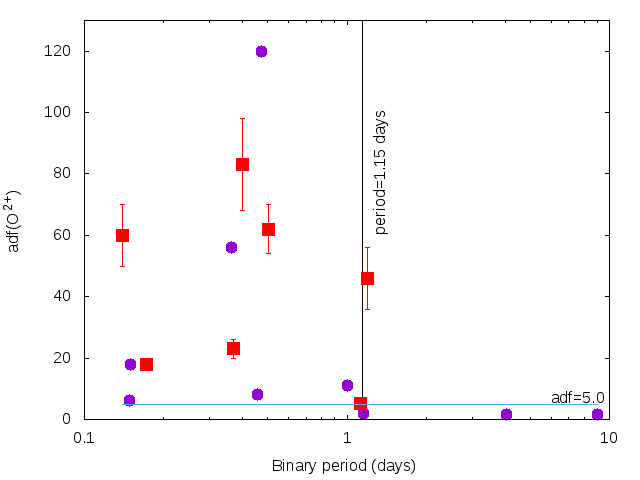}
	\caption{Abundance discrepancy for O$^{2+}$ plotted against binary period for the 15 objects where both are known: 9 literature values (purple dots) and 6 from this study (red squares). Also plotted is a point for the Necklace, which has a period of 1.16~d and an unmeasured abundance discrepancy but with an upper limit reported to be low and plotted here as a factor of 3.0. A horizontal line indicates an \textit{adf} of 5.0, which we consider the dividing line between ``normal" and ``elevated", and a vertical line indicates the period of 1.15~d which roughly divides objects with low and extreme \textit{adfs}.}
	\label{adf-period-figure}
\end{figure}

We therefore note that all objects with a binary period of less than 1.15~d have an extreme abundance discrepancy, while 3 of 4 objects with a binary period longer than 1.15~d do not. This finding is subject to a significant selection effect, because known binary central star periods are intrinsically strongly concentrated at periods of $\lesssim$1~d (due to detection biases; \citealt{jones2017}); the absence of high \textit{adf} objects with long periods may simply be due to the very small sample of longer period objects. However, the probability of the two lowest \textit{adfs} coinciding with the two longest orbital periods by chance alone is only 0.95\%. Additionally, after the submission of the first draft of this paper, \citet{miszalski2018} reported the discovery of a binary central star in the nebula MyCn~18, with a period of 18.15~d; the abundance discrepancy in this nebula was measured to be 1.8 by \citet{tsamis2004}. This third example of a low-adf nebula surrounding a longer-period central star argues strongly against a coincidencental relationship.
\subsubsection{Stellar abundances}

Correlations between central star abundances and nebular abundance discrepancies have been proposed in the past; \citet{ercolano2004} noted that the central star of NGC~1501 was hydrogen-deficient, in common with several other then-recently identified extreme \textit{adf} objects. However, they also noted that other objects with H-rich stellar atmospheres were also associated with extreme \textit{adfs}, and concluded that no clear relationship existed between the phenomena. \citet{garcia-rojas2013} measured the \textit{adf} in several objects with known H-deficient central stars, and did not find any elevated or extreme abundance discrepancies.

Given the evidence that extreme \textit{adfs} are caused by cold hydrogen-deficient clumps embedded in hot gas of normal composition, a source for these clumps needs to be identified. Two possibilities have been discussed; first, a very late thermal pulse (VLTP), in which a single star experiences a thermal pulse after having begun its descent of the white dwarf cooling track. The second scenario is a nova-like eruption relying on a binary central star. Additional and more complex scenarios are possible: \citet{lau2011} suggested that some combination of VLTP and nova in which the former triggered the latter could explain the properties of the hydrogen-deficient knot in Abell~58. Given the lack of observational constraints on such scenarios, we consider only these two relatively simple cases.

The two scenarios make contrasting predictions for the central star abundances. The VLTP scenario would result in a hydrogen-deficient central star, and indeed that scenario is commonly invoked specifically as a mechanism for creating such stars. Meanwhile, the nova-like scenario is as yet ill-defined. An eruption in which some hydrogen is neither burned nor ejected would be required to leave behind a hydrogen-rich post-nova object.

We have searched the literature for central star classifications, and find that 33 nebulae have both a measurement of the \textit{adf}, and a published classification of their central star indicating that they are [WR] or PG-1159 stars (\citealt{aller1987}, \citealt{weidmann2011}, \citealt{akras2015}, \citealt{todt2015}). A further 31 are classified as \textit{wels} stars; we exclude these from consideration due to the high likelihood of many of these classifications being either contamination by nebular emission or irradiation of the secondary (\citealt{miszalski2011b}). The left-hand panel of Figure~\ref{qqplots} shows a Q-Q plot comparing the quantiles of the distributions of \textit{adfs}, for nebulae with H-deficient central stars and those without. In such a plot, for two different data sets, if the two sets are drawn from the same underlying probability distribution, the points will lie close to the line of $y=x$. Populations drawn from differing probability distributions will diverge from the 1:1 relation. In this case, the points indeed mostly lie close to $y=x$. In the right-hand panel of the figure, we show the Q-Q plot for planetary nebulae with binary central stars against those without a known binary central star. The large deviations from the $y=x$ line indicate that the \textit{adfs} of nebulae with binary central stars come from a strongly differing distribution to those without.

Some caveats apply to the surface chemistry determinations: many literature values have been obtained from relatively poor-quality spectra at low resolution, where nebular and stellar features may be difficult to distinguish. \citet{basurah2016} showed that for at least some stars classified as ``weak emission line" objects and suspected to be hydrogen-deficient, the weak emission lines were of nebular origin. Another potential issue is that the light of the secondary star may be bright enough to affect the classification - a particularly strong possibility given that many binary central stars have been discovered through their large irradiation effects.

\begin{figure*}
	\includegraphics[width=0.47\textwidth]{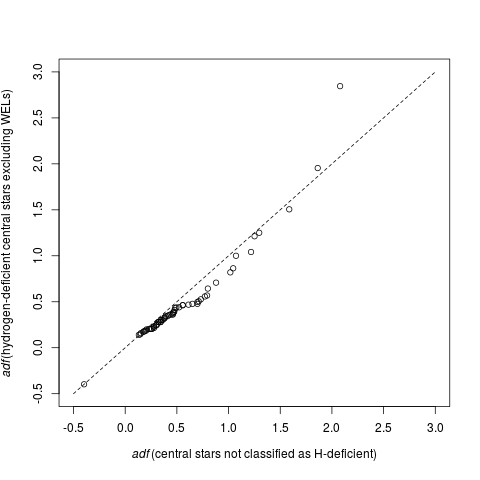}
	\includegraphics[width=0.47\textwidth]{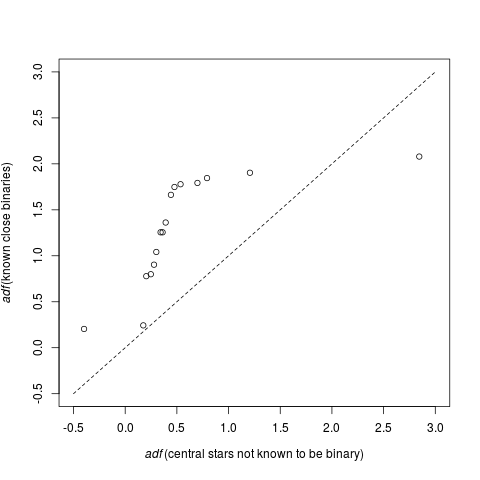}
	\caption{Q--Q plots for \textit{adf} for (l) nebulae with hydrogen-deficient central stars against those without; (r) nebulae with close binary central stars against those without. When samples are drawn from the same underlying distribution, points in a Q--Q plot will lie close to $y=x$.}
	\label{qqplots}
\end{figure*}

However, the available data on central star surface chemistries argues against a VLTP, while a nova-like outburst as the source of hydrogen-deficient material in planetary nebulae remains plausible. This is of course consistent with the strong association of close binary central stars with extreme \textit{adfs}. The apparent relationship between binary period and \textit{adf} may imply that binaries with periods longer than the $\sim$1.15~d threshold never experience the event which leads to the ejection of hydrogen-deficient material into the nebula; it could alternatively suggest that in the shorter period objects, the event occurs at the time of or very soon after the ejection of the nebula, while in the longer period objects it may occur later. This could potentially account for objects such as Abell 30 and Abell 78, where hydrogen-deficient material is directly observed, with kinematic observations showing that it was ejected thousands of years after the outer nebula into which it is expanding (\citealt{borkowski1995}).

\subsubsection{Weak emission lines}

As noted by e.g. \citet{basurah2016}, nebular recombination lines may be mistaken for stellar emission lines, leading to a misclassification. The opposite may also be true; close binary stars with an irradiated secondary typically show strong stellar emission at the wavelengths of nebular recombination lines. If stellar emission contaminated the nebular spectrum, the \textit{adf} could be significantly overestimated. However, we do not believe that this is likely for several reasons: we excluded bright central star emission from our integrated spectra (see Figure~\ref{forsimages}), and in any case several extreme-\textit{adf} objects have central stars that show no sign of irradiation lines in our spectra. Direct excitation of the nebular lines by the central star emission could contribute to the nebular emission line fluxes, but arguing against this possibility is the evidence for significant recombination excitation of the [O~{\sc ii}] $\lambda$3727, 3729 and [O~{\sc iii}] $\lambda$4363 lines, which cannot be emitted by the central star. The relations between \textit{adf} and nebular density discussed below would also not be expected to arise if some significant fraction of the \textit{adf} could be attributed to central star rather than nebular emission.

\subsection{Relationship of \textit{adfs} with nebular properties}

To assess the possible relationships between abundance discrepancies and other nebular properties, we have compiled data on each of the 15 PNe that have known recombination line abundances and a close binary central star. Previous authors have considered relationships between nebular parameters and abundance discrepancies (e.g. \citealt{tsamis2003}, \citealt{wesson2005}); however, such comparisons included two separate populations of objects, the post-CE PNe and the ``normal" PNe, and thus any trends that differ between the two populations would have been incorrectly estimated. Here, we are restricting our investigations to definite post-CE PNe. In Table~\ref{sampleproperties}, we present nebular parameters of the sample objects, and we discuss them below.

\begin{table*}
\begin{tabular}{lllllll}
\hline
Object          & \textit{adf}   & period & N/H   & O/H   & electron density      & reference \\
                &       & (days) &       &       & (cm$^{-3}$) \\
\hline
A~46            & 120             & 0.47  & 6.77  & 7.93  & 1590  & \citet{corradi2015} \\
A~63            & 8               & 0.46  & 7.46  & 8.59  & 1560  & \citet{corradi2015} \\
Fg~1            & 46$^{+10}_{-8}$ & 1.2   & 7.84  & 8.26  & 500   & This work \\
Hen~2-283       & 5.1$\pm$0.5     & 1.15  & 8.56  & 8.75  & 3200  & This work \\
Hen~2-155       & 6.3             & 0.148 & 7.85  & 8.21  & 1300  & \citet{jones2015a} \\
Hen~2-161       & 11              & 1.01? & 8.06  & 8.34  & 1600  & \citet{jones2015a} \\
Hf~2-2          & 83$\pm$15       & 0.4   & 7.60  & 7.92  & 700   & \citet{liu2006}; This work \\
IC~4776         & 1.75            & 9     & 7.80  & 8.57  & $\sim$20000 & \citet{sowicka2017} \\
MPA~1759        & 62$\pm$8        & 0.5   & 7.52  & 8.25  & 740   & This work \\
NGC~5189        & 1.6             & 4.04  & 8.60  & 8.77  & 1000  & \citet{garcia-rojas2013} \\
NGC~6326        & 23$\pm$3        & 0.37  & 7.10  & 7.43  & 820   & This work \\
NGC~6337        & 18$\pm$2        & 0.173 & 8.31  & 8.40  & 500   & This work \\
NGC~6778        & 18              & 0.153 & 8.61  & 8.45  & 600   & \citet{jones2016} \\
Pe~1-9          & 60$\pm$10       & 0.140 & 7.85  & 7.98  & 740   & This work \\
Ou~5            & 56              & 0.36  & 7.58  & 8.40  & 560   & \citet{corradi2015} \\
\hline
\end{tabular}
\caption{Properties of the 15 nebulae with close binary central stars and a measured \textit{adf}}
\label{sampleproperties}
\end{table*}

\subsubsection{Electron density}
\label{section:electrondensity}
We have established that close binary central stars and extreme abundance discrepancies are very strongly linked, to the extent that it is surprising to find two or three objects that have close binary central stars and yet a moderate abundance discrepancy. It is of interest, then, to try to determine why these objects depart from the otherwise strong relationship. In addition to the finding that the two nebulae with the lowest \textit{adfs} have the central stars with the longest periods, we find several other properties that seem distinct from the extreme objects.

First, there is a correlation with electron density. In the current sample, all the objects except Hen~2-283 have electron densities estimated (from the average of available diagnostics) at $<$1000\,cm$^{-3}$. Hen~2-283 has an electron density about five times higher than the rest of the sample, at 3200\,cm$^{-3}$ compared to the average for the rest of the sample of 720\,cm$^{-3}$. It also has the lowest \textit{adf} of the sample, only just exceeding the threshold of a factor of 5 which we consider ``elevated". In the full sample, the two post-CE objects with moderate discrepancies are NGC~5189 and IC~4776. These too have higher densities than the rest of the nebulae; the median density for the sample of 15 is 820\,cm$^{-3}$, while the densities of NGC~5189 and IC~4776 are 1000 and $>$10\,000 respectively, the latter being by far the highest among the sample.

Considering individual diagnostics, all 15 binary+\textit{adf} objects have a density estimated from both [O~{\sc ii}] and [S~{\sc ii}]; [Ar~{\sc iv}] and [Cl~{\sc iii}] densities are available for five and seven of the nebulae, respectively. Figure~\ref{oii_sii_comparison} shows n$_e$([O~{\sc ii}]) against n$_e$([S~{\sc ii}]) for the 15 objects, together with the 1:1 relation. This shows that these diagnostics closely agree in 8 objects, but significantly differ in 7 objects, with [O~{\sc ii}] giving a higher density than [S~{\sc ii}] in every case. This is in contrast to the findings of \citet{wang2004}, who examined density diagnostics for a sample of over 100 planetary nebulae, and found close agreement between these two diagnostics over densities ranging from 100 to 10$^4$\,cm$^{-3}$. While we used different atomic data for O$^{2+}$ to that used by \citet{wang2004}, we have verified that this has a negligible effect by recalculating electron densities using their tabulated values of the line ratios, with the newer atomic data; no values differ by more than 5\%. The difference can thus be clearly attributed to sample selection: their sample of 102 nebulae included only 2 of the 15 binary+\textit{adf} nebulae and thus consisted almost entirely of objects where this effect would be small.

The objects where the density diagnostics agree all have densities estimated from [S~{\sc ii}] which are higher than 600\,cm$^{-3}$; this group has an average \textit{adf} of 16.2. The objects where the diagnostics disagree have densities lower than 600\,cm$^{-3}$, and an average \textit{adf} of 54. This is further illustrated by Figure~\ref{adf_densityratio}, which shows the clear relation between \textit{adf} and density diagnostic divergence.

\begin{figure}
	\includegraphics[width=0.47\textwidth]{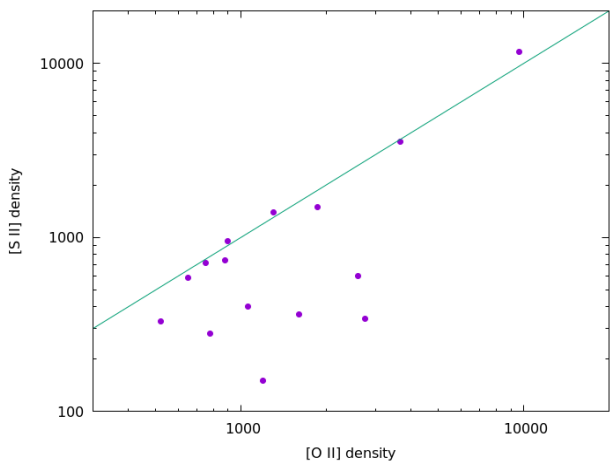}
	\caption{The relation between densities derived from [O~{\sc ii}] and [S~{\sc ii}] for the 15 nebulae with a close binary central star and a measured \textit{adf}}
	\label{oii_sii_comparison}
\end{figure}

We therefore conclude that low-density nebulae show the most extreme abundance discrepancy factors, and propose that in these extreme-\textit{adf} nebulae, recombination of O$^+$ significantly affects the ratio of the $\lambda$3727/$\lambda$3729 lines. The discrepancy between [O~{\sc ii}] and [S~{\sc ii}] densities will depend on the amount of H-deficient material and its density; where the two diagnostics agree, it could be because there is not enough H-deficient material to affect the [O~{\sc ii}] lines, or that the material has a similar density to the ambient gas. The latter appears to be the case in MPA~1759, for example, where O~{\sc ii} recombination line ratios, which should arise predominantly from the H-deficient clumps, indicate a density similar to the CEL value. The fact that [O~{\sc ii}] is always higher when the two densities do not agree indicates that the H-deficient material is generally denser than the ambient gas.

\citet{barlow2003} measured recombination line abundances of magnesium in a sample of planetary nebulae including several with large abundance discrepancies for O$^{2+}$, finding Mg/H abundances close to solar in all cases. They noted that nucleosynthetic processes which might enhance CNONe abundances would not be expected to affect the abundances of third-row elements. Thus, if the recombination lines of oxygen are emitted in cold knots formed in a nova-like outburst, recombination of S$^{+}$ would not be expected to significantly affect the [S~{\sc ii}] density diagnostic. The difference in abundance discrepancies between second and third row elements also rules out any mechanism for extreme abundance discrepancies that assumes a chemically homogeneous gas; temperature fluctuations (\citealt{peimbert1967}, \citealt{bautista2017}), quasi-neutral X-ray irradiated regions (\citealt{ercolano2009}) and $\kappa$-distributed electrons (\citealt{nicholls2012}) could not account for this, although they could still potentially account for non-extreme \textit{adfs} in H~{\sc ii} regions and PNe.

This finding also suggests that an [O~{\sc ii}] density significantly exceeding the [S~{\sc ii}] density may be used to infer an extreme-\textit{adf} nebula and hence a close binary central star, even if recombination lines are not detected. For the objects where we obtained only upper limits to the abundance discrepancy, two have this property: H\,2-29 and Sab\,41 (Table~\ref{upperlimittable}). We thus predict that deeper spectra of these two objects in particular should confirm a particularly large \textit{adf}.

\begin{figure}
	\includegraphics[width=0.47\textwidth]{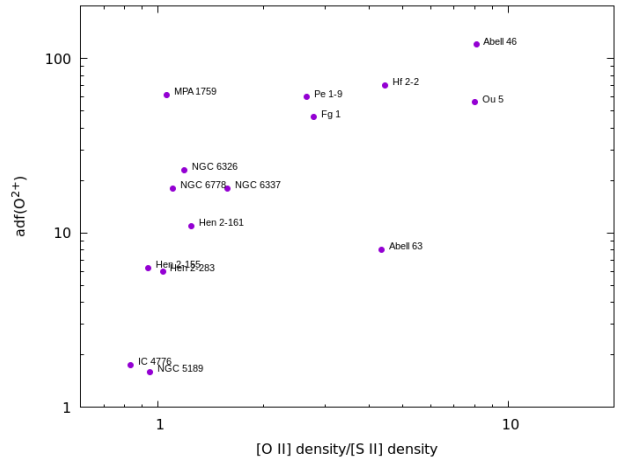}
	\caption{The relation between \textit{adf} and the ratio of density diagnostics. Recombination to the low-lying levels of [O~{\sc ii}] results in an overestimate of the density from this diagnostic, where the density from [S~{\sc ii}] is unaffected.}
	\label{adf_densityratio}
\end{figure}

Figure~\ref{density_adf_relation} shows \textit{adf}(O$^{2+}$) against electron densities from [O~{\sc ii}] and [S~{\sc ii}]. These plots are also consistent with [O~{\sc ii}] giving an overestimated density for high-\textit{adf} nebulae; the plot for [S~{\sc ii}] shows a strong correlation between density and \textit{adf}, with a Pearson correlation coefficient of $-$0.76 between the logarithmic values of each ($-$0.84 if the outlier NGC~5189 is excluded). The plot for [O~{\sc ii}] shows a much weaker correlation, with r=$-$0.38, or $-$0.54 if NGC~5189 is excluded. This nebula may be an outlier due to it being observed with an aperture that covered a small part of the nebula in a brighter and possibly denser region (\citealt{garcia-rojas2013}) whereas the other values were obtained from long-slit observations covering a larger fraction of the nebula.

\begin{figure*}
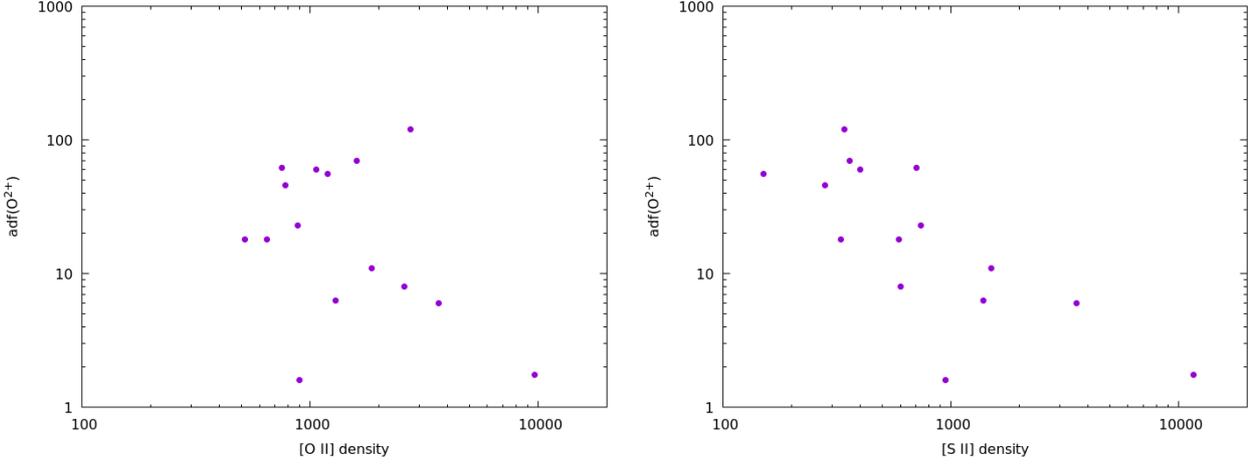

\includegraphics[width=0.47\textwidth]{figures/binaries_adf_oii.png}
\includegraphics[width=0.47\textwidth]{figures/binaries_adf_sii.png}
\caption{The relationship between \textit{adf}(O$^{2+}$) and (l) n$_e$([O~{\sc ii}]); (r) n$_e$([S~{\sc ii}])}
\label{density_adf_relation}
\end{figure*}


To investigate this correlation further, we have compiled the published electron densities for all the objects with measured \textit{adfs} shown in Figure~\ref{adfcompilation}. Figure~\ref{adf_density_all} shows the relation between \textit{adf} and electron density for the four most commonly used diagnostics, [O~{\sc ii}], [S~{\sc ii}], [Cl~{\sc iii}] and [Ar~{\sc iv}]. Several things are immediately apparent from this figure. First, H~{\sc ii} regions and planetary nebulae occupy distinctly different regions of the diagram. Secondly, there is a clearly delineated triangular region that contains almost all objects, showing that the electron density effectively places an upper limit on the abundance discrepancy factor. The lowest \textit{adf} values measured are around 1.3, while the highest, excluding a couple of outliers, have values up to \textit{adf}$_{\rm max} < $1.2$\times$10$^4$n$_e^{-0.8}$, indicated on the plots with a dashed line. \citet{robertsontessi2005} previously suggested a correlation between \textit{adf} and electron density, with a linear fit to a sample of 24 objects giving \textit{adf}=7.76$\times$10$^3$n$_e^{-0.635}$. And thirdly, the number of objects lying outside the triangular region is much higher for [O~{\sc ii}] than for the other diagnostics. This supports our conclusion that the [O~{\sc ii}] density is less reliable than estimates from third-row elements.

It further appears that planetary nebulae with binary central stars may occupy a region of this parameter space distinct from other planetary nebulae. Because the central star status of many PNe is not known, this relation is much less clear than the distinction between PNe and H~{\sc ii} regions in the plot, but tentatively we suggest that whereas H~{\sc ii} regions have low densities and low \textit{adfs}, ``normal" planetary nebulae have densities covering a large range and low \textit{adfs}, while PNe with binary central stars have densities covering a large range, and \textit{adf} roughly inversely proportional to the density. The two populations of planetary nebulae become indistinguishable at high densities, but based on this plot, a number of objects very likely to have close binary central stars are readily identifiable. 

\begin{figure*}
\includegraphics[width=0.47\textwidth]{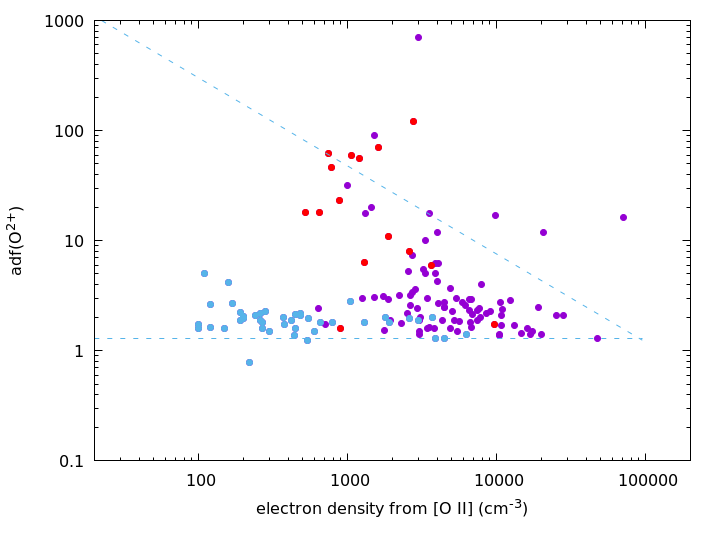}
\includegraphics[width=0.47\textwidth]{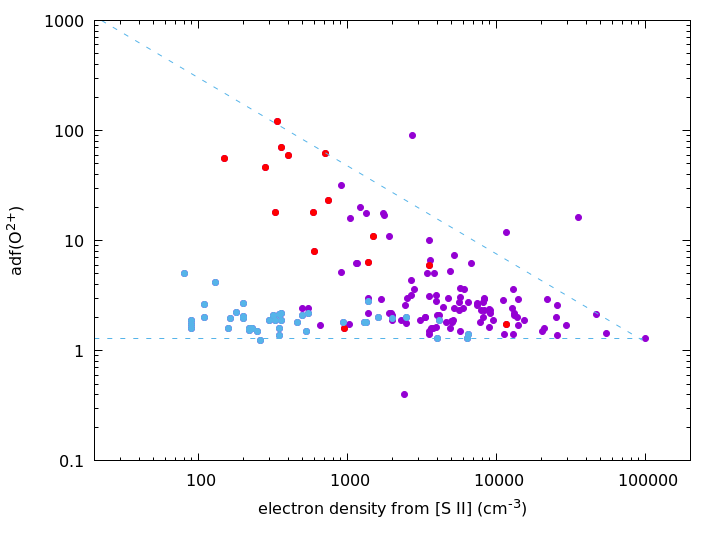}
\includegraphics[width=0.47\textwidth]{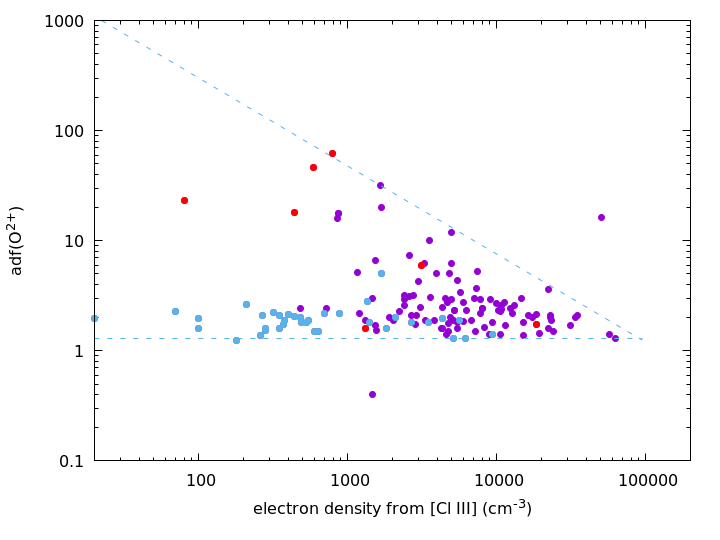}
\includegraphics[width=0.47\textwidth]{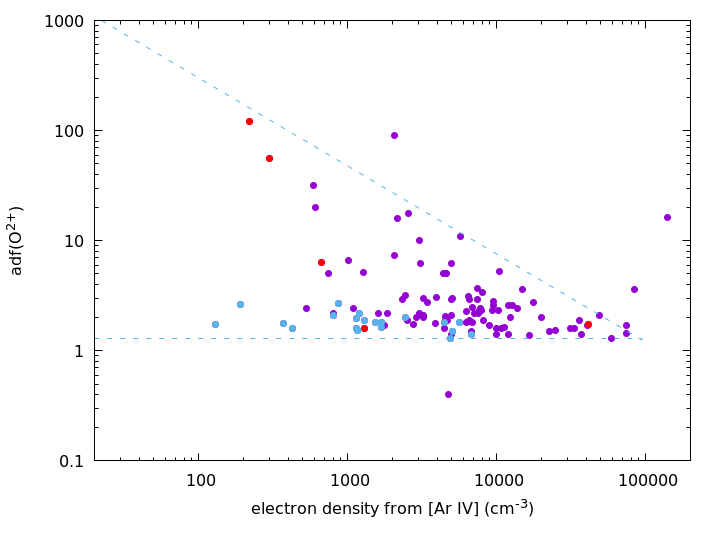}
\caption{\textit{adf} against electron density estimated from [O~{\sc ii}] (top left), [S~{\sc ii}] (top right), [Cl~{\sc iii}] (bottom left), and [Ar~{\sc iv}] (bottom right) line ratios. Planetary nebulae with binary central stars are shown with red points, other planetary nebulae are shown with purple points, and H~{\sc ii} regions are shown with light blue points.}
\label{adf_density_all}
\end{figure*}

\subsubsection{Chemical abundances}
\label{subsec:abundances}

Models of nebular abundances based on single star evolution predict a strong relationship between N/O and C/O and the mass of the progenitor star. Initially O-rich, AGB stars bring more carbon formed via helium fusion to their surfaces with each successive thermal pulse due to convective overshooting, leading ultimately to C/O ratios exceeding unity for stars with initial masses between about 1.5 and 3\,M$_\odot$. Above this, hot-bottom burning (HBB) can burn carbon into nitrogen, leading to lower C/O and higher N/O ratios (\citealt{herwig2005}, \citealt{karakas2014}). High N/O ratios indicate Type~{\sc i} planetary nebulae, which are thus believed to originate from higher mass progenitor stars (\citealt{peimbert1983}, \citealt{kingsburgh1994}).

These relationships break down for post-common envelope objects; the evolution of the surface abundances is terminated when the system enters the common envelope phase, which may occur at any point on the ascent of the AGB. Thus, for post-CE nebulae, a low C/O ratio could be due to early termination of the AGB, or higher initial mass and conversion of C to N via HBB, the value of N/O depending on which is the case. The post-CE population should then, broadly speaking, show systematically lower C/O ratios but have any range of N/O ratio, compared to the overall PN population (\citealt{demarco2009}). Hen~2-11 is an example of a system with very low C/O and N/O ratios, consistent with an early termination of post-AGB evolution by the CE phase (\citealt{jones2014a}).

In the full sample of 15 post-CE objects, we find a correlation between abundance ratios and the magnitude of the \textit{adf}: the lower the value of N/H or O/H, the higher the observed \textit{adf} (see Figure~\ref{abundances_adf_figure}). In the sample analysed in this work, Hen~2-283 has the highest abundance of both, and the lowest abundance discrepancy. In the full sample, the 5 objects with 12+log(O/H)$>$8.5 have a mean \textit{adf} of 7.1, while the 10 with 12+log(O/H)$<$8.5 have a mean \textit{adf} of 48.23. For N/H, the Pearson correlation coefficient is -0.66, with a $p$-value of 0.008, indicating a very low probability of the observed correlation arising by chance; for O/H, the Pearson correlation coefficient is -0.53 with a $p$-value of 0.043 when all 15 objects are included, and -0.79 with a $p$-value of 0.0009 if NGC~6326 is excluded.

\begin{figure*}
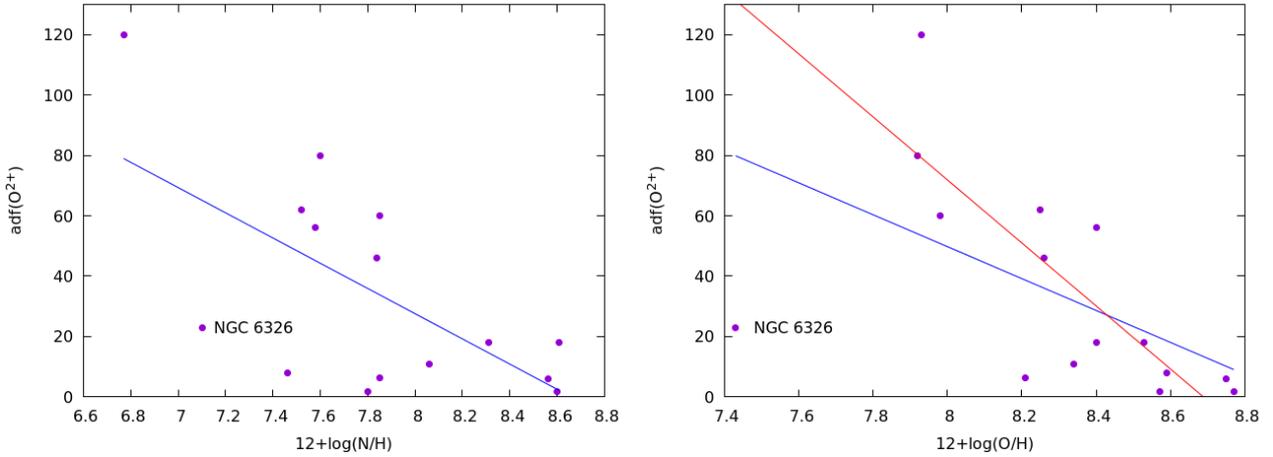

\includegraphics[width=0.47\textwidth]{figures/n_adf_relation.png}
\includegraphics[width=0.47\textwidth]{figures/o_adf_relation.png}
\caption{The anticorrelation between heavy element abundances determined from collisionally excited lines and the abundance discrepancy. (l) \textit{adf} versus N/H (r) \textit{adf} versus O/H. Linear least-squares fit lines are shown in blue; for O/H, the fit excluding the outlier NGC~6326 is shown in red.}
\label{abundances_adf_figure}
\end{figure*}

In terms of their space distribution, the extreme objects appear fairly similar to the whole Galactic population: 13 of the 15 objects (83\%) lie within $\pm$60$^{\circ}$ of the Galactic centre, compared to 2629 of 3545 (74\%) of all PNe in the HASH catalogue (\citealt{parker2016}; Fig.~\ref{spatialdistribution}). However, their heavy element abundances as measured from CELs are almost all sub-solar, and lower than those generally found for nebulae inside the Solar circle: \citet{kingsburgh1994} found average logarithmic abundances of N and O of 8.35 and 8.68 respectively, whereas for the sample of 15 extreme-\textit{adf} PNe we find 12+log(|N/H|)=8.11, and 12+log(|O/H|)=8.34. \citet{wesson2005} found values of 8.08 and 8.54 respectively, for a sample of northern PNe predominantly outside the solar circle.

\begin{figure*}
\includegraphics[width=\textwidth]{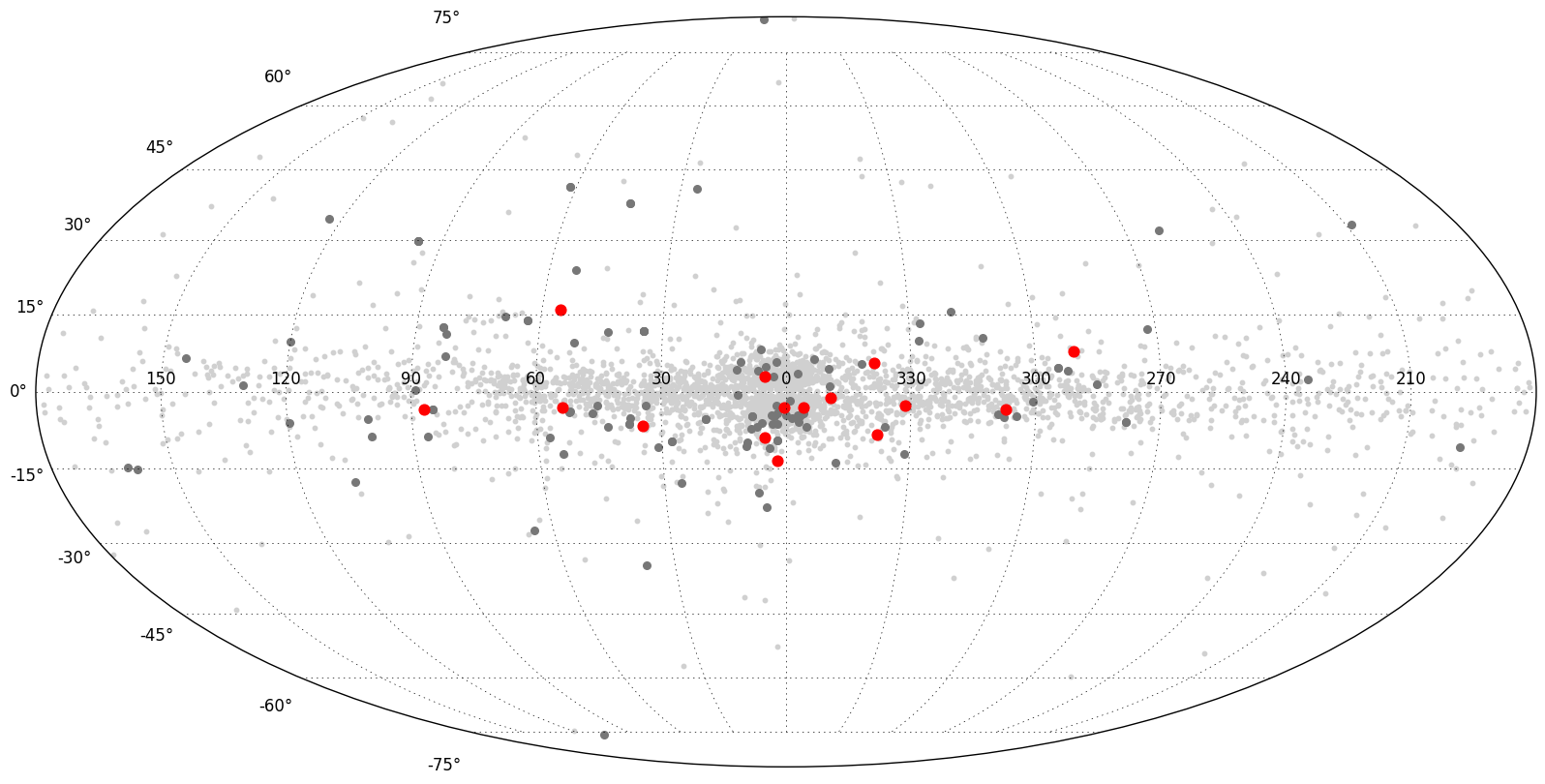}
\caption{The Galactic distribution of all planetary nebulae (small grey dots), those with measured \textit{adfs} (large grey dots), and the close binary+\textit{adf} objects (red dots)}
\label{spatialdistribution}
\end{figure*}

Values of N/O in the sample of 15 may be weakly anticorrelated with \textit{adf}: as shown in Figure~\ref{adf_n_o}, lower values of N/O are associated with higher \textit{adfs}. The five objects with log(N/O)$<-$0.4 have an average \textit{adf} of 57.15 (though this group includes IC~4776 with its extremely low \textit{adf} of 1.75), while the 10 objects with log(N/O)$>-$0.4 have an average \textit{adf} of 27.0 (though this group includes Hf~2-2 and Pe~1-9, both among the extremest of the extremes). The Pearson correlation coefficient for adf against log(N/O) is $-$0.41, or $-$0.75 if the outliers Abell~63 and IC~4776 are excluded. The $p$-values of these correlations are 0.13 and 0.003 respectively, the former implying a non-negligible probability of the apparent correlation having arisen by chance from uncorrelated data. The validity of the latter correlation would require the identification of factors clearly dividing both A~63 and IC~4776 from the other 13 objects; currently no such factors are apparent.

The large scatter about a vague anti-correlation may be consistent with the picture in which N/O in post-CE objects is a function of both the mass of the progenitor and the time of termination of the post-AGB evolution. However, important systematic uncertainties in determinations of N/O can arise from the recombination contribution to [O~{\sc iii}] $\lambda$4363, which would result in an overestimated temperature and underestimated CEL abundances; and also the recombination contribution to [O~{\sc ii}] $\lambda$3727/$\lambda$3729, which would result in an overestimate of the O$^{+}$ abundance. This would then cause a moderate overestimate of O/H and a much larger overestimate of N/H. We have recalculated N/O abundances in our sample applying corrections to the line fluxes affected by recombinations, and we find that these weak correlations are only slightly affected; points on the O/H versus \textit{adf} plot move in the direction of the correlation, while points on the N/O and N/H versus \textit{adf} plots do not move so much as to obscure the correlation found when corrections are not applied (Fig.~\ref{correctionsplot}). Finally, the ionization correction factor proposed by \citet{delgado-inglada2014} for N/H was subsequently found to introduce some correlation with ionization degree (\citealt{delgado-inglada2015}), which is not seen with the earlier \citet{kingsburgh1994} ICF. We investigated the effect of the choice of ICF on N/H for our objects and find that this may result in a difference of up to $\sim$20\% in the value of N/H.

\begin{figure}
\includegraphics[width=0.47\textwidth]{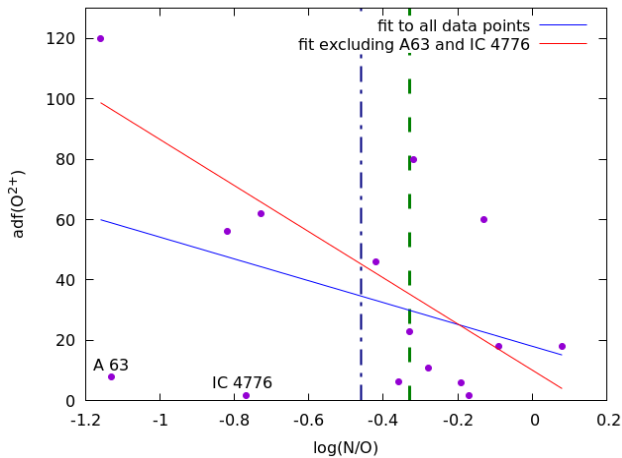}
\caption{Relation between N/O and \textit{adf}. Linear fits are shown both for the whole sample of 15 objects, and when the outliers A~63 and IC~4776 are excluded, which results in a better correlation and a steeper gradient to the fit. The dashed green line indicates the average value for nebulae analysed by \citet{kingsburgh1994}, while the dot-dashed blue line indicates the average for nebulae in \citet{wesson2005}.}
\label{adf_n_o}
\end{figure}

\begin{figure}
\includegraphics[width=0.47\textwidth]{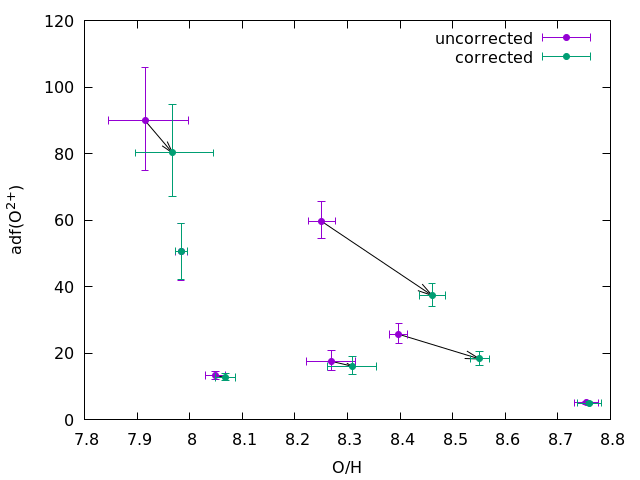}
\caption{An anticorrelation is found between \textit{adf}(O$^{2+}$) and O/H; plotted here are values of both when recombination contribution to $\lambda$4363 is not corrected for, and when it is corrected for using recombination line O$^{2+}$/H$^+$ abundances estimated from recombination lines. The resulting movement of points is often larger than the estimated statistical uncertainties, but is predominantly in the direction of the correlation.}
\label{correctionsplot}
\end{figure}


The Ne/O ratio in planetary nebulae is not expected to be affected by nucleosynthetic processes, and is observed to be approximately constant in a wide range of environments (e.g. \citealt{henry1990}). However, in our sample of 15, considerable scatter is seen, and Ne/O exceeds the solar value in 13 of the objects. The average value of Ne/O for the 15 PNe of our sample is $-$0.47, compared to the solar value of $-$0.61 (Figure~\ref{ne_o_adf}). This may be due to the probable underestimate of O/H abundances in high \textit{adf} objects discussed earlier. However, Ne abundances should also be underestimated for the same reason, and we note that large masses of neon were reported in the hydrogen-deficient knots of Abell~30 and Abell~58, comparable to the amounts seen in neon novae (\citealt{wesson2003},\citealt{wesson2008a}).

We also investigate the C/O ratio; however, where all other abundance ratios discussed in this section are derived from collisionally excited lines, we restrict our discussion of C/O to recombination lines only. Ultraviolet spectra are required to derive C/H from collisionally excited lines; a systematic uncertainty will then result from the inevitable combination of data covering different parts of the nebula. Recombination line ratios may also be subject to a systematic uncertainty when comparing them to CEL ratios, given that the two line types arise from physically distinct regions, but ratios of second row elements to each other appear similar whether derived from RLs or CELs. Figure~\ref{c_o_adf} shows the relation between C/O from RLs and \textit{adf}, with vertical lines indicating the average values from \citet{kingsburgh1994} (a sample of nebulae predominantly interior to the solar circle) and \citet{wesson2005} (predominantly exterior to the solar circle). There is no evident correlation between C/O and \textit{adf}, but most values are below the average values of larger samples of PNe. For N/O, an equal number of the sample have abundance ratios above and below the average of the \citet{kingsburgh1994} sample.

\begin{figure}
	\includegraphics[width=0.47\textwidth]{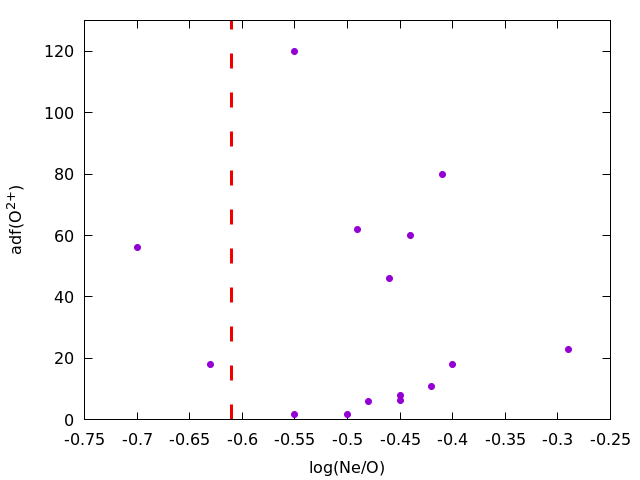}
	\caption{Ne/O plotted against \textit{adf}(O$^{2+}$). The vertical dashed red line indicates the solar value.}
	\label{ne_o_adf}
\end{figure}

\begin{figure}
	\includegraphics[width=0.47\textwidth]{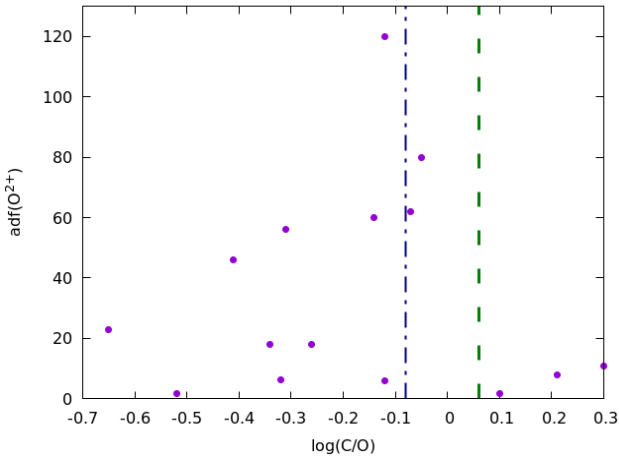}
	\caption{C/O plotted against \textit{adf}(O$^{2+}$). The dashed green line indicates the average value for nebulae analysed by \citet{kingsburgh1994}, while the dot-dashed blue line indicates the average for nebulae in \citet{wesson2005}.}
	\label{c_o_adf}
\end{figure}

Empirical abundance analysis methods assume a chemically homogenous nebula, and their application to two-phase nebulae introduces systematic uncertainties. To properly assess the extent to which the abundance trends we have found in this section are affected by this will require the creation of tailored three-dimensional photoionization models.

\subsubsection{Nebular morphology}

As would be expected, given that we are discussing planetary nebulae with binary central stars, the vast majority of the sample with measured \textit{adfs} are bipolar, with morphologies ranging from canonical bipolars (IC~4776, Ou~5, A~63, Hen~2-283) through to ring nebulae (Fg~1, Necklace, Hen~2-161) and more irregular bipolars (Hen~2-155, NGC~6778). NGC~6326 is particularly irregular, with multiple filamentary structures that make discerning the overall morphology difficult, although the imagery presented by \cite{miszalski2011b} shows a clear extension in roughly the East-West direction consistent with an elliptical or bipolar structure viewed at an intermediate inclination. The morphology of Abell~46 is somewhat unclear but is hypothesized by \cite{bond1990b} and \cite{pollacco1994} to be most likely a bipolar which has undergone some interaction with the interstellar medium. The morphology of MPA~J1759-3007 is similarly unclear, with the imaging of \cite{miszalski2009b} indicating the possible presence of bipolar lobes protruding from a central overdensity - unfortunately our shallow acquisition imagery shows only the bright central regions preventing us from confirming its bipolarity.  As such, only Hf~2-2 and Pe~1-9 represent the possible deviations from bipolarity present in the sample, with both showing multiple shells with a roughly circular projection. However, \cite{miszalski2009b} concluded that they too are most likely bipolars just viewed pole-on as is the case for NGC~6337, the end-on nature of which has been confirmed spectroscopically \citep{garcia-diaz2009}.

As such, given that all (or almost all) of the sample display somewhat bipolar morphologies typical of central star binarity, it is not possible to draw any more specific connection between \textit{adf} value and nebular morphology.

\section{Conclusions}
\label{sec:conclusions}
We have carried out a survey of planetary nebulae selected for having known close binary central stars. We have presented in this paper six new discoveries of elevated or extreme abundance discrepancies; the rest of the sample yielded only weak upper limits for the abundance discrepancy. This nearly doubles the sample of objects known to have both an extreme abundance discrepancy and a close binary central star. The selection strategy of our survey alone clearly demonstrates that close binarity and extreme abundance discrepancies are closely linked. We have carried out further statistical analysis of the complete sample of objects with measured \textit{adfs}, and have found that the population of nebulae with close binary central stars has a distribution of \textit{adfs} very different from that of the whole population, while for nebulae with hydrogen-deficient central stars, the distribution of \textit{adfs} is statistically indistinguishable from that of the whole population. This argues strongly against VLTPs as a source of hydrogen-deficient material in planetary nebulae, and in favour of nova-like eruptions.

Given that the relationship between pre- and post-CE orbital parameters is extremely uncertain \citep{zorotovic2010,demarco2011}, it is particularly intriguing that high \textit{adf}s present an apparent cut-off in post-CE orbital period at around 1.15~d.  This cut-off offers support for the hypothesis that CE phases that occur later in the AGB evolution (when the envelope is less bound and thus will require less orbital energy to unbind it) are less likely to lead to whatever process gives rise to the high \textit{adf}s observed in short-period, post-CE PNe.  Furthermore, we find an anticorrelation between N/O and \textit{adf}\footnote{It is worth noting the calculated N/O ratios are rather uncertain, as they rely on the use of ICFs (see section \ref{subsec:abundances}).}. Again, this could be seen as support for the hypothesis that early CE phases give rise to the highest \textit{adf}s as the N/O ratio is expected to rise throughout the progenitor's AGB evolution.  However, it is important to note that there is significant evidence that the PNe observed around post-CE central stars may not represent the entirety of the envelope - with many having measured masses significantly lower than would be expected \citep{corradi2011b,corradi2014}. This is perhaps indicative that the envelope is ejected in multiple phases \citep[known as a grazing envelope evolution][]{soker2015}, with only the most recent being observed as a PN. As such, the correlation between short periods and high \textit{adf}s may be less likely due to the evolutionary phase at which the binary entered the common envelope, but rather a reflection that shorter post-CE periods are required in order for the binary to inject high-metallicity material into the nebula.  This would be consistent with the idea that a nova-like process is the origin of these high \textit{adf}s, with the eruption being driven by material transferred back from the companion \citep[many of which have been observed to be highly inflated and near Roche-lobe-filling][]{jones2015a} - something which would occur only in the shortest period systems.

The observed anti-correlation between nebular density and \textit{adf} may be a result of nebular age, if the density contrast between H-deficient clumps and the `normal' nebula increases with time. This might be expected if the H-deficient material is photo-evaporating from dense neutral cores. Literature estimates of nebular age are only available for Fg~1 (5000 years; \citealt{boffin2012}) and NGC~6337 (20--24\,000 years; \citealt{hillwig2010}); despite their very different ages, these two nebulae have similar electron densities, and Fg~1 has a much higher \textit{adf}, which is inconsistent with a simple time evolution of \textit{adf}.  Another possibility is that the \textit{adf} and the nebular density are both determined by the nature of the CE ejection - a very poorly understood process \citep[see e.g.][]{ivanova2013,ohlmann2016,grichener2018}. Furthermore, it is a complex interplay between ionization and shocks which leads to the formation of the nebular shell from the ejected envelope \citep[see e.g.][]{garcia-segura2018}, and as such, it is not only the ejection process but also the post-CE evolution of the central star that causes the nebula to form.  In any case, the results presented here are indicative that an earlier termination of the AGB tends to result in both a higher \textit{adf} and a lower average electron density.

A possible link between the evolutionary state of the primary upon entering the CE phase and a high \textit{adf} was previously highlighted by \citet{jones2016}, who pointed out that the highest known \textit{adf} was observed in Abell~46, a PN where the mass of the central star star is found to lie on the limit between red giant branch (RGB) and AGB core masses. As such, the CE phase in that system must have occurred either on the RGB or on the very early AGB which, given the findings presented here, is unlikely to be coincidental.

We thus conclude that abundance discrepancy factors are a reliable way to infer the presence of a binary central star; any object with an extreme abundance discrepancy should contain one. However, in the contrary hypothesis, binarity is a necessary but not sufficient condition for extreme \textit{adfs}; the CE event must also occur not too close to the end of the AGB evolution. The latter case would give rise to the post-CE nebulae that have a high electron density and a `normal' \textit{adf}. Given a planetary nebula with a low measured \textit{adf} and high electron density but with morphology indicative of binarity, we would predict that the binary period should be much longer than 1~d. A fortuitous test of this latter prediction is possible thanks to the publication, shortly after we first submitted this paper, of the discovery of a binary in the hourglass-shaped nebula planetary nebula MyCn~18 (\citealt{miszalski2018}). Based on its measured \textit{adf} of 1.8, and densities of 5\,000--10\,000\,cm$^{-3}$ (depending on the diagnostic used; \citealt{tsamis2004}), we would have predicted a period much longer than 1~d, and indeed, \citet{miszalski2018} report a period of 18.15~d.

\section*{Acknowledgements}

We thank the anonymous referee for their detailed and helpful review. RW was supported by European Research Grant SNDUST 694520. JGR acknowledges support from an Advanced Fellowship from the Severo Ochoa excellence program (SEV-2015-0548). This research has been supported by the Spanish Ministry of Economy and Competitiveness (MINECO) under the grant AYA2017-83383-P. This work is based on observations made with ESO Telescopes at the La Silla Paranal Observatory under programme IDs 087.D-0174(A), 088.D-0573(A), 089.D-0357(A), 090.D-0449(A), 090.D-0693(A), 091.D-0475(A), 092.D-0449(A), 093.D-0038(A), 094.D-0091(A), 096.D-0080(A), 096.D-0234(A), and 097.D-0351(A).

This research made use of NASA's Astrophysics Data System; the SIMBAD database, operated at CDS, Strasbourg, France; APLpy, an open-source plotting package for \sc{python} hosted at \href{http://aplpy.github.io}{http://aplpy.github.io}; Astropy, a community-developed core \sc{python} package for Astronomy \citep{astropycollaboration2013}; matplotlib, a \sc{python} library for publication quality graphics \citep{barrett2005}; CHIANTI, a collaborative project involving George Mason University, the University of Michigan (USA), and the University of Cambridge (UK).




\bibliographystyle{mnras}
\bibliography{references} 

\begin{thebibliography}{}
\makeatletter
\relax
\def\mn@urlcharsother{\let\do\@makeother \do\$\do\&\do\#\do\^\do\_\do\%\do\~}
\def\mn@doi{\begingroup\mn@urlcharsother \@ifnextchar [ {\mn@doi@}
  {\mn@doi@[]}}
\def\mn@doi@[#1]#2{\def\@tempa{#1}\ifx\@tempa\@empty \href
  {http://dx.doi.org/#2} {doi:#2}\else \href {http://dx.doi.org/#2} {#1}\fi
  \endgroup}
\def\mn@eprint#1#2{\mn@eprint@#1:#2::\@nil}
\def\mn@eprint@arXiv#1{\href {http://arxiv.org/abs/#1} {{\tt arXiv:#1}}}
\def\mn@eprint@dblp#1{\href {http://dblp.uni-trier.de/rec/bibtex/#1.xml}
  {dblp:#1}}
\def\mn@eprint@#1:#2:#3:#4\@nil{\def\@tempa {#1}\def\@tempb {#2}\def\@tempc
  {#3}\ifx \@tempc \@empty \let \@tempc \@tempb \let \@tempb \@tempa \fi \ifx
  \@tempb \@empty \def\@tempb {arXiv}\fi \@ifundefined
  {mn@eprint@\@tempb}{\@tempb:\@tempc}{\expandafter \expandafter \csname
  mn@eprint@\@tempb\endcsname \expandafter{\@tempc}}}

\bibitem[\protect\citeauthoryear{{Akras}, {Boumis}, {Meaburn}, {Alikakos},
  {L{\'o}pez}  \& {Gon{\c c}alves}}{{Akras} et~al.}{2015}]{akras2015}
{Akras} S.,  {Boumis} P.,  {Meaburn} J.,  {Alikakos} J.,  {L{\'o}pez} J.~A.,
  {Gon{\c c}alves} D.~R.,  2015, \mn@doi [\mnras] {10.1093/mnras/stv1468},
  \href {http://adsabs.harvard.edu/abs/2015MNRAS.452.2911A} {452, 2911}

\bibitem[\protect\citeauthoryear{{Aller} \& {Keyes}}{{Aller} \&
  {Keyes}}{1987}]{aller1987}
{Aller} L.~H.,  {Keyes} C.~D.,  1987, \mn@doi [\apjs] {10.1086/191230}, \href
  {http://adsabs.harvard.edu/abs/1987ApJS...65..405A} {65, 405}

\bibitem[\protect\citeauthoryear{{Appenzeller} et~al.,}{{Appenzeller}
  et~al.}{1998}]{appenzeller1998}
{Appenzeller} I.,  et~al., 1998, The Messenger, \href
  {http://adsabs.harvard.edu/abs/1998Msngr..94....1A} {94, 1}

\bibitem[\protect\citeauthoryear{{Astropy Collaboration} et~al.,}{{Astropy
  Collaboration} et~al.}{2013}]{astropycollaboration2013}
{Astropy Collaboration} et~al., 2013, \mn@doi [\aap]
  {10.1051/0004-6361/201322068}, \href
  {http://adsabs.harvard.edu/abs/2013A%26A...558A..33A} {558, A33}

\bibitem[\protect\citeauthoryear{{Barlow}, {Liu}, {.~P{\'e}quignot}, {Storey},
  {Tsamis}  \& {Morisset}}{{Barlow} et~al.}{2003}]{barlow2003}
{Barlow} M.~J.,  {Liu} X.-W.,  {.~P{\'e}quignot} D.,  {Storey} P.~J.,  {Tsamis}
  Y.~G.,   {Morisset} C.,  2003, in {Kwok} S.,  {Dopita} M.,   {Sutherland} R.,
   eds,  IAU Symposium Vol. 209, Planetary Nebulae: Their Evolution and Role in
  the Universe. p.~373

\bibitem[\protect\citeauthoryear{{Barrett}, {Hunter}, {Miller}, {Hsu}  \&
  {Greenfield}}{{Barrett} et~al.}{2005}]{barrett2005}
{Barrett} P.,  {Hunter} J.,  {Miller} J.~T.,  {Hsu} J.-C.,   {Greenfield} P.,
  2005, in {Shopbell} P.,  {Britton} M.,   {Ebert} R.,  eds,  Astronomical
  Society of the Pacific Conference Series Vol. 347, Astronomical Data Analysis
  Software and Systems XIV. p.~91

\bibitem[\protect\citeauthoryear{{Basurah}, {Ali}, {Dopita}, {Alsulami}, {Amer}
   \& {Alruhaili}}{{Basurah} et~al.}{2016}]{basurah2016}
{Basurah} H.~M.,  {Ali} A.,  {Dopita} M.~A.,  {Alsulami} R.,  {Amer} M.~A.,
  {Alruhaili} A.,  2016, \mnras, \href
  {http://adsabs.harvard.edu/abs/2016MNRAS.458.2694B} {458, 2694}

\bibitem[\protect\citeauthoryear{{Bautista} \& {Ahmed}}{{Bautista} \&
  {Ahmed}}{2017}]{bautista2017}
{Bautista} M.~A.,  {Ahmed} E.~E.,  2017, preprint, \href
  {http://adsabs.harvard.edu/abs/2017arXiv170907945B} {} (\mn@eprint {arXiv}
  {1709.07945})

\bibitem[\protect\citeauthoryear{{Boffin}, {Miszalski}, {Rauch}, {Jones},
  {Corradi}, {Napiwotzki}, {Day-Jones}  \& {K{\"o}ppen}}{{Boffin}
  et~al.}{2012a}]{boffin2012}
{Boffin} H.~M.~J.,  {Miszalski} B.,  {Rauch} T.,  {Jones} D.,  {Corradi}
  R.~L.~M.,  {Napiwotzki} R.,  {Day-Jones} A.~C.,   {K{\"o}ppen} J.,  2012a,
  \mn@doi [Science] {10.1126/science.1225386}, \href
  {http://adsabs.harvard.edu/abs/2012Sci...338..773B} {338, 773}

\bibitem[\protect\citeauthoryear{{Boffin}, {Miszalski}  \& {Jones}}{{Boffin}
  et~al.}{2012b}]{boffin2012a}
{Boffin} H.~M.~J.,  {Miszalski} B.,   {Jones} D.,  2012b, \mn@doi [\aap]
  {10.1051/0004-6361/201218934}, \href
  {http://adsabs.harvard.edu/abs/2012A%26A...545A.146B} {545, A146}

\bibitem[\protect\citeauthoryear{{Bond}}{{Bond}}{2000}]{bond2000}
{Bond} H.~E.,  2000, in {Kastner} J.~H.,  {Soker} N.,   {Rappaport} S.,  eds,
  Astronomical Society of the Pacific Conference Series Vol. 199, Asymmetrical
  Planetary Nebulae II: From Origins to Microstructures. p.~115 (\mn@eprint {}
  {astro-ph/9909516})

\bibitem[\protect\citeauthoryear{{Bond} \& {Ciardullo}}{{Bond} \&
  {Ciardullo}}{1990}]{bond1990b}
{Bond} H.~E.,  {Ciardullo} R.,  1990, in {Cacciari} C.,  {Clementini} G.,  eds,
   Astronomical Society of the Pacific Conference Series Vol. 11, Confrontation
  Between Stellar Pulsation and Evolution. pp 529--533

\bibitem[\protect\citeauthoryear{{Borkowski}, {Harrington}  \&
  {Tsvetanov}}{{Borkowski} et~al.}{1995}]{borkowski1995}
{Borkowski} K.~J.,  {Harrington} J.~P.,   {Tsvetanov} Z.~I.,  1995, \mn@doi
  [\apjl] {10.1086/309643}, \href
  {http://adsabs.harvard.edu/abs/1995ApJ...449L.143B} {449, L143}

\bibitem[\protect\citeauthoryear{{Corradi} et~al.,}{{Corradi}
  et~al.}{2011}]{corradi2011b}
{Corradi} R.~L.~M.,  et~al., 2011, \mn@doi [\mnras]
  {10.1111/j.1365-2966.2010.17523.x}, \href
  {http://adsabs.harvard.edu/abs/2011MNRAS.410.1349C} {410, 1349}

\bibitem[\protect\citeauthoryear{{Corradi} et~al.,}{{Corradi}
  et~al.}{2014}]{corradi2014}
{Corradi} R.~L.~M.,  et~al., 2014, \mn@doi [\mnras] {10.1093/mnras/stu703},
  \href {http://adsabs.harvard.edu/abs/2014MNRAS.441.2799C} {441, 2799}

\bibitem[\protect\citeauthoryear{{Corradi}, {Garc{\'{\i}}a-Rojas}, {Jones}  \&
  {Rodr{\'{\i}}guez-Gil}}{{Corradi} et~al.}{2015}]{corradi2015}
{Corradi} R.~L.~M.,  {Garc{\'{\i}}a-Rojas} J.,  {Jones} D.,
  {Rodr{\'{\i}}guez-Gil} P.,  2015, \mn@doi [\apj]
  {10.1088/0004-637X/803/2/99}, \href
  {http://adsabs.harvard.edu/abs/2015ApJ...803...99C} {803, 99}

\bibitem[\protect\citeauthoryear{{Davey}, {Storey}  \& {Kisielius}}{{Davey}
  et~al.}{2000}]{davey2000}
{Davey} A.~R.,  {Storey} P.~J.,   {Kisielius} R.,  2000, \mn@doi [\aaps]
  {10.1051/aas:2000139}, \href
  {http://adsabs.harvard.edu/abs/2000A%26AS..142...85D} {142, 85}

\bibitem[\protect\citeauthoryear{{De Marco}}{{De Marco}}{2009}]{demarco2009}
{De Marco} O.,  2009, \mn@doi [\pasp] {10.1086/597765}, \href
  {http://adsabs.harvard.edu/abs/2009PASP..121..316D} {121, 316}

\bibitem[\protect\citeauthoryear{{De Marco}, {Passy}, {Moe}, {Herwig}, {Mac
  Low}  \& {Paxton}}{{De Marco} et~al.}{2011}]{demarco2011}
{De Marco} O.,  {Passy} J.-C.,  {Moe} M.,  {Herwig} F.,  {Mac Low} M.-M.,
  {Paxton} B.,  2011, \mn@doi [\mnras] {10.1111/j.1365-2966.2010.17891.x},
  \href {http://adsabs.harvard.edu/abs/2011MNRAS.411.2277D} {411, 2277}

\bibitem[\protect\citeauthoryear{{De Marco}, {Passy}, {Frew}, {Moe}  \&
  {Jacoby}}{{De Marco} et~al.}{2013}]{demarco2013}
{De Marco} O.,  {Passy} J.-C.,  {Frew} D.~J.,  {Moe} M.,   {Jacoby} G.~H.,
  2013, \mn@doi [\mnras] {10.1093/mnras/sts180}, \href
  {http://adsabs.harvard.edu/abs/2013MNRAS.428.2118D} {428, 2118}

\bibitem[\protect\citeauthoryear{{Delgado-Inglada}, {Morisset}  \&
  {Stasi{\'n}ska}}{{Delgado-Inglada} et~al.}{2014}]{delgado-inglada2014}
{Delgado-Inglada} G.,  {Morisset} C.,   {Stasi{\'n}ska} G.,  2014, \mn@doi
  [\mnras] {10.1093/mnras/stu341}, \href
  {http://adsabs.harvard.edu/abs/2014MNRAS.440..536D} {440, 536}

\bibitem[\protect\citeauthoryear{{Delgado-Inglada}, {Rodr{\'{\i}}guez},
  {Peimbert}, {Stasi{\'n}ska}  \& {Morisset}}{{Delgado-Inglada}
  et~al.}{2015}]{delgado-inglada2015}
{Delgado-Inglada} G.,  {Rodr{\'{\i}}guez} M.,  {Peimbert} M.,  {Stasi{\'n}ska}
  G.,   {Morisset} C.,  2015, \mn@doi [\mnras] {10.1093/mnras/stv388}, \href
  {http://adsabs.harvard.edu/abs/2015MNRAS.449.1797D} {449, 1797}

\bibitem[\protect\citeauthoryear{{Dere}, {Landi}, {Mason}, {Monsignori Fossi}
  \& {Young}}{{Dere} et~al.}{1997}]{dere1997}
{Dere} K.~P.,  {Landi} E.,  {Mason} H.~E.,  {Monsignori Fossi} B.~C.,   {Young}
  P.~R.,  1997, \mn@doi [\aaps] {10.1051/aas:1997368}, \href
  {http://adsabs.harvard.edu/abs/1997A%26AS..125..149D} {125, 149}

\bibitem[\protect\citeauthoryear{{Ercolano}}{{Ercolano}}{2009}]{ercolano2009}
{Ercolano} B.,  2009, \mn@doi [\mnras] {10.1111/j.1745-3933.2009.00686.x},
  \href {http://adsabs.harvard.edu/abs/2009MNRAS.397L..69E} {397, L69}

\bibitem[\protect\citeauthoryear{{Ercolano}, {Wesson}, {Zhang}, {Barlow}, {De
  Marco}, {Rauch}  \& {Liu}}{{Ercolano} et~al.}{2004}]{ercolano2004}
{Ercolano} B.,  {Wesson} R.,  {Zhang} Y.,  {Barlow} M.~J.,  {De Marco} O.,
  {Rauch} T.,   {Liu} X.-W.,  2004, \mn@doi [\mnras]
  {10.1111/j.1365-2966.2004.08218.x}, \href
  {http://adsabs.harvard.edu/abs/2004MNRAS.354..558E} {354, 558}

\bibitem[\protect\citeauthoryear{{Escalante} \& {Victor}}{{Escalante} \&
  {Victor}}{1990}]{escalante1990}
{Escalante} V.,  {Victor} G.~A.,  1990, \mn@doi [\apjs] {10.1086/191479}, \href
  {http://adsabs.harvard.edu/abs/1990ApJS...73..513E} {73, 513}

\bibitem[\protect\citeauthoryear{{Escalante}, {Morisset}  \&
  {Georgiev}}{{Escalante} et~al.}{2012}]{escalante2012}
{Escalante} V.,  {Morisset} C.,   {Georgiev} L.,  2012, \mn@doi [\mnras]
  {10.1111/j.1365-2966.2012.21862.x}, \href
  {http://adsabs.harvard.edu/abs/2012MNRAS.426.2318E} {426, 2318}

\bibitem[\protect\citeauthoryear{{Esteban}, {Garc{\'{\i}}a-Rojas}, {Carigi},
  {Peimbert}, {Bresolin}, {L{\'o}pez-S{\'a}nchez}  \& {Mesa-Delgado}}{{Esteban}
  et~al.}{2014}]{esteban2014}
{Esteban} C.,  {Garc{\'{\i}}a-Rojas} J.,  {Carigi} L.,  {Peimbert} M.,
  {Bresolin} F.,  {L{\'o}pez-S{\'a}nchez} A.~R.,   {Mesa-Delgado} A.,  2014,
  \mn@doi [\mnras] {10.1093/mnras/stu1177}, \href
  {http://adsabs.harvard.edu/abs/2014MNRAS.443..624E} {443, 624}

\bibitem[\protect\citeauthoryear{{Ferland}}{{Ferland}}{1992}]{ferland1992}
{Ferland} G.~J.,  1992, \mn@doi [\apjl] {10.1086/186349}, \href
  {http://adsabs.harvard.edu/abs/1992ApJ...389L..63F} {389, L63}

\bibitem[\protect\citeauthoryear{{Garc{\'{\i}}a-D{\'{\i}}az}, {Clark},
  {L{\'o}pez}, {Steffen}  \& {Richer}}{{Garc{\'{\i}}a-D{\'{\i}}az}
  et~al.}{2009}]{garcia-diaz2009}
{Garc{\'{\i}}a-D{\'{\i}}az} M.~T.,  {Clark} D.~M.,  {L{\'o}pez} J.~A.,
  {Steffen} W.,   {Richer} M.~G.,  2009, \mn@doi [\apj]
  {10.1088/0004-637X/699/2/1633}, \href
  {http://adsabs.harvard.edu/abs/2009ApJ...699.1633G} {699, 1633}

\bibitem[\protect\citeauthoryear{{Garc{\'{\i}}a-Rojas} \&
  {Esteban}}{{Garc{\'{\i}}a-Rojas} \& {Esteban}}{2007}]{garcia-rojas2007}
{Garc{\'{\i}}a-Rojas} J.,  {Esteban} C.,  2007, \mn@doi [\apj]
  {10.1086/521871}, \href {http://adsabs.harvard.edu/abs/2007ApJ...670..457G}
  {670, 457}

\bibitem[\protect\citeauthoryear{{Garc{\'{\i}}a-Rojas}, {Pe{\~n}a}, {Morisset},
  {Delgado-Inglada}, {Mesa-Delgado}  \& {Ruiz}}{{Garc{\'{\i}}a-Rojas}
  et~al.}{2013}]{garcia-rojas2013}
{Garc{\'{\i}}a-Rojas} J.,  {Pe{\~n}a} M.,  {Morisset} C.,  {Delgado-Inglada}
  G.,  {Mesa-Delgado} A.,   {Ruiz} M.~T.,  2013, \mn@doi [\aap]
  {10.1051/0004-6361/201322354}, \href
  {http://adsabs.harvard.edu/abs/2013A%26A...558A.122G} {558, A122}

\bibitem[\protect\citeauthoryear{{Garc{\'{\i}}a-Rojas}, {Corradi}, {Monteiro},
  {Jones}, {Rodr{\'{\i}}guez-Gil}  \& {Cabrera-Lavers}}{{Garc{\'{\i}}a-Rojas}
  et~al.}{2016}]{garcia-rojas2016}
{Garc{\'{\i}}a-Rojas} J.,  {Corradi} R.~L.~M.,  {Monteiro} H.,  {Jones} D.,
  {Rodr{\'{\i}}guez-Gil} P.,   {Cabrera-Lavers} A.,  2016, \mn@doi [\apjl]
  {10.3847/2041-8205/824/2/L27}, \href
  {http://adsabs.harvard.edu/abs/2016ApJ...824L..27G} {824, L27}

\bibitem[\protect\citeauthoryear{{Garc{\'{\i}}a-Segura}, {Ricker}  \&
  {Taam}}{{Garc{\'{\i}}a-Segura} et~al.}{2018}]{garcia-segura2018}
{Garc{\'{\i}}a-Segura} G.,  {Ricker} P.~M.,   {Taam} R.~E.,  2018, \mn@doi
  [\apj] {10.3847/1538-4357/aac08c}, \href
  {http://adsabs.harvard.edu/abs/2018ApJ...860...19G} {860, 19}

\bibitem[\protect\citeauthoryear{{Grichener}, {Sabach}  \& {Soker}}{{Grichener}
  et~al.}{2018}]{grichener2018}
{Grichener} A.,  {Sabach} E.,   {Soker} N.,  2018, \mn@doi [\mnras]
  {10.1093/mnras/sty1178}, \href
  {http://adsabs.harvard.edu/abs/2018MNRAS.478.1818G} {478, 1818}

\bibitem[\protect\citeauthoryear{{Henry}}{{Henry}}{1990}]{henry1990}
{Henry} R.~B.~C.,  1990, \mn@doi [\apj] {10.1086/168833}, \href
  {http://adsabs.harvard.edu/abs/1990ApJ...356..229H} {356, 229}

\bibitem[\protect\citeauthoryear{{Herwig}}{{Herwig}}{2005}]{herwig2005}
{Herwig} F.,  2005, \mn@doi [\araa] {10.1146/annurev.astro.43.072103.150600},
  \href {http://adsabs.harvard.edu/abs/2005ARA%26A..43..435H} {43, 435}

\bibitem[\protect\citeauthoryear{{Hillwig}, {Bond}, {Af{\c s}ar}  \& {De
  Marco}}{{Hillwig} et~al.}{2010}]{hillwig2010}
{Hillwig} T.~C.,  {Bond} H.~E.,  {Af{\c s}ar} M.,   {De Marco} O.,  2010,
  \mn@doi [\aj] {10.1088/0004-6256/140/2/319}, \href
  {http://adsabs.harvard.edu/abs/2010AJ....140..319H} {140, 319}

\bibitem[\protect\citeauthoryear{{Howarth}}{{Howarth}}{1983}]{howarth1983}
{Howarth} I.~D.,  1983, \mn@doi [\mnras] {10.1093/mnras/203.2.301}, \href
  {http://adsabs.harvard.edu/abs/1983MNRAS.203..301H} {203, 301}

\bibitem[\protect\citeauthoryear{{Huckvale} et~al.,}{{Huckvale}
  et~al.}{2013}]{huckvale2013}
{Huckvale} L.,  et~al., 2013, \mn@doi [\mnras] {10.1093/mnras/stt1109}, \href
  {http://adsabs.harvard.edu/abs/2013MNRAS.434.1505H} {434, 1505}

\bibitem[\protect\citeauthoryear{{Ivanova} et~al.,}{{Ivanova}
  et~al.}{2013}]{ivanova2013}
{Ivanova} N.,  et~al., 2013, \mn@doi [\aapr] {10.1007/s00159-013-0059-2}, \href
  {http://adsabs.harvard.edu/abs/2013A%26ARv..21...59I} {21, 59}

\bibitem[\protect\citeauthoryear{{Jones} \& {Boffin}}{{Jones} \&
  {Boffin}}{2017}]{jones2017}
{Jones} D.,  {Boffin} H.~M.~J.,  2017, \mn@doi [Nature Astronomy]
  {10.1038/s41550-017-0117}, \href
  {http://adsabs.harvard.edu/abs/2017NatAs...1E.117J} {1, 0117}

\bibitem[\protect\citeauthoryear{{Jones}, {Boffin}, {Miszalski}, {Wesson},
  {Corradi}  \& {Tyndall}}{{Jones} et~al.}{2014}]{jones2014a}
{Jones} D.,  {Boffin} H.~M.~J.,  {Miszalski} B.,  {Wesson} R.,  {Corradi}
  R.~L.~M.,   {Tyndall} A.~A.,  2014, \mn@doi [\aap]
  {10.1051/0004-6361/201322797}, \href
  {http://adsabs.harvard.edu/abs/2014A%26A...562A..89J} {562, A89}

\bibitem[\protect\citeauthoryear{{Jones}, {Boffin}, {Rodr{\'{\i}}guez-Gil},
  {Wesson}, {Corradi}, {Miszalski}  \& {Mohamed}}{{Jones}
  et~al.}{2015}]{jones2015a}
{Jones} D.,  {Boffin} H.~M.~J.,  {Rodr{\'{\i}}guez-Gil} P.,  {Wesson} R.,
  {Corradi} R.~L.~M.,  {Miszalski} B.,   {Mohamed} S.,  2015, \mn@doi [\aap]
  {10.1051/0004-6361/201425454}, \href
  {http://adsabs.harvard.edu/abs/2015A%26A...580A..19J} {580, A19}

\bibitem[\protect\citeauthoryear{{Jones}, {Wesson}, {Garc{\'{\i}}a-Rojas},
  {Corradi}  \& {Boffin}}{{Jones} et~al.}{2016}]{jones2016}
{Jones} D.,  {Wesson} R.,  {Garc{\'{\i}}a-Rojas} J.,  {Corradi} R.~L.~M.,
  {Boffin} H.~M.~J.,  2016, \mn@doi [\mnras] {10.1093/mnras/stv2519}, \href
  {http://adsabs.harvard.edu/abs/2016MNRAS.455.3263J} {455, 3263}

\bibitem[\protect\citeauthoryear{{Karakas} \& {Lattanzio}}{{Karakas} \&
  {Lattanzio}}{2014}]{karakas2014}
{Karakas} A.~I.,  {Lattanzio} J.~C.,  2014, \mn@doi [\pasa]
  {10.1017/pasa.2014.21}, \href
  {http://adsabs.harvard.edu/abs/2014PASA...31...30K} {31, e030}

\bibitem[\protect\citeauthoryear{{Kingsburgh} \& {Barlow}}{{Kingsburgh} \&
  {Barlow}}{1994}]{kingsburgh1994}
{Kingsburgh} R.~L.,  {Barlow} M.~J.,  1994, \mn@doi [\mnras]
  {10.1093/mnras/271.2.257}, \href
  {http://adsabs.harvard.edu/abs/1994MNRAS.271..257K} {271, 257}

\bibitem[\protect\citeauthoryear{{Kisielius}, {Storey}, {Davey}  \&
  {Neale}}{{Kisielius} et~al.}{1998}]{kisielius1998}
{Kisielius} R.,  {Storey} P.~J.,  {Davey} A.~R.,   {Neale} L.~T.,  1998,
  \mn@doi [\aaps] {10.1051/aas:1998319}, \href
  {http://adsabs.harvard.edu/abs/1998A%26AS..133..257K} {133, 257}

\bibitem[\protect\citeauthoryear{{Landi}, {Del Zanna}, {Young}, {Dere}  \&
  {Mason}}{{Landi} et~al.}{2012}]{landi2012}
{Landi} E.,  {Del Zanna} G.,  {Young} P.~R.,  {Dere} K.~P.,   {Mason} H.~E.,
  2012, \mn@doi [\apj] {10.1088/0004-637X/744/2/99}, \href
  {http://adsabs.harvard.edu/abs/2012ApJ...744...99L} {744, 99}

\bibitem[\protect\citeauthoryear{{Lau}, {De Marco}  \& {Liu}}{{Lau}
  et~al.}{2011}]{lau2011}
{Lau} H.~H.~B.,  {De Marco} O.,   {Liu} X.-W.,  2011, \mn@doi [\mnras]
  {10.1111/j.1365-2966.2010.17568.x}, \href
  {http://adsabs.harvard.edu/abs/2011MNRAS.410.1870L} {410, 1870}

\bibitem[\protect\citeauthoryear{{Liu}}{{Liu}}{2006}]{liu2006b}
{Liu} X.-W.,  2006, in {Barlow} M.~J.,  {M{\'e}ndez} R.~H.,  eds,  IAU
  Symposium Vol. 234, Planetary Nebulae in our Galaxy and Beyond. pp 219--226
  (\mn@eprint {} {astro-ph/0605082}), \mn@doi{10.1017/S1743921306003000}

\bibitem[\protect\citeauthoryear{{Liu}, {Storey}, {Barlow}  \& {Clegg}}{{Liu}
  et~al.}{1995}]{liu1995}
{Liu} X.-W.,  {Storey} P.~J.,  {Barlow} M.~J.,   {Clegg} R.~E.~S.,  1995,
  \mn@doi [\mnras] {10.1093/mnras/272.2.369}, \href
  {http://adsabs.harvard.edu/abs/1995MNRAS.272..369L} {272, 369}

\bibitem[\protect\citeauthoryear{{Liu}, {Storey}, {Barlow}, {Danziger}, {Cohen}
   \& {Bryce}}{{Liu} et~al.}{2000}]{liu2000}
{Liu} X.-W.,  {Storey} P.~J.,  {Barlow} M.~J.,  {Danziger} I.~J.,  {Cohen} M.,
   {Bryce} M.,  2000, \mn@doi [\mnras] {10.1046/j.1365-8711.2000.03167.x},
  \href {http://adsabs.harvard.edu/abs/2000MNRAS.312..585L} {312, 585}

\bibitem[\protect\citeauthoryear{{Liu}, {Barlow}, {Zhang}, {Bastin}  \&
  {Storey}}{{Liu} et~al.}{2006a}]{liu2006a}
{Liu} X.-W.,  {Barlow} M.~J.,  {Zhang} Y.,  {Bastin} R.~J.,   {Storey} P.~J.,
  2006a, \mn@doi [\mnras] {10.1111/j.1365-2966.2006.10283.x}, \href
  {http://adsabs.harvard.edu/abs/2006MNRAS.368.1959L} {368, 1959}

\bibitem[\protect\citeauthoryear{{Liu}, {Leggett}, {Golimowski}, {Chiu}, {Fan},
  {Geballe}, {Schneider}  \& {Brinkmann}}{{Liu} et~al.}{2006b}]{liu2006}
{Liu} M.~C.,  {Leggett} S.~K.,  {Golimowski} D.~A.,  {Chiu} K.,  {Fan} X.,
  {Geballe} T.~R.,  {Schneider} D.~P.,   {Brinkmann} J.,  2006b, \mn@doi [\apj]
  {10.1086/505561}, \href {http://adsabs.harvard.edu/abs/2006ApJ...647.1393L}
  {647, 1393}

\bibitem[\protect\citeauthoryear{{Lopez}, {Meaburn}  \& {Palmer}}{{Lopez}
  et~al.}{1993}]{lopez1993a}
{Lopez} J.~A.,  {Meaburn} J.,   {Palmer} J.~W.,  1993, \mn@doi [\apjl]
  {10.1086/187051}, \href {http://adsabs.harvard.edu/abs/1993ApJ...415L.135L}
  {415, L135}

\bibitem[\protect\citeauthoryear{{Manick}, {Miszalski}  \& {McBride}}{{Manick}
  et~al.}{2015}]{manick2015}
{Manick} R.,  {Miszalski} B.,   {McBride} V.,  2015, \mn@doi [\mnras]
  {10.1093/mnras/stv074}, \href
  {http://adsabs.harvard.edu/abs/2015MNRAS.448.1789M} {448, 1789}

\bibitem[\protect\citeauthoryear{{McNabb}, {Fang}  \& {Liu}}{{McNabb}
  et~al.}{2016}]{mcnabb2016}
{McNabb} I.~A.,  {Fang} X.,   {Liu} X.-W.,  2016, \mn@doi [\mnras]
  {10.1093/mnras/stw1405}, \href
  {http://adsabs.harvard.edu/abs/2016MNRAS.461.2818M} {461, 2818}

\bibitem[\protect\citeauthoryear{{Mendez} \& {Niemela}}{{Mendez} \&
  {Niemela}}{1981}]{mendez1981}
{Mendez} R.~H.,  {Niemela} V.~S.,  1981, \mn@doi [\apj] {10.1086/159368}, \href
  {http://adsabs.harvard.edu/abs/1981ApJ...250..240M} {250, 240}

\bibitem[\protect\citeauthoryear{{Mendoza} \& {Zeippen}}{{Mendoza} \&
  {Zeippen}}{1982}]{mendoza1982a}
{Mendoza} C.,  {Zeippen} C.~J.,  1982, \mn@doi [\mnras]
  {10.1093/mnras/199.4.1025}, \href
  {http://adsabs.harvard.edu/abs/1982MNRAS.199.1025M} {199, 1025}

\bibitem[\protect\citeauthoryear{{Mendoza} \& {Zeippen}}{{Mendoza} \&
  {Zeippen}}{1983}]{mendoza1983}
{Mendoza} C.,  {Zeippen} C.~J.,  1983, \mn@doi [\mnras]
  {10.1093/mnras/202.4.981}, \href
  {http://adsabs.harvard.edu/abs/1983MNRAS.202..981M} {202, 981}

\bibitem[\protect\citeauthoryear{{Mesa-Delgado}, {Esteban},
  {Garc{\'{\i}}a-Rojas}, {Reyes-P{\'e}rez}, {Morisset}  \&
  {Bresolin}}{{Mesa-Delgado} et~al.}{2014}]{mesa-delgado2014}
{Mesa-Delgado} A.,  {Esteban} C.,  {Garc{\'{\i}}a-Rojas} J.,  {Reyes-P{\'e}rez}
  J.,  {Morisset} C.,   {Bresolin} F.,  2014, \mn@doi [\apj]
  {10.1088/0004-637X/785/2/100}, \href
  {http://adsabs.harvard.edu/abs/2014ApJ...785..100M} {785, 100}

\bibitem[\protect\citeauthoryear{{Miszalski}}{{Miszalski}}{2009}]{miszalski2009}
{Miszalski} B.,  2009, PhD thesis, Department of Physics, Macquarie University,
  NSW 2109, Australia

\bibitem[\protect\citeauthoryear{{Miszalski}, {Acker}, {Moffat}, {Parker}  \&
  {Udalski}}{{Miszalski} et~al.}{2009a}]{miszalski2009b}
{Miszalski} B.,  {Acker} A.,  {Moffat} A.~F.~J.,  {Parker} Q.~A.,   {Udalski}
  A.,  2009a, \mn@doi [\aap] {10.1051/0004-6361/200811380}, \href
  {http://adsabs.harvard.edu/abs/2009A%26A...496..813M} {496, 813}

\bibitem[\protect\citeauthoryear{{Miszalski}, {Acker}, {Parker}  \&
  {Moffat}}{{Miszalski} et~al.}{2009b}]{miszalski2009a}
{Miszalski} B.,  {Acker} A.,  {Parker} Q.~A.,   {Moffat} A.~F.~J.,  2009b,
  \mn@doi [\aap] {10.1051/0004-6361/200912176}, \href
  {http://adsabs.harvard.edu/abs/2009A%26A...505..249M} {505, 249}

\bibitem[\protect\citeauthoryear{{Miszalski}, {Acker}, {Parker}, {Boffin},
  {Frew}, {Mikolajewska}, {Moffat}  \& {Napiwotzki}}{{Miszalski}
  et~al.}{2011a}]{miszalski2011a}
{Miszalski} B.,  {Acker} A.,  {Parker} Q.~A.,  {Boffin} H.~M.~J.,  {Frew}
  D.~J.,  {Mikolajewska} J.,  {Moffat} A.~F.~J.,   {Napiwotzki} R.,  2011a, in
  Asymmetric Planetary Nebulae 5 Conference.

\bibitem[\protect\citeauthoryear{{Miszalski}, {Jones}, {Rodr{\'{\i}}guez-Gil},
  {Boffin}, {Corradi}  \& {Santander-Garc{\'{\i}}a}}{{Miszalski}
  et~al.}{2011b}]{miszalski2011b}
{Miszalski} B.,  {Jones} D.,  {Rodr{\'{\i}}guez-Gil} P.,  {Boffin} H.~M.~J.,
  {Corradi} R.~L.~M.,   {Santander-Garc{\'{\i}}a} M.,  2011b, \mn@doi [\aap]
  {10.1051/0004-6361/201117084}, \href
  {http://adsabs.harvard.edu/abs/2011A%26A...531A.158M} {531, A158}

\bibitem[\protect\citeauthoryear{{Miszalski}, {Manick}, {Miko{\l}ajewska}, {Van
  Winckel}  \& {I{\l}kiewicz}}{{Miszalski} et~al.}{2018}]{miszalski2018}
{Miszalski} B.,  {Manick} R.,  {Miko{\l}ajewska} J.,  {Van Winckel} H.,
  {I{\l}kiewicz} K.,  2018, \mn@doi [\pasa] {10.1017/pasa.2018.23}, \href
  {http://adsabs.harvard.edu/abs/2018PASA...35...27M} {35, e027}

\bibitem[\protect\citeauthoryear{{Nicholls}, {Dopita}  \&
  {Sutherland}}{{Nicholls} et~al.}{2012}]{nicholls2012}
{Nicholls} D.~C.,  {Dopita} M.~A.,   {Sutherland} R.~S.,  2012, \mn@doi [\apj]
  {10.1088/0004-637X/752/2/148}, \href
  {http://adsabs.harvard.edu/abs/2012ApJ...752..148N} {752, 148}

\bibitem[\protect\citeauthoryear{{Ohlmann}, {R{\"o}pke}, {Pakmor}  \&
  {Springel}}{{Ohlmann} et~al.}{2016}]{ohlmann2016}
{Ohlmann} S.~T.,  {R{\"o}pke} F.~K.,  {Pakmor} R.,   {Springel} V.,  2016,
  \mn@doi [\apjl] {10.3847/2041-8205/816/1/L9}, \href
  {http://adsabs.harvard.edu/abs/2016ApJ...816L...9O} {816, L9}

\bibitem[\protect\citeauthoryear{{Palmer}, {Lopez}, {Meaburn}  \&
  {Lloyd}}{{Palmer} et~al.}{1996}]{palmer1996}
{Palmer} J.~W.,  {Lopez} J.~A.,  {Meaburn} J.,   {Lloyd} H.~M.,  1996, \aap,
  \href {http://adsabs.harvard.edu/abs/1996A%26A...307..225P} {307, 225}

\bibitem[\protect\citeauthoryear{{Parker}, {Boji{\v c}i{\'c}}  \&
  {Frew}}{{Parker} et~al.}{2016}]{parker2016}
{Parker} Q.~A.,  {Boji{\v c}i{\'c}} I.~S.,   {Frew} D.~J.,  2016, in Journal of
  Physics Conference Series. p. 032008 (\mn@eprint {arXiv} {1603.07042}),
  \mn@doi{10.1088/1742-6596/728/3/032008}

\bibitem[\protect\citeauthoryear{{Peimbert}}{{Peimbert}}{1967}]{peimbert1967}
{Peimbert} M.,  1967, \mn@doi [\apj] {10.1086/149385}, \href
  {http://adsabs.harvard.edu/abs/1967ApJ...150..825P} {150, 825}

\bibitem[\protect\citeauthoryear{{Peimbert} \& {Peimbert}}{{Peimbert} \&
  {Peimbert}}{2013}]{peimbert2013}
{Peimbert} A.,  {Peimbert} M.,  2013, \mn@doi [\apj]
  {10.1088/0004-637X/778/2/89}, \href
  {http://adsabs.harvard.edu/abs/2013ApJ...778...89P} {778, 89}

\bibitem[\protect\citeauthoryear{{Peimbert} \& {Torres-Peimbert}}{{Peimbert} \&
  {Torres-Peimbert}}{1983}]{peimbert1983}
{Peimbert} M.,  {Torres-Peimbert} S.,  1983, in {Flower} D.~R.,  ed.,  IAU
  Symposium Vol. 103, Planetary Nebulae. pp 233--241

\bibitem[\protect\citeauthoryear{{Pequignot}, {Petitjean}  \&
  {Boisson}}{{Pequignot} et~al.}{1991}]{pequignot1991}
{Pequignot} D.,  {Petitjean} P.,   {Boisson} C.,  1991, \aap, \href
  {http://adsabs.harvard.edu/abs/1991A%26A...251..680P} {251, 680}

\bibitem[\protect\citeauthoryear{{Pollacco} \& {Bell}}{{Pollacco} \&
  {Bell}}{1994}]{pollacco1994}
{Pollacco} D.~L.,  {Bell} S.~A.,  1994, \mn@doi [\mnras]
  {10.1093/mnras/267.2.452}, \href
  {http://adsabs.harvard.edu/abs/1994MNRAS.267..452P} {267, 452}

\bibitem[\protect\citeauthoryear{{Porter}, {Ferland}, {Storey}  \&
  {Detisch}}{{Porter} et~al.}{2012}]{porter2012}
{Porter} R.~L.,  {Ferland} G.~J.,  {Storey} P.~J.,   {Detisch} M.~J.,  2012,
  \mn@doi [\mnras] {10.1111/j.1745-3933.2012.01300.x}, \href
  {http://adsabs.harvard.edu/abs/2012MNRAS.425L..28P} {425, L28}

\bibitem[\protect\citeauthoryear{{Porter}, {Ferland}, {Storey}  \&
  {Detisch}}{{Porter} et~al.}{2013}]{porter2013}
{Porter} R.~L.,  {Ferland} G.~J.,  {Storey} P.~J.,   {Detisch} M.~J.,  2013,
  \mn@doi [\mnras] {10.1093/mnrasl/slt049}, \href
  {http://adsabs.harvard.edu/abs/2013MNRAS.433L..89P} {433, L89}

\bibitem[\protect\citeauthoryear{{Pradhan}}{{Pradhan}}{1976}]{pradhan1976}
{Pradhan} A.~K.,  1976, \mn@doi [\mnras] {10.1093/mnras/177.1.31}, \href
  {http://adsabs.harvard.edu/abs/1976MNRAS.177...31P} {177, 31}

\bibitem[\protect\citeauthoryear{{Robertson-Tessi} \&
  {Garnett}}{{Robertson-Tessi} \& {Garnett}}{2005}]{robertsontessi2005}
{Robertson-Tessi} M.,  {Garnett} D.~R.,  2005, \mn@doi [\apjs]
  {10.1086/427905}, \href {http://adsabs.harvard.edu/abs/2005ApJS..157..371R}
  {157, 371}

\bibitem[\protect\citeauthoryear{{Sarazin}, {Melnick}, {Navarrete}  \&
  {Lombardi}}{{Sarazin} et~al.}{2008}]{sarazin2008}
{Sarazin} M.,  {Melnick} J.,  {Navarrete} J.,   {Lombardi} G.,  2008, The
  Messenger, \href {http://adsabs.harvard.edu/abs/2008Msngr.132...11S} {132,
  11}

\bibitem[\protect\citeauthoryear{{Sharpee}, {Williams}, {Baldwin}  \& {van
  Hoof}}{{Sharpee} et~al.}{2003}]{sharpee2003}
{Sharpee} B.,  {Williams} R.,  {Baldwin} J.~A.,   {van Hoof} P.~A.~M.,  2003,
  \mn@doi [\apjs] {10.1086/378321}, \href
  {http://adsabs.harvard.edu/abs/2003ApJS..149..157S} {149, 157}

\bibitem[\protect\citeauthoryear{{Shortridge}, {Meyerdierks}, {Currie},
  {Davenhall}, {Jenness}  \& {Clayton}}{{Shortridge}
  et~al.}{2014}]{shortridge2014}
{Shortridge} K.,  {Meyerdierks} H.,  {Currie} M.~J.,  {Davenhall} C.,
  {Jenness} T.,   {Clayton} M.,  2014, {Starlink Figaro: Starlink version of
  the Figaro data reduction software package}, Astrophysics Source Code Library
  (\mn@eprint {ascl} {1411.022})

\bibitem[\protect\citeauthoryear{{Soker}}{{Soker}}{2015}]{soker2015}
{Soker} N.,  2015, \mn@doi [\apj] {10.1088/0004-637X/800/2/114}, \href
  {http://adsabs.harvard.edu/abs/2015ApJ...800..114S} {800, 114}

\bibitem[\protect\citeauthoryear{{Sowicka}, {Jones}, {Corradi}, {Wesson},
  {Garc{\'{\i}}a-Rojas}, {Santander-Garc{\'{\i}}a}, {Boffin}  \&
  {Rodr{\'{\i}}guez-Gil}}{{Sowicka} et~al.}{2017}]{sowicka2017}
{Sowicka} P.,  {Jones} D.,  {Corradi} R.~L.~M.,  {Wesson} R.,
  {Garc{\'{\i}}a-Rojas} J.,  {Santander-Garc{\'{\i}}a} M.,  {Boffin} H.~M.~J.,
   {Rodr{\'{\i}}guez-Gil} P.,  2017, \mn@doi [\mnras] {10.1093/mnras/stx1697},
  \href {http://adsabs.harvard.edu/abs/2017MNRAS.471.3529S} {471, 3529}

\bibitem[\protect\citeauthoryear{{Storey}}{{Storey}}{1994}]{storey1994}
{Storey} P.~J.,  1994, \aap, \href
  {http://adsabs.harvard.edu/abs/1994A%26A...282..999S} {282, 999}

\bibitem[\protect\citeauthoryear{{Storey} \& {Hummer}}{{Storey} \&
  {Hummer}}{1995}]{storey1995}
{Storey} P.~J.,  {Hummer} D.~G.,  1995, \mn@doi [\mnras]
  {10.1093/mnras/272.1.41}, \href
  {http://adsabs.harvard.edu/abs/1995MNRAS.272...41S} {272, 41}

\bibitem[\protect\citeauthoryear{{Storey}, {Sochi}  \& {Bastin}}{{Storey}
  et~al.}{2017}]{storey2017}
{Storey} P.~J.,  {Sochi} T.,   {Bastin} R.,  2017, \mn@doi [\mnras]
  {10.1093/mnras/stx1189}, \href
  {http://adsabs.harvard.edu/abs/2017MNRAS.470..379S} {470, 379}

\bibitem[\protect\citeauthoryear{{Todt} et~al.,}{{Todt}
  et~al.}{2015}]{todt2015}
{Todt} H.,  et~al., 2015, in {Dufour} P.,  {Bergeron} P.,   {Fontaine} G.,
  eds,  Astronomical Society of the Pacific Conference Series Vol. 493, 19th
  European Workshop on White Dwarfs. p.~539

\bibitem[\protect\citeauthoryear{{Toribio San Cipriano},
  {Dom{\'{\i}}nguez-Guzm{\'a}n}, {Esteban}, {Garc{\'{\i}}a-Rojas},
  {Mesa-Delgado}, {Bresolin}, {Rodr{\'{\i}}guez}  \&
  {Sim{\'o}n-D{\'{\i}}az}}{{Toribio San Cipriano}
  et~al.}{2017}]{toribiosancipriano2017}
{Toribio San Cipriano} L.,  {Dom{\'{\i}}nguez-Guzm{\'a}n} G.,  {Esteban} C.,
  {Garc{\'{\i}}a-Rojas} J.,  {Mesa-Delgado} A.,  {Bresolin} F.,
  {Rodr{\'{\i}}guez} M.,   {Sim{\'o}n-D{\'{\i}}az} S.,  2017, \mn@doi [\mnras]
  {10.1093/mnras/stx328}, \href
  {http://adsabs.harvard.edu/abs/2017MNRAS.467.3759T} {467, 3759}

\bibitem[\protect\citeauthoryear{{Torres-Peimbert}, {Peimbert}  \&
  {Pena}}{{Torres-Peimbert} et~al.}{1990}]{torres-peimbert1990}
{Torres-Peimbert} S.,  {Peimbert} M.,   {Pena} M.,  1990, \aap, \href
  {http://adsabs.harvard.edu/abs/1990A%26A...233..540T} {233, 540}

\bibitem[\protect\citeauthoryear{{Tsamis}, {Barlow}, {Liu}, {Danziger}  \&
  {Storey}}{{Tsamis} et~al.}{2003}]{tsamis2003}
{Tsamis} Y.~G.,  {Barlow} M.~J.,  {Liu} X.-W.,  {Danziger} I.~J.,   {Storey}
  P.~J.,  2003, \mn@doi [\mnras] {10.1046/j.1365-8711.2003.06081.x}, \href
  {http://adsabs.harvard.edu/abs/2003MNRAS.338..687T} {338, 687}

\bibitem[\protect\citeauthoryear{{Tsamis}, {Barlow}, {Liu}, {Storey}  \&
  {Danziger}}{{Tsamis} et~al.}{2004}]{tsamis2004}
{Tsamis} Y.~G.,  {Barlow} M.~J.,  {Liu} X.-W.,  {Storey} P.~J.,   {Danziger}
  I.~J.,  2004, \mn@doi [\mnras] {10.1111/j.1365-2966.2004.08140.x}, \href
  {http://adsabs.harvard.edu/abs/2004MNRAS.353..953T} {353, 953}

\bibitem[\protect\citeauthoryear{{Viegas} \& {Clegg}}{{Viegas} \&
  {Clegg}}{1994}]{viegas1994}
{Viegas} S.~M.,  {Clegg} R.~E.~S.,  1994, \mn@doi [\mnras]
  {10.1093/mnras/271.4.993}, \href
  {http://adsabs.harvard.edu/abs/1994MNRAS.271..993V} {271, 993}

\bibitem[\protect\citeauthoryear{{Wang}, {Liu}, {Zhang}  \& {Barlow}}{{Wang}
  et~al.}{2004}]{wang2004}
{Wang} W.,  {Liu} X.-W.,  {Zhang} Y.,   {Barlow} M.~J.,  2004, \mn@doi [\aap]
  {10.1051/0004-6361:20041470}, \href
  {http://adsabs.harvard.edu/abs/2004A%26A...427..873W} {427, 873}

\bibitem[\protect\citeauthoryear{{Warren-Smith}, {Draper}, {Taylor}  \&
  {Allan}}{{Warren-Smith} et~al.}{2014}]{warren-smith2014}
{Warren-Smith} R.~F.,  {Draper} P.~W.,  {Taylor} M.,   {Allan} A.,  2014,
  {CCDPACK: CCD Data Reduction Package}, Astrophysics Source Code Library
  (\mn@eprint {ascl} {1403.021})

\bibitem[\protect\citeauthoryear{{Weidmann} \& {Gamen}}{{Weidmann} \&
  {Gamen}}{2011}]{weidmann2011}
{Weidmann} W.~A.,  {Gamen} R.,  2011, \mn@doi [\aap]
  {10.1051/0004-6361/200913984}, \href
  {http://adsabs.harvard.edu/abs/2011A%26A...526A...6W} {526, A6}

\bibitem[\protect\citeauthoryear{{Wesson}}{{Wesson}}{2016}]{wesson2016a}
{Wesson} R.,  2016, \mn@doi [\mnras] {10.1093/mnras/stv2946}, \href
  {http://adsabs.harvard.edu/abs/2016MNRAS.456.3774W} {456, 3774}

\bibitem[\protect\citeauthoryear{{Wesson}, {Liu}  \& {Barlow}}{{Wesson}
  et~al.}{2003}]{wesson2003}
{Wesson} R.,  {Liu} X.-W.,   {Barlow} M.~J.,  2003, \mn@doi [\mnras]
  {10.1046/j.1365-8711.2003.06289.x}, \href
  {http://adsabs.harvard.edu/abs/2003MNRAS.340..253W} {340, 253}

\bibitem[\protect\citeauthoryear{{Wesson}, {Liu}  \& {Barlow}}{{Wesson}
  et~al.}{2005}]{wesson2005}
{Wesson} R.,  {Liu} X.-W.,   {Barlow} M.~J.,  2005, \mn@doi [\mnras]
  {10.1111/j.1365-2966.2005.09325.x}, \href
  {http://adsabs.harvard.edu/abs/2005MNRAS.362..424W} {362, 424}

\bibitem[\protect\citeauthoryear{{Wesson}, {Barlow}, {Liu}, {Storey},
  {Ercolano}  \& {De Marco}}{{Wesson} et~al.}{2008}]{wesson2008a}
{Wesson} R.,  {Barlow} M.~J.,  {Liu} X.-W.,  {Storey} P.~J.,  {Ercolano} B.,
  {De Marco} O.,  2008, \mn@doi [\mnras] {10.1111/j.1365-2966.2007.12683.x},
  \href {http://adsabs.harvard.edu/abs/2008MNRAS.383.1639W} {383, 1639}

\bibitem[\protect\citeauthoryear{{Wesson}, {Stock}  \& {Scicluna}}{{Wesson}
  et~al.}{2012}]{wesson2012}
{Wesson} R.,  {Stock} D.~J.,   {Scicluna} P.,  2012, \mn@doi [\mnras]
  {10.1111/j.1365-2966.2012.20863.x}, \href
  {http://adsabs.harvard.edu/abs/2012MNRAS.422.3516W} {422, 3516}

\bibitem[\protect\citeauthoryear{{Wyse}}{{Wyse}}{1942}]{wyse1942}
{Wyse} A.~B.,  1942, \mn@doi [\apj] {10.1086/144409}, \href
  {http://adsabs.harvard.edu/abs/1942ApJ....95..356W} {95, 356}

\bibitem[\protect\citeauthoryear{{Zeippen}}{{Zeippen}}{1982}]{zeippen1982}
{Zeippen} C.~J.,  1982, \mn@doi [\mnras] {10.1093/mnras/198.1.111}, \href
  {http://adsabs.harvard.edu/abs/1982MNRAS.198..111Z} {198, 111}

\bibitem[\protect\citeauthoryear{{Zorotovic}, {Schreiber}, {G{\"a}nsicke}  \&
  {Nebot G{\'o}mez-Mor{\'a}n}}{{Zorotovic} et~al.}{2010}]{zorotovic2010}
{Zorotovic} M.,  {Schreiber} M.~R.,  {G{\"a}nsicke} B.~T.,   {Nebot
  G{\'o}mez-Mor{\'a}n} A.,  2010, \mn@doi [\aap] {10.1051/0004-6361/200913658},
  \href {https://ui.adsabs.harvard.edu/#abs/2010A&A...520A..86Z} {520}

\bibitem[\protect\citeauthoryear{{van Dokkum}}{{van
  Dokkum}}{2001}]{vandokkum2001}
{van Dokkum} P.~G.,  2001, \mn@doi [\pasp] {10.1086/323894}, \href
  {http://adsabs.harvard.edu/abs/2001PASP..113.1420V} {113, 1420}

\makeatother
\end{thebibliography}



\appendix

\section{Observing logs}
\label{sec:tables}

\begin{table*}
	\centering
	\caption{Observing log}
	\label{tab:obs}
	\begin{tabular}{llllll} 
		\hline
		PN & Position & Grism & Exposure & Modified & Airmass\\
		 & angle ($\degr$) & & time (s)&Julian Date\\
		\hline
A~41                   &    	20             &         	1200B &                 	1200	 &56774.161 & 1.805\\
A~41                   &    	20             &         	1200R &                 	120 	 &56774.176 & 1.605\\
A~63                   &    	--35            &         	1200B &                 	1200	 &56822.300 & 1.344\\
A~63                   &    	--35            &         	1200R &                 	120 	 &56822.315 & 1.362\\
A~65                   &    	45             &         	1200B &                 	600 	 &56827.352 & 1.100\\
A~65                   &    	45             &         	1200R &                 	60  	 &56827.360 & 1.125\\
Bl~3-15                &    	0              &         	1200B &                 	1200	 &56817.032 & 2.093\\
Bl~3-15                &    	0              &         	1200R &                 	120 	 &56817.047 & 1.831\\
BMP~1800-3407               &    	0              &         	1200B &                 	1200	 &56772.246 & 1.220\\
BMP~1800-3407               &    	0              &         	1200R &                 	120 	 &56772.261 & 1.161\\
BMP~1801-2947               &    	0              &         	1200B &                 	1200	 &56772.347 & 1.005\\
BMP~1801-2907               &    	0              &         	1200R &                 	120 	 &56772.362 & 1.005\\
DS~1                   &    	--45            &          	1200B &                 	1250	 &57385.237 & 1.453\\
DS~1                   &    	--45            &          	600RI &                 		120 	 &57385.253 & 1.365\\
Fg~1  &  --14  &  1200B  &  300  &  57378.327  &  1.268 \\
Fg~1  &  --14  &  600RI  &  120  &  57378.324  &  1.284 \\
HaTr~4                 &    	0              &         	1200B &                 	1200	 &56731.319 & 1.255\\
HaTr~4                 &    	0              &         	1200R &                 	120 	 &56731.334 & 1.213\\
H~2-29                 &    	--30            &         	1200B &                 	1200	 &56818.172 & 1.052\\
H~2-29                 &    	--30            &         	1200R &                 	120 	 &56818.187 & 1.031\\
Hen~2-11               &   	25             &          	1200B &                 	2500	 &57312.336 & 1.456\\
Hen~2-11               &   	25             &          	600RI &                 		120 	 &57312.366 & 1.274\\
Hen~2-283              &    	0              &         	1200B &                 	1200	 &56789.303 & 1.014\\
Hen~2-283              &    	0              &         	1200R &                 	120 	 &56789.318 & 1.020\\
Hen~2-428              &    	15             &         	1200B &                 	1200	 &56822.269 & 1.312\\
Hen~2-428              &    	15             &         	1200R &                 	120 	 &56822.290 & 1.328\\
Hf~2-2                 &    	0              &         	1200B &                 	120	 &56827.298 & 1.090\\
Hf~2-2                 &    	0              &         	1200R &                 	120 	 &56827.313 & 1.135\\
JaSt~66                &    	0              &         	1200B &                 	1200	 &56806.368 & 1.250\\
JaSt~66                &    	0              &         	1200R &                 	120 	 &56806.383 & 1.337\\
K~1-2                  &   		--90            &          	1200B &                 	2500	 &57357.192 & 1.899\\
K~1-2                  &   		--90            &          	600RI &                 		120 	 &57357.222 & 1.522\\
K~6-34                 &    	--167          &         	1200B &                 	120	 &56827.257 & 1.062\\
K~6-34                 &    	--167          &         	1200R &                 	120 	 &56827.272 & 1.096\\
Lo~16                  &    	0              &         	1200B &                 	1800	 &56822.215 & 1.041\\
Lo~16                  &    	0              &         	1200R &                 	300 	 &56822.237 & 1.058\\
M~2-19                 &    	80             &         	1200B &                 	1200	 &56818.199 & 1.014\\
M~2-19                 &    	80             &         	1200B &                 	300 	 &56818.217 & 1.005\\
M~3-16                 &    	40             &         	1200B &                 	1500	 &56845.010 & 1.402\\
M~3-16                 &    	40             &         	1200R &                 	180 	 &56845.031 & 1.267\\
MPA~1759-3007               &    	0              &         	1200B &                 	1200	 &56770.377 & 1.010\\
MPA~1759-3007               &    	0              &         	1200R &                 	120 	 &56770.392 & 1.023\\
MPA~1508-6455               &    	0              &         	1200B &                 	1200	 &56735.355 & 1.316\\
MPA~1508-6455               &    	0              &         	1200R &                 	120 	 &56735.370 & 1.327\\
NGC~2346               &    	15             &          	1200B &                 	2500	 &57309.278 & 1.892\\
NGC~2346               &    	15             &          	600RI &                 		120 	 &57309.309 & 1.517\\
NGC~6026               &    	25             &         	1200B &                 	600 	 &56731.252 & 1.344\\
NGC~6026               &    	25             &         	1200R &                 	60  	 &56731.260 & 1.297\\
NGC~6326               &    	--45            &         	1200B &                 	600 	 &56735.383 & 1.152\\
NGC~6326               &    	--45            &         	1200R &                 	60  	 &56735.391 & 1.142\\
NGC~6337               &    	0              &         	1200B &                 	600 	 &56749.300 & 1.162\\
NGC~6337               &    	0              &         	1200R &                 	60  	 &56749.309 & 1.137\\
Pe~1-9                 &    	0              &         	1200B &                 	1200	 &56806.157 & 1.165\\
Pe~1-9                 &    	0              &         	1200R &                 	120 	 &56806.172 & 1.113\\
PHR~1744-3355               &    	0              &         	1200B &                 	1200	 &56774.193 & 1.432\\
PHR~1744-3355               &    	0              &         	1200R &                 	120 	 &56774.208 & 1.332\\
PHR~1756-3342               &    	0              &         	1200B &                 	1200	 &56770.257 & 1.185\\
PHR~1756-3342               &    	0              &         	1200R &                 	1200  &56770.272 & 1.133\\
PHR~1757-2824               &    	0              &         	1200B &                 	1200	 &56770.306 & 1.045\\
PHR~1757-2824               &    	0              &         	1200R &                 	120 	 &56770.321 & 1.023\\
PHR~1759-2915               &    	0              &         	1200B &                 	1200	 &56770.349 & 1.005\\
PHR~1759-2915               &    	0              &         	1200R &                 	120 	 &56770.364 & 1.004\\
PHR~1801-2947               &    	0              &         	1200B &                 	1200	 &56772.370 & 1.005\\
PHR~1801-2947               &    	0              &         	1200R &                 	120 	 &56772.385 & 1.016\\
		\hline
	\end{tabular}
\end{table*}

\begin{table*}
	\centering
	\contcaption{Observing log}
	\begin{tabular}{llllll}
		\hline
		PN & Position & Grism & Exposure & Modified & Airmass\\
		 & angle ($\degr$) & & time (s)&Julian Date\\
		\hline
PHR~1804-2645               &    	0              &         	1200B &                 	1200	 &56772.397 & 1.027\\
PHR~1804-2645              &    	0              &         	1200R &                 	120 	 &56772.412 & 1.051\\
PM~1-23                &   	-55            &          	1200B &                 	2500	 &57297.346 & 1.291\\
PM~1-23                &   	-55            &          	600RI &                 		1200 	 &57297.376 & 1.153\\
PPA~1747-3435               &    	0              &         	1200B &                 	1200	 &56827.230 & 1.035\\
PPA~1747-3435               &    	0              &         	1200R &                 	120 	 &56827.245 & 1.055\\
PPA~1759-2834               &    	0              &         	1200B &                 	1200	 &56770.398 & 1.028\\
PPA~1759-2834               &    	0              &         	1200R &                 	120 	 &56770.413 & 1.052\\
Sab~41                 &    	30             &         	1200B &                 	1200	 &56792.329 & 1.041\\
Sab~41                 &    	30             &         	1200R &                 	120 	 &56792.344 & 1.063\\
Sp~1                   &    	0              &         	1200B &                 	600 	 &56740.371 & 1.126\\
Sp~1                   &    	0              &         	1200R &                 	60  	 &56740.379 & 1.132\\	
		\hline
	\end{tabular}
\end{table*}

\section{Spatially integrated line fluxes}
\begin{table*}
 \begin{tabular}{lrlrllllllll}
 \hline
 $ \lambda $ & Ion & $F \left( \lambda \right) $ && $I \left( \lambda \right) $ & Ion & Multiplet & Lower term & Upper term & g$_1$ & g$_2$ \\
 \hline
  3677.94 &   3676.36 &    2.663& $\pm$   0.656&   3.038& $^{  +0.926}_{  -0.986}$ & H~{\sc i}        & H22        & 2p+ 2P*    & 22d+ 2D    &          8 &        *    \\
  3723.23 &   3721.63 &    2.628& $\pm$   0.450&   3.581& $^{  +0.360}_{  -0.400}$ & [S~{\sc iii}]    & F2         & 3p2 3P     & 3p2 1S     &          3 &        1    \\
        * &   3721.94 &  *     &             & *     &         & H~{\sc i}        & H14        & 2p+ 2P*    & 14d+ 2D    &          8 &        *    \\
  3727.63 &   3726.03 &    4.488& $\pm$   0.801&   7.170& $^{  +1.210}_{  -1.450}$ & [O~{\sc ii}]     & F1         & 2p3 4S*    & 2p3 2D*    &          4 &        4    \\
  3730.42 &   3728.82 &    5.011& $\pm$   0.740&   6.260& $^{  +1.100}_{  -1.340}$ & [O~{\sc ii}]     & F1         & 2p3 4S*    & 2p3 2D*    &          4 &        6    \\
  3751.76 &   3750.15 &    4.208& $\pm$   0.786&   7.030& $^{  +1.180}_{  -1.420}$ & H~{\sc i}        & H12        & 2p+ 2P*    & 12d+ 2D    &          8 &        *    \\
  3761.48 &   3759.87 &    2.186& $\pm$   0.563&   3.055& $^{  +0.794}_{  -0.845}$ & O~{\sc iii}      & V2         & 3s 3P*     & 3p 3D      &          5 &        7    \\
\hline
 \end{tabular}
 \caption{Observed and dereddened line fluxes for Fg~1 (complete tables for all objects are available online)}
 \label{tab:python/Fg1.dat_lines_linelist}
 \end{table*}
\bsp	
\label{lastpage}
\end{document}